\documentclass[aps, prd, fleqn, nofootinbib, 12pt]{revtex4-2}

\usepackage{
	amsmath, 
	amssymb, 
	graphicx,
	hyperref, 
	float, 
	enumerate, 
	pgffor,
    longtable,
    xcolor,
	multirow,
	mathrsfs,
	bm,
	slashed,
    lineno,
    soul,
    pdfpages,
    breqn,
    courier,
    xparse,
    fvextra,
    array,
    pdflscape,
    adjustbox
}
\usepackage{booktabs, siunitx}

\hypersetup{
	colorlinks=true
}

\usepackage{tikz}

\def\beq{\begin{equation}}
\def\eeq{\end{equation}}
\def\beqa{\begin{eqnarray}}
\def\eeqa{\end{eqnarray}}

\def\pp#1{\left(#1\right)}

\def\cc#1{\left\{#1\right\}}

\def\CC#1{\Big\{#1\Big\}}

\def\Parr#1{\renewcommand{\arraystretch}{1.5}\left(\begin{array}{l}#1\end{array}\right)}

\def\nn{\nonumber \\ &}

\def\d#1{\mathop{{\rm d}#1}}

\def\vec#1{\bm{#1}}

\allowdisplaybreaks

\def\GeV{{\rm GeV}}
\def\TeV{{\rm TeV}}
\def\fb{{\rm fb}}

\DeclareMathAlphabet{\mymathbb}{U}{BOONDOX-ds}{m}{n}

\def\bar{\overline}
\def\tilde{\widetilde}

\DefineVerbatimEnvironment{verbatim}{Verbatim}{breaklines=true}

\def\smeftatnlo{{\small \sc SMEFT@NLO}}
\def\madgraph{{\small \sc MadGraph5\_aMC@NLO}}
\def\feynarts{{\small \sc FeynArts}}
\def\feyncalc{{\small \sc FeynCalc}}
\def\lhapdf{{\small \sc LHAPDF}}
\def\vegas{{\small \sc Vegas}}

\def\matrix{{\small \sc MATRIX}}

\usepackage{listings}
\lstdefinelanguage{mypython}{
  language=Python,
  keywords={def, return, if, elif, else, for, while, in, import, from, as, with, try, except, raise, class, self, None, True, False},
  keywordstyle=\color{blue}\bfseries,
  ndkeywords={int, float, str, list, dict, set, tuple},
  ndkeywordstyle=\color{magenta},
  identifierstyle=\color{black},
  comment=[l]{\#},
  commentstyle=\color{gray}\ttfamily,
  stringstyle=\color{orange},
  sensitive=true
}
\makeatletter
\AtBeginDocument{\let\LS@rot\@undefined}
\makeatother 

\begin{document}

\title{$t\bar t$ production as a probe of dimension-six SMEFT at higher orders}
\author{Nikolaos Kidonakis and Kaan \c{S}im\c{s}ek \\
\textsl{Department of Physics, Kennesaw State University, Kennesaw, Georgia 30144, USA}}

\begin{abstract}
	We study top-antitop pair production in proton-proton collisions within the Standard Model effective field theory (SMEFT) at dimension six. We focus on the top chromomagnetic operator $C_{tG}$ together with the four-quark operators relevant for unpolarized $t{\bar t}$ production. We analyze the top-quark single-differential transverse-momentum ($p_T$) and rapidity ($y$) distributions and the double-differential distributions in $p_T$ and $y$ ($p_T\times y$) at 13~TeV, and we present projections for 13.6~TeV using the same binning. Our highest-order setup combines next-to-next-to-leading-order (NNLO) Standard Model (SM) predictions with approximate NNLO SM-SMEFT interference corrections. We find that higher-order QCD effects are essential for a stable SMEFT interpretation, yielding substantially more robust bounds and milder parameter degeneracies. The strongest sensitivity is obtained for $C_{tG}$. In the combined highest-order analysis of the 13-TeV results and 13.6-TeV projections, we obtain sensitivity to effective scales up to $3.9~\TeV$. Our results show that higher-order differential top-pair production provides a precise and theoretically controlled probe of the top chromomagnetic interaction.
\end{abstract}

\maketitle

\section{Introduction\label{sec:intro}} 
The Standard Model effective field theory (SMEFT) provides a systematic, model-independent framework for parametrizing heavy new physics through higher-dimensional operators built from the Standard Model (SM) fields. A complete nonredundant basis of dimension-six operators, the Warsaw basis, was formulated in Ref.~\cite{Grzadkowski:2010es}, with equivalent alternative organizations developed subsequently~\cite{Pomarol:2013zra, Falkowski:2015fla}. The operator program has been extended to dimension eight and beyond~\cite{Murphy:2020rsh, Li:2020gnx, Harlander:2023psl, Henning:2015daa, Henning:2015alf, Graf:2020yxt, Graf:2022rco}, while the practical toolkit has matured through automated implementations of Feynman rules~\cite{Dedes:2017zog, Dedes:2019uzs, Dedes:2023zws, Brivio:2017btx, Brivio:2020onw}, extensive work on renormalization and operator mixing~\cite{Jenkins:2013zja, Jenkins:2013wua, Alonso:2013hga}, positivity bounds~\cite{Baker:2019sli, Zhang:2020jyn, deRham:2022hpx}, and very recently the complete one-loop threshold corrections at dimension six~\cite{Biekotter:2026dlb}. For a broad overview, see Ref.~\cite{Brivio:2017vri, Isidori:2023pyp, Aebischer:2025qhh}. 
\par 
On the phenomenological side, SMEFT analyses have progressed from sector-specific studies of electroweak precision data, namely Higgs, and gauge-boson observables, and low-energy probes~\cite{Han:2004az, Cirigliano:2012ab, Chen:2013kfa, Ellis:2014dva, Wells:2014pga, Hartmann:2016pil, Cirigliano:2016nyn, Falkowski:2017pss, Boughezal:2022pmb, Bissolotti:2023vdw, Bissolotti:2023pjh, Petriello:2025lur} toward global fits combining Higgs, electroweak, and top-quark information across many Wilson coefficients~\cite{deBlas:2016ojx, Hartland:2019bjb, Biekotter:2018ohn, Grojean:2018dqj, Baglio:2020oqu, Boughezal:2020uwq, Boughezal:2020klp, Ethier:2021ydt, Carrazza:2019sec, Ellis:2018gqa, Dawson:2018dxp}. Effects beyond the leading dimension-six level have also received growing attention~\cite{Alioli:2020kez, Boughezal:2021tih, Cohen:2020qvb, Li:2022rag, Dawson:2022cmu, Boughezal:2022nof, Allwicher:2022gkm, Degrande:2023iob, Boughezal:2023nhe, ElFaham:2025fow, Petriello:2025lur}, and the SMEFT is now routinely employed by experiments in direct effective-field-theory (EFT) interpretations of LHC measurements~\cite{CMS:2022ubq}. For realistic phenomenology, however, the central limitation is often theoretical precision, namely meaningful constraints on SMEFT parameters require predictions consistently and with sufficient perturbative accuracy. This has driven substantial progress in next-to-leading-order (NLO) SMEFT calculations~\cite{Grober:2015cwa, Zhang:2016omx, Passarino:2016pzb, Englert:2019rga, Dawson:2024pft, Bellafronte:2025jbk, Bellafronte:2026mhp}, and automated NLO QCD calculations can be performed in \madgraph~\cite{Alwall:2014hca} using the \smeftatnlo~implementation~\cite{Degrande:2020evl}.
\par 
Top-quark processes provide a particularly important arena for such studies. In particular, top-pair production at the LHC is both experimentally well measured and theoretically well developed, making it a natural process in which to test the interplay between perturbative QCD control and SMEFT sensitivity. Higher-order corrections from soft-gluon contributions can be calculated via the framework of threshold resummation. We use the resummation formalism developed in \cite{Kidonakis:1996aq,Kidonakis:1997gm,Kidonakis:2000ui,Kidonakis:2009ev,Kidonakis:2010dk}, which has been applied to top-quark anomalous couplings in \cite{Belyaev:2001hf,Kidonakis:2003sc,Kidonakis:2014dua,Kidonakis:2017mfy,Kidonakis:2018ybz,Forslund:2018qcp,Guzzi:2019ucs} and, in particular, to top-quark processes in the SMEFT in \cite{Kidonakis:2023htm,Kidonakis:2026fle}. 

The top-quark chromomagnetic operator is one of the most prominent dimension-six directions affecting $t\bar t$ production, and approximate next-to-next-to-leading order (aNNLO) results for this operator in top-pair production were presented in Ref.~\cite{Kidonakis:2023htm}. In the present work, we use higher-order calculations for differential $t\bar t$ production to constrain dimension-six SMEFT effects through the single-differential top-quark $p_T$ and $y$ distributions and the double-differential $p_T\times y$ distribution. Rather than attempting the most comprehensive global determination of all relevant top-sector operators, our goal is to establish perturbative-QCD control in this channel and to quantify how higher-order corrections, theoretical uncertainties, and correlations with additional four-quark directions affect the extraction of the chromomagnetic coefficient.
\par 
Our analysis is performed in a four-parameter setup. In addition to $C_{tG}$, we include three combinations of four-quark operators, denoted by $C_u^+$, $C_d^+$, and $C_b^+$, which arise as the relevant linear combinations in the \smeftatnlo\ top basis for unpolarized top-pair production. These directions are not introduced because the present dataset is expected to provide competitive stand-alone bounds on each of them separately, but because they furnish the natural correlated directions against which the robustness of the extracted $C_{tG}$ constraint should be tested. In this sense, the main phenomenological target of the paper is the chromomagnetic interaction, while the additional four-quark directions are included to assess the extent to which flat or weakly constrained combinations can affect its determination.
\par 
To this end, we analyze the available 13-TeV differential measurements and also construct projections for 13.6~TeV using the same binning. Our highest-order predictions combine next-to-next-to-leading order (NNLO) SM results with aNNLO SM-SMEFT interference corrections, allowing us to examine directly how the fitted Wilson-coefficient sensitivity changes from leading order (LO) to NLO and then to our highest-order setup. This comparison is particularly important in view of the fact that, at insufficient perturbative order, a fit can partially absorb missing SM QCD corrections into apparent SMEFT effects. A central question of this work is therefore not only how strong the resulting bound on $C_{tG}$ can become, but also how stable and interpretable that bound remains once additional operator directions and realistic uncertainty estimates are included.
\par 
This paper is organized as follows. In Sec.~\ref{sec:analytical}, we describe $t\bar t$ production at the LHC within the SMEFT framework and define the set of dimension-six operators and Wilson coefficients relevant to our analysis. In Sec.~\ref{sec:resum}, we briefly describe the formalism for the calculation of the soft-gluon corrections used to construct the aNNLO predictions. In Sec.~\ref{sec:numerical}, we present the numerical setup, the SM and SMEFT distributions, and our statistical treatment and uncertainty modeling. In Sec.~\ref{sec:results}, we present the SMEFT fit results. We conclude in Sec.~\ref{sec:conclusion}.

\section{$t\bar t$ production at the LHC within the SMEFT\label{sec:analytical}} 
The SMEFT provides a model-independent description of possible new-physics effects beyond the SM. One augments the SM with higher-dimensional operators $O_k^{(n)}$ of mass dimension $n>4$, constructed from the SM fields and suppressed by powers of a UV scale $\Lambda$ which is higher than accessible collider energy. The corresponding Wilson coefficients $C_k^{(n)}$ parametrize the strength of these effective interactions. Schematically, the SMEFT Lagrangian is given by
\begin{gather}
    \mathcal L = \mathcal L_{\rm SM} + \sum_{n>4}\frac{1}{\Lambda^{n-4}}\sum_k C_k^{(n)}\,O_k^{(n)}.
\end{gather}
These terms modify the SM vertices in a gauge-invariant way and preserve the symmetries of the SM. In this work, we focus on the dimension-six ($n=6$) operators. For the input parameters, we adopt the $\{G_F,m_Z,m_W\}$ scheme, and we work in the five-flavor scheme, following the conventions of the \smeftatnlo~\cite{Degrande:2020evl} implementation whose Feynman rules we use throughout. 
\par
\begin{figure} 
    [H] 
    \centering 
    \includegraphics[width=.6\linewidth]{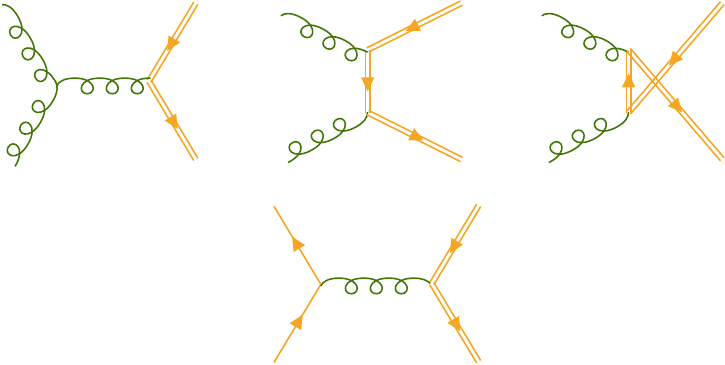} 
    \caption{The Feynman diagrams for the partonic process contributing to $p p \to t \bar t$ at leading order within the SM: gluon fusion (top) and pair annihilation (bottom).} 
    \label{fig:lo_diagrams_sm} 
\end{figure}

\begin{figure} 
    [H] 
    \centering 
    \includegraphics[width=.8\linewidth]{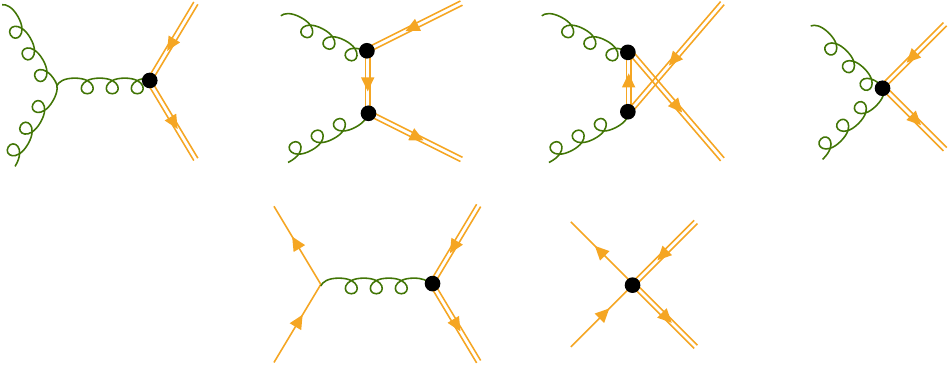} 
    \caption{The same as Fig.~\ref{fig:lo_diagrams_sm} but within the SMEFT. The blobs denote SMEFT operator insertions.} 
    \label{fig:lo_diagrams_smeft} 
\end{figure}

 
We consider the hadronic process $p p \to t \bar t$. At leading order, the underlying partonic processes are gluon fusion $gg\to t\bar t$ and pair annihilation $q\bar q\to t\bar t$ with $q=d,u,s,c,b$. The Feynman diagrams at leading order within the SM are pictured in Fig.~\ref{fig:lo_diagrams_sm}. The top quark is denoted by the double lines with an arrow.
There are two classes of dimension-six SMEFT operators of interest from the Warsaw basis~\cite{Grzadkowski:2010es} that interfere constructively with the SM, presented in Table~\ref{tab:ops}. First, we have the operator that modifies the 3-point top-quark coupling to the gluon, also introducing the 4-point $ttgg$ vertex. This is the chromomagnetic coupling. The second class of operators is the ones that introduce the 4-point $ttqq$ vertex (see Fig.~\ref{fig:lo_diagrams_smeft}). Here, the blobs indicate SMEFT operator insertions. We emphasize that the triple gluon operator is excluded in this work because it is already strongly constrained by dijet angular or multijet data~\cite{Goldouzian:2020wdq, Atkinson:2021jnj, Elmer:2023wtr, Aoude:2025jzc, deBlas:2025xhe, Haisch:2025lvd}.
\begin{table}
    [H]
    \centering
    \caption{The dimension-six SMEFT operators from the Warsaw basis~\cite{Grzadkowski:2010es} that affect the observable at leading order in Wilson coefficients.}
    \label{tab:ops}
    {\renewcommand{\arraystretch}{1}\begin{tabular}{|c|c|c|c|}
        \hline
        \multicolumn{2}{|c|}{Top-gluon couplings ($ttg$, $ttgg$)} & \multicolumn{2}{c|}{Four-quark contact ($ttqq$)} \\
        \hline
        $C_{uG}{}^{[pr]}$ & $O_{uG} = [\bar q_p \sigma^{\mu\nu} T^A u_r] \widetilde \varphi G_{\mu\nu}^A + {\rm h.c.}$ & $C_{qq}^{(1)}{}^{[prst]}$ & $O_{qq}^{(1)} = [\bar q_p \gamma_\mu q_r] [\bar q_s \gamma^\mu q_t]$ \\ 
        & & $C_{qq}^{(3)}{}^{[prst]}$ & $O_{qq}^{(3)} = [\bar q_p \gamma_\mu \tau^I q_r] [\bar q_s \gamma^\mu \tau^I q_t]$ \\ 
        & & $C_{uu}{}^{[prst]}$ & $O_{uu} = [\bar u_p \gamma_\mu u_r] [\bar u_s \gamma^\mu u_t]$ \\ 
        & & $C_{ud}^{(1)}{}^{[prst]}$ & $O_{ud}^{(1)} = [\bar u_p \gamma_\mu u_r] [\bar d_s \gamma^\mu d_t]$ \\ 
        & & $C_{ud}^{(8)}{}^{[prst]}$ & $O_{ud}^{(8)} = [\bar u_p \gamma_\mu T^A u_r] [\bar d_s \gamma^\mu T^A d_t]$ \\ 
        & & $C_{qu}^{(1)}{}^{[prst]}$ & $O_{qu}^{(1)} = [\bar q_p \gamma_\mu q_r] [\bar u_s \gamma^\mu u_t]$ \\ 
        & & $C_{qu}^{(8)}{}^{[prst]}$ & $O_{qu}^{(8)} = [\bar q_p \gamma_\mu T^A q_r] [\bar u_s \gamma^\mu T^A u_t]$ \\ 
        & & $C_{qd}^{(1)}{}^{[prst]}$ & $O_{qd}^{(1)} = [\bar q_p \gamma_\mu q_r] [\bar d_s \gamma^\mu d_t]$ \\ 
        & & $C_{qd}^{(8)}{}^{[prst]}$ & $O_{qd}^{(8)} = [\bar q_p \gamma_\mu T^A q_r] [\bar d_s \gamma^\mu T^A d_t]$ \\ 
        \hline
    \end{tabular}}
\end{table}
In Table~\ref{tab:ops}, $q$ denotes the left-handed quark $\mathrm{SU}(2)_L$ doublet and $u$ ($d$) the right-handed up-type (down-type) quark singlet, with $\bar\psi\equiv \psi^\dagger\gamma^0$. $\varphi$ is the Higgs doublet, and $\tilde\varphi \equiv i\tau^2\varphi^\ast$ is its charge conjugate. Color is carried by the $\mathrm{SU}(3)_c$ generators $T^A$, and $G_{\mu\nu}^A$ is the gluon field-strength tensor. The fermion bilinears involve $\gamma^\mu$ and $\sigma^{\mu\nu}\equiv \tfrac{i}{2}[\gamma^\mu,\gamma^\nu]$; square brackets indicate gauge- and Lorentz-invariant contractions within each current/bilinear with all indices left implicit. The superscripts $[pr]$ and $[prst]$ label quark flavor indices, with $p,r,s,t\in\{1,2,3\}$. For a two-fermion operator, e.g., $O_{uG}^{[pr]}\sim[\bar q_p\cdots u_r]\tilde\varphi\,G$, the indices $p$ and $r$ identify the generations of $q$ and $u$. For a four-fermion operator, e.g.\ $O_{qu}^{(1)[prst]}=[\bar q_p\gamma_\mu q_r][\bar u_s\gamma^\mu u_t]$, $(p,r)$ belong to the first bilinear and $(s,t)$ to the second.
\par
\begin{figure}
    [H]
    \centering
    \includegraphics[width=1\linewidth]{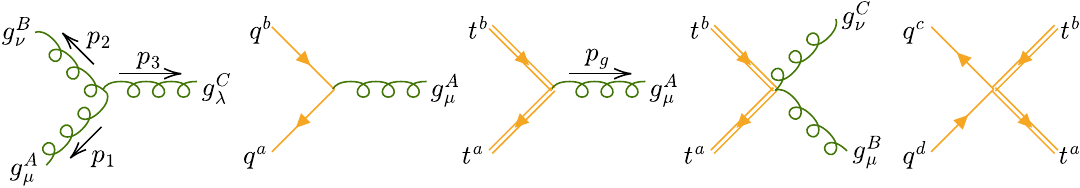}
    \caption{The relevant SMEFT Feynman rules for the partonic process of interest, adopted from~\cite{Degrande:2020evl}.}
    \label{fig:smeft_feynman_rules}
\end{figure}

 
Following the \smeftatnlo~model, we actually work in the \textit{top basis}. The operator that gives us the $ttg$ and $ttgg$ vertices reads
\begin{gather}
    O_{tG} = i g_s [\bar Q \tau^{\mu\nu} T_A t] \widetilde \varphi G^A_{\mu\nu} + {\rm h.c.},
\end{gather}
so the main differences are that the strong coupling is factored out and that $\tau^{\mu\nu} = \sigma^{\mu\nu}/i$, and the four-quark Wilson coefficients of interest are defined as
\begin{gather}
    C_{Qq}^{1,1} = C_{qq}^{(1)}{}^{[ii33]} + \frac16 C_{qq}^{(1)}{}^{[i33i]} + \frac12 C_{qq}^{(3)}{}^{[i33i]}, \quad 
    C_{Qq}^{1,8} = C_{qq}^{(1)}{}^{[i33i]} + 3 C_{qq}^{(3)}{}^{[i33i]}, \\
    C_{Qq}^{3,1} = C_{qq}^{(3)}{}^{[ii33]} + \frac16 C_{qq}^{(1)}{}^{[i33i]} - \frac16 C_{qq}^{(3)}{}^{[i33i]}, \quad 
    C_{Qq}^{3,8} = C_{qq}^{(1)}{}^{[i33i]} - C_{qq}^{(3)}{}^{[i33i]}, \\
    C_{QQ}^1 = 2 C_{qq}^{(1)}{}^{[3333]} - \frac23 C_{qq}^{(3)}{}^{[3333]}, \quad
    C_{QQ}^8 = 8 C_{qq}^{(3)}{}^{[3333]}, \\
    C_{Qu}^1 = C_{qu}^{(1)}{}^{[33ii]}, \quad 
    C_{Qu}^8 = C_{qu}^{(8)}{}^{[33ii]}, \\ 
    C_{Qd}^1 = C_{qd}^{(1)}{}^{[33ii]}, \quad 
    C_{Qd}^8 = C_{qd}^{(8)}{}^{[33ii]}, \\ 
    C_{Qt}^1 = C_{qu}^{(1)}{}^{[3333]}, \quad 
    C_{Qt}^8 = C_{qu}^{(8)}{}^{[3333]}, \\ 
    C_{tu}^1 = C_{uu}{}^{[ii33]} + \frac13 C_{uu}{}^{[i33i]}, \quad
    C_{tu}^8 = 2 C_{uu}{}^{[i33i]}, \\
    C_{td}^1 = C_{ud}^{(1)}{}^{[33ii]}, \quad 
    C_{td}^8 = C_{ud}^{(8)}{}^{[33ii]}, \\
    C_{tq}^1 = C_{qu}^{(1)}{}^{[ii33]}, \quad
    C_{tq}^8 = C_{qu}^{(8)}{}^{[ii33]}, 
\end{gather}
with $i=1,2$. In total, we have 19 Wilson coefficients of interest. The relevant Feynman rules from~\cite{Degrande:2020evl} can be summarized as follows:
\begin{align}
    V_{ggg}^{\mu\nu\lambda} &= f^{ABC} \pp{(p_1-p_2)^\lambda g^{\mu\nu} + (p_2-p_3)^\mu g^{\nu\lambda} + (p_3-p_1)^\nu g^{\lambda\mu}} C_{10}, \\ 
    V_{qqg}^\mu &= T^A_{ab} \gamma^\mu C_{11}, \\ 
    V_{ttg}^\mu &= T^A_{ab} \pp{[\gamma^\mu, \slashed p_g] C_{358} + \gamma^\mu C_{11}}, \\ 
    V_{ttgg}^{\mu\nu} &= f^{ABC} T^A_{ab} [\gamma^\mu, \gamma^\nu] C_{360}, \\ 
    V_{ttuu} = V_{ttcc} &= \delta_{ab} \delta_{cd} \Parr{
        [\gamma^\alpha P_L]_t [\gamma^\alpha P_L]_u C_{31} + [\gamma^\alpha P_R]_t [\gamma^\alpha P_L]_u C_{109} \\
        + [\gamma^\alpha P_L]_t [\gamma^\alpha P_R]_u C_{87} + [\gamma^\alpha P_R]_t [\gamma^\alpha P_R]_u C_{112}
    }
    \nn + T^A_{ab} T^A_{cd} \Parr{
        [\gamma^\alpha P_L]_t [\gamma^\alpha P_L]_u C_{35} + [\gamma^\alpha P_R]_t [\gamma^\alpha P_L]_u C_{110} \\
        + [\gamma^\alpha P_L]_t [\gamma^\alpha P_R]_u C_{88} + [\gamma^\alpha P_R]_t [\gamma^\alpha P_R]_u C_{113}
    }, \\ 
    V_{ttdd} = V_{ttss} &= \delta_{ab} \delta_{cd} \Parr{
        [\gamma^\alpha P_L]_d [\gamma^\alpha P_L]_t C_{30} + [\gamma^\alpha P_R]_d [\gamma^\alpha P_L]_t C_{71} \\
        + [\gamma^\alpha P_L]_d [\gamma^\alpha P_R]_t C_{109} + [\gamma^\alpha P_R]_d [\gamma^\alpha P_R]_t C_{89}
    }
    \nn + T^A_{ab} T^A_{cd} \Parr{
        [\gamma^\alpha P_L]_d [\gamma^\alpha P_L]_t C_{34} + [\gamma^\alpha P_R]_d [\gamma^\alpha P_L]_t C_{72} \\
        + [\gamma^\alpha P_L]_d [\gamma^\alpha P_R]_t C_{110} + [\gamma^\alpha P_R]_d [\gamma^\alpha P_R]_t C_{90}
    }, \\ 
    V_{ttbb} &= \delta_{ad} \delta_{bc} [\gamma^\alpha P_L]_\bullet [\gamma^\alpha P_L]_\bullet C_{83}
    \nn + \delta_{ab} \delta_{cd} \Parr{
        [\gamma^\alpha P_L]_t [\gamma^\alpha P_L]_b C_{32} + [\gamma^\alpha P_R]_t [\gamma^\alpha P_L]_b C_{85} \\
        + [\gamma^\alpha P_R]_t [\gamma^\alpha P_R]_b C_{89} + [\gamma^\alpha P_L]_t [\gamma^\alpha P_R]_b C_{71}
    }
    \nn + T^A_{ab} T^A_{cd} \Parr{
        [\gamma^\alpha P_R]_t [\gamma^\alpha P_L]_b C_{86} \\
        + [\gamma^\alpha P_R]_t [\gamma^\alpha P_R]_b C_{90} + [\gamma^\alpha P_L]_t [\gamma^\alpha P_R]_b C_{72}
    },
\end{align}
where $\slashed p = \gamma^\mu p_\mu$, $P_{R/L} = (1 \pm \gamma_5)/2$, $\delta_{ab}$ is the Kronecker delta in the fundamental representation, $[\gamma^\alpha P_L]_\bullet [\gamma^\alpha P_L]_\bullet$ represents the coupling of flavor-changing vector $tb$ currents, and we have defined
\begin{gather}
    C_{10} = -g_s, \quad
    C_{11} = i g_s, \\
    C_{30} = \frac{i}{\Lambda^{2}}\left( C_{Qq}^{1,1} - C_{Qq}^{3,1} \right), \quad
    C_{31} = \frac{i}{\Lambda^{2}}\left( C_{Qq}^{1,1} + C_{Qq}^{3,1} \right), \\
    C_{32} = \frac{i}{\Lambda^{2}}\left( C_{QQ}^{1} - \frac{1}{6} C_{QQ}^{8} \right), \\
    C_{34} = \frac{i}{\Lambda^{2}}\left( C_{Qq}^{1,8} - C_{Qq}^{3,8} \right), \quad
    C_{35} = \frac{i}{\Lambda^{2}}\left( C_{Qq}^{1,8} + C_{Qq}^{3,8} \right), \\
    C_{71} = \frac{i}{\Lambda^{2}} C_{Qd}^{1}, \quad 
    C_{72} = \frac{i}{\Lambda^{2}} C_{Qd}^{8}, \\
    C_{83} = \frac{i}{2 \Lambda^{2}} C_{QQ}^{8}, \\
    C_{85} = \frac{i}{\Lambda^{2}} C_{Qt}^{1}, \quad
    C_{86} = \frac{i}{\Lambda^{2}} C_{Qt}^{8}, \\
    C_{87} = \frac{i}{\Lambda^{2}} C_{Qu}^{1}, \quad
    C_{88} = \frac{i}{\Lambda^{2}} C_{Qu}^{8}, \\
    C_{89} = \frac{i}{\Lambda^{2}} C_{td}^{1}, \quad
    C_{90} = \frac{i}{\Lambda^{2}} C_{td}^{8}, \\
    C_{109} = \frac{i}{\Lambda^{2}} C_{tq}^{1}, \quad
    C_{110} = \frac{i}{\Lambda^{2}} C_{tq}^{8}, \\
    C_{112} = \frac{i}{\Lambda^{2}} C_{tu}^{1}, \quad
    C_{113} = \frac{i}{\Lambda^{2}} C_{tu}^{8}, \\
    C_{358} = \frac{i g_s v_0}{\sqrt{2} \Lambda^{2}} C_{tG}, \quad 
    C_{360} = -\frac{g_s{}^{2} v_0}{\sqrt{2} \Lambda^{2}} C_{tG},
\end{gather}
where $v_0$ is the Higgs vacuum expectation value in the SM. The Feynman rules are illustrated in Fig.~\ref{fig:smeft_feynman_rules}. 
The hadronic cross section at the $pp$ center-of-mass (c.m.) energy $\sqrt{s}$ can be written as
\begin{gather}
    \sigma(s) = \sum_{i=-N_f}^{N_f} \sum_{j=-N_f}^{N_f} \int \d{x_1} \d{x_2} f_i(x_1, \mu_F) f_j(x_2, \mu_F) \sigma_{ij}(\hat s),
\end{gather}
with $N_f$ denoting the number of active flavors. The indices $i$ and $j$ label the parton flavors, where negative values correspond to antiquarks and 0 to the gluon. The variables $x_1$ and $x_2$ are the Bjorken-$x$ parameters of the incoming partons, and $f_i$ and $f_j$ are the parton distribution functions (PDFs) evaluated at the factorization scale $\mu_F$. The quantity $\sigma_{ij}$ is the partonic cross section for the subprocess $f_i f_j \to t \bar t$, and $\sqrt{\hat s}$ is the c.m. energy of the partonic reaction.
\par
We parametrize the cross section in terms of SMEFT parameters as
\begin{gather}
    \sigma = \sigma_{\rm SM} + \sum_k C_k \sigma_k + \sum_{k\leq k'} C_k C_{k'} \sigma_{kk'}
\end{gather}
at all perturbative orders. In order to obtain the leading-order amplitudes, we use \feynarts~\cite{Hahn:2000kx} and \feyncalc~\cite{Shtabovenko:2016sxi, Shtabovenko:2020gxv, Shtabovenko:2023idz, Mertig:1990an}. For the gluon fusion,
\begin{gather}
    g(p_1) + g(p_2) \to t(p_3) + \bar t(p_4),
\end{gather}
the spin and color-averaged squared SM amplitude is
\begin{align}
    &\bar{|\mathcal A_{gg}|^2}_{\rm SM} = {1 \over 48}{\pi^2 \alpha_s{}^2 \over \hat s^2} D_t^{(t)} D_t^{(u)} \CC{
        -(7 \hat{s}^2+9 (\hat{t}-\hat{u})^2) \nn\times 
   (12 \hat{s}^3 (\hat{t}+\hat{u})+4 \hat{s}^2 (2 \hat{t} \hat{u}+3 \hat{t}^2+3 \hat{u}^2)+4 \hat{s}
   (\hat{t}-\hat{u})^2 (\hat{t}+\hat{u})+3 \hat{s}^4+(\hat{t}-\hat{u})^4)
    },
\end{align}
and for the pair annihilation,
\begin{gather}
    q(p_1) + \bar q(p_2) \to t(p_3) + \bar t(p_4), \quad q = d, u, s, c, b,
\end{gather}
it is 
\begin{gather}
    \bar{|\mathcal A_{q\bar q}|^2}_{\rm SM} = {32 \over 9}{\pi^2 \alpha_s{}^2 \over \hat s^2} \CC{4 \hat{s} m_t{}^2+\hat{s}^2+\left(\hat{t}-\hat{u}\right)^2}.
\end{gather}
Here, $m_t$ is the mass of the top quark, the top-quark propagators are given by
\begin{gather}
    D_t^{(t)} = {1 \over (\hat t - m_t{}^2)^2},  \quad 
    D_t^{(u)} = {1 \over (\hat u - m_t{}^2)^2}, 
\end{gather}
and we have defined $\hat s = (p_1 + p_2)^2$, $\hat t = (p_1 - p_3)^2$, and $\hat u = (p_1 - p_4)^2$, which satisfy $\hat s + \hat t + \hat u = 2m_t{}^2$. 
\par 
Next, we calculate the linear SMEFT corrections to the squared amplitude. The gluon fusion receives corrections characterized by $C_{tG}$ only:
\begin{align}
    \bar{|\mathcal A_{gg}|^2}_{C_{tG}} &= {1 \over 3 \times 2^{3/4}} {\pi^2 \alpha_s{}^2 m_t \over \sqrt{G_F} \Lambda^2} D_t^{(t)} D_t^{(u)} \CC{
        2 \hat{s}^2(\hat{t}-\hat{u})^2+7 \hat{s}^4-9 (\hat{t}-\hat{u})^4
    }.
\end{align}
All the pair annihilation channels receive the same linear correction characterized by $C_{tG}$:
\begin{align}
    \bar{|\mathcal A_{q\bar q}|^2}_{C_{tG}} = \frac{256 \sqrt[4]{2} \pi ^2 m_t \alpha _s{}^2}{9 \Lambda ^2 \sqrt{G_F}}, \quad q = d,u,s,c,b.
\end{align}
There are also corrections that can be classified as even and odd in $\hat t - \hat u$:
\begin{gather}
    \bar{|\mathcal A_{q\bar q}|^2}_{C_q^+} = {8 \pi \alpha_s \over 9 \Lambda^2 \hat s} \cc{4 \hat s m_t{}^2 + \hat s^2 + (\hat t - \hat u)^2}, \\
    \bar{|\mathcal A_{q\bar q}|^2}_{C_q^-} = {16 \pi \alpha_s \over 9 \Lambda^2} \cc{\hat t - \hat u},
\end{gather}
for $q=d,u,s,c,b$, where we have defined
\begin{gather}
    C_d^+ = C_s^+ = \frac{1}{2} \left(C_{Qq}^{\text{1,8}}-C_{Qq}^{\text{3,8}}+C_{Qd}^8+C_{td}^8+C_{tq}^8\right), \\
    C_d^- = C_s^- = \frac{1}{2} \left(C_{Qq}^{\text{1,8}}-C_{Qq}^{\text{3,8}}-C_{Qd}^8+C_{td}^8-C_{tq}^8\right), \\
    C_u^+ = C_c^+ = \frac{1}{2} \left(C_{Qq}^{\text{1,8}}+C_{Qq}^{\text{3,8}}+C_{tq}^8+C_{Qu}^8+C_{tu}^8\right), \\
    C_u^- = C_c^- = \frac{1}{2} \left(C_{Qq}^{\text{1,8}}+C_{Qq}^{\text{3,8}}-C_{tq}^8-C_{Qu}^8+C_{tu}^8\right), \\
    C_b^+ = \frac{1}{2} \left(C_{Qd}^8+C_{td}^8+C_{Qt}^8+C_{QQ}^8\right), \\
    C_b^- = \frac{1}{2} \left(-C_{Qd}^8+C_{td}^8-C_{Qt}^8+C_{QQ}^8\right).
\end{gather}
The $\bar q q$ channels are given by $\hat u \to \hat t$. We note that the SM and $C_{tG}$ linear contributions are even in $\hat t - \hat u$, and that with the inclusive cross section computed with symmetrical acceptance cuts, the terms odd in $\hat t - \hat u$ drop. This leaves us with only four Wilson coefficients of interest: $C_{tG}$, $C_d^+$, $C_u^+$, and $C_b^+$.
\par 
Finally, we restrict ourselves to the linear SMEFT corrections, namely we keep only the $\mathcal O(\Lambda^{-2})$ interference with the SM. The quadratic dimension-six terms are $\mathcal O(\Lambda^{-4})$ and, in the present case, involve many additional Wilson-coefficient structures that do not reorganize into the compact $C_q^\pm$ combinations. Including them would substantially enlarge the parameter space and the computational workload, so we omit $\mathcal O(\Lambda^{-4})$ effects in this study.
\par
For the NLO QCD predictions, we use the \smeftatnlo\ model~\cite{Degrande:2020evl} with \madgraph~\cite{Alwall:2014hca}. We simply generate $pp\to t\bar t$ at NLO in QCD and we exclude electroweak corrections.

\section{Soft-gluon corrections\label{sec:resum}}

In this section, we briefly review the soft-gluon resummation formalism that we use for the calculation of soft-gluon corrections through aNNLO for $t{\bar t}$ production \cite{Kidonakis:1996aq,Kidonakis:1997gm,Kidonakis:2000ui,Kidonakis:2009ev,Kidonakis:2010dk}. In addition to the kinematical variables ${\hat s}$, ${\hat t}$, and ${\hat u}$ that were defined for the partonic processes in Sec.~II, we introduce the variable $s_4={\hat s}+{\hat t}+{\hat u}-2m_t^2$ which vanishes at partonic threshold where there is no available energy for extra radiation. The soft-gluon contributions in the perturbative series take the form of plus distributions of logarithms of $s_4$, i.e., $[(\ln^k(s_4/m_t^2))/s_4]_+$, where in the $n$th-order corrections the integer $k$ takes values from 0 to $2n-1$.

The resummation of soft-gluon contributions is derived via the factorization of the (in general, differential) cross section into functions that describe soft and collinear emission in the partonic process. Using one-particle-inclusive kinematics, we take Laplace transforms of the partonic cross section,
${\hat \sigma}(N)=\int (ds_4/s) \;  e^{-N s_4/s} {\hat \sigma}(s_4)$, with $N$ the
transform variable, and we write a factorized expression,
\beq
\sigma_{ij \to t{\bar t}}(N)=
\psi_i\left (N_i \right) \;
\psi_j\left (N_j \right) \;
{\rm tr}\left[H_{ij \to t{\bar t}} \; S_{ij \to t{\bar t}}\left(\frac{m_t}{N \mu}\right)\right] \, ,
\label{factorized}
\eeq
where $\mu$ is the scale, $\psi_i$ and $\psi_j$ are functions which describe soft and collinear emission from the incoming partons $i$ and $j$, $H_{ij \to t{\bar t}}$ is the hard-scattering function, and $S_{ij \to t{\bar t}}$ is the soft-gluon function that encompasses noncollinear soft-gluon emission. The hard and soft functions are matrices in the space of color exchanges, and the trace of their product is taken in the above equation.
The soft function $S_{ij \to t{\bar t}}$ requires renormalization, and its $N$-dependence can be resummed via renormalization group evolution (RGE). The resummed partonic cross section follows from the solution of the RGE. Details and explicit results can be found in Refs. \cite{Kidonakis:1996aq,Kidonakis:1997gm,Kidonakis:2000ui,Kidonakis:2009ev,Kidonakis:2010dk}
.

In this paper, we expand the $N$-space resummed cross section to NNLO and, then, invert back to momentum space. Thus, we derive results for the second-order soft-gluon corrections without any need for a prescription. We calculate aNNLO cross sections that are matched to NLO as described in the next section.

\section{Numerical analysis\label{sec:numerical}}
In this section, we provide details of our calculations for the numerical analysis. We carry out the calculations at LO and for the soft-gluon corrections using inhouse codes. We use \lhapdf~\cite{Buckley:2014ana} and \vegas~\cite{Lepage:2020tgj, Lepage:vegas} for Python, and at NLO we use \madgraph~\cite{Alwall:2014hca, Mastrolia:2012bu, Peraro:2014cba, Hirschi:2016mdz, Denner:2016kdg}. The aNNLO results are calculated by adding the second-order soft-gluon corrections to the exact NLO. We validate our SM calculations at LO and NLO with \matrix~\cite{Grazzini:2017mhc, Catani:2019iny, Catani:2019hip, Denner:2016kdg, Cascioli:2011va, Buccioni:2019sur, Buccioni:2017yxi, Barnreuther:2013qvf, Catani:2012qa, Catani:2007vq} and obtain the NNLO distributions with it, as well.
\par 
The numerical values of the parameters are given by
\begin{gather}
    G_F = 1.1663788 \times 10^{-5} \ \GeV^{-2}, \\
    m_Z = 91.1880 \ \GeV, \\
    m_W = 80.3692 \ \GeV, \\
    m_t = 172.5 \ \GeV.
\end{gather}
The renormalization and factorization scales are fixed at $\mu_R = \mu_F = \mu_0 = m_t$ and we consider variations $\mu_R = f_R \mu_0$ and $\mu_F = f_F \mu_0$ with $f_R, f_F \in \cc{0.5, 1, 2}$ and choose the 7-point variations as the basis for scale variations in our work. We use the MSHT20 NLO and NNLO PDFs~\cite{Bailey:2020ooq}. We assume $N_f=5$, and the running of the strong coupling $\alpha_s$ is provided from the PDF set. We set $\Lambda = 1 \ \TeV$. We consider the 2025 CMS 13 TeV single differential $p_T$ and $y$ distributions with the bins~\cite{CMS:2024ybg},
\begin{gather}
    p_T(t) \in [0, 55, 100, 165, 240, 330, 440, 600] \ \GeV, \\
    y(t) \in [-2.6, -1.8, -1.35, -0.9, -0.45, 0, 0.45, 0.9, 1.35, 1.8, 2.6],
\end{gather}
as well as the 2018 CMS 13 TeV double differential $p_T \times y$ distribution with the bins~\cite{CMS:2018htd},
\begin{gather}
    p_T(t) \in [0, 40, 80, 120, 160, 200, 240, 280, 330, 380, 450, 800] \ \GeV, \\
    |y(t)| \in [0, 0.5, 1, 1.5, 2.5].
\end{gather}
The double-differential data is first sorted by rapidity and then transverse momentum. We denote the region $0<|y|<0.5$ by I, $0.5<|y|<1$ by II, $1<|y|<1.5$ by III, and $1.5<|y|<2.5$ by IV. 
\par
We note that the 2018 CMS double-differential measurement is defined at the parton level in the $\ell+\text{jets}$ decay channel, where $\ell = e,\mu$, and explicitly excludes $\tau+\text{jets}$ contributions. While no kinematic restriction is applied to the parton-level top quarks, the observable is therefore not fully inclusive in decay modes. As a result, the integrated normalization of the double-differential distribution corresponds to the semileptonic branching fraction of $t\bar t$ production rather than the total inclusive cross section. To enable a consistent comparison with inclusive theoretical predictions, we convert the measurement to an approximate inclusive normalization by dividing out the corresponding branching fraction,
\begin{gather}
    \mathcal B = \mathcal B_{\ell+\text{jets},\ e/\mu,\text{ no }\tau} = 2 [\mathcal B(W\to e\nu)+\mathcal B(W\to \mu\nu)] \mathcal B(W\to qq') = 0.2877,
\end{gather}
where the factor of $2$ accounts for either the top or antitop decaying leptonically. This procedure assumes factorization between production and decay in the narrow-width approximation. The same rescaling is applied to the covariance matrix,
\begin{gather}
    \mathcal E^{\text{incl.}} = \frac{1}{\mathcal B^2} \mathcal E^{\text{CMS}},
\end{gather}
while preserving the correlation structure. We emphasize that this defines an approximate inclusive proxy of the measurement, which we use for consistency with our inclusive theoretical framework.
\par 
For the 13 TeV analysis, the integrated luminosity is given to be $\mathcal L = 138 \ \fb^{-1}$ for the single-differential bins and $\mathcal L = 35.8\ \fb^{-1}$ for the double-differential case. Experimental uncertainties are borrowed from the published data. For the single-differential distributions, we have five classes of uncertainties: statistical, jet energy scale (JES), lepton efficiency, background, and other experimental uncertainties. Since the unfolded correlation matrix is not provided yet, we assume all uncertainties to be bin-by-bin uncorrelated. For the double-differential data, we use the provided covariance matrix as is, which we generally refer to as the experimental uncertainties in what follows.  
\par 
For the 13.6 TeV projections, we assume $\mathcal L = 300 \ \fb^{-1}$. Since the data is not available yet, we assume the same set of $p_T$, $y$, and $p_T \times y$ bins as the 13-TeV measurements. As for the uncertainties, we introduce uncorrelated statistical uncertainties calculated with the bin-by-bin event count, 5\% uncorrelated global uncertainties motivated by the typical size of experimental systematics at 13 TeV~\cite{CMS:2017xio}, and 1.7\% fully correlated luminosity uncertainties consistent with ATLAS Run II calibration~\cite{ATLAS:2022hro}. 
\par 
For both studies, we introduce fully correlated PDF uncertainties and uncorrelated scale variations as theoretical uncertainties. 
\par
\begin{figure}
    [H]
    \centering
    \includegraphics[width=0.32\linewidth]{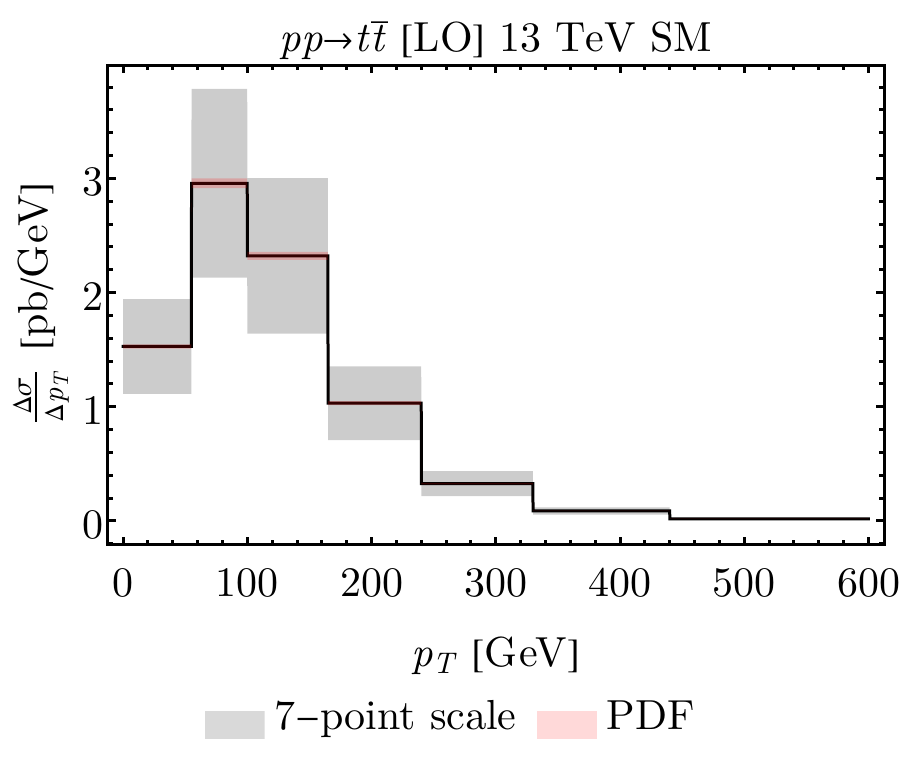}
    \includegraphics[width=0.32\linewidth]{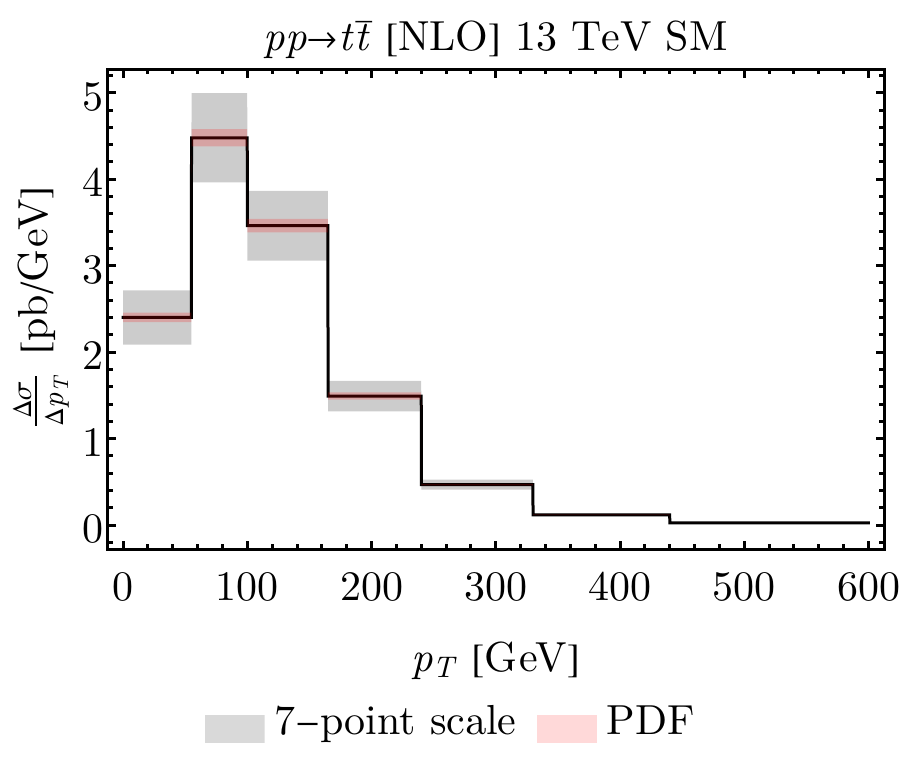}
    \includegraphics[width=0.32\linewidth]{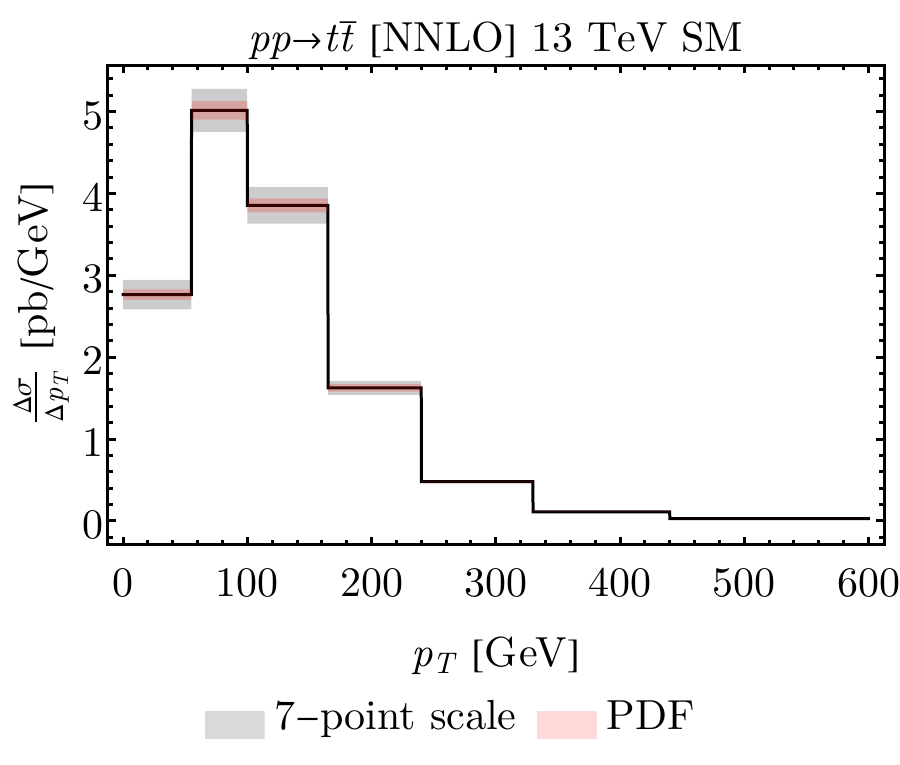}
    \includegraphics[width=0.32\linewidth]{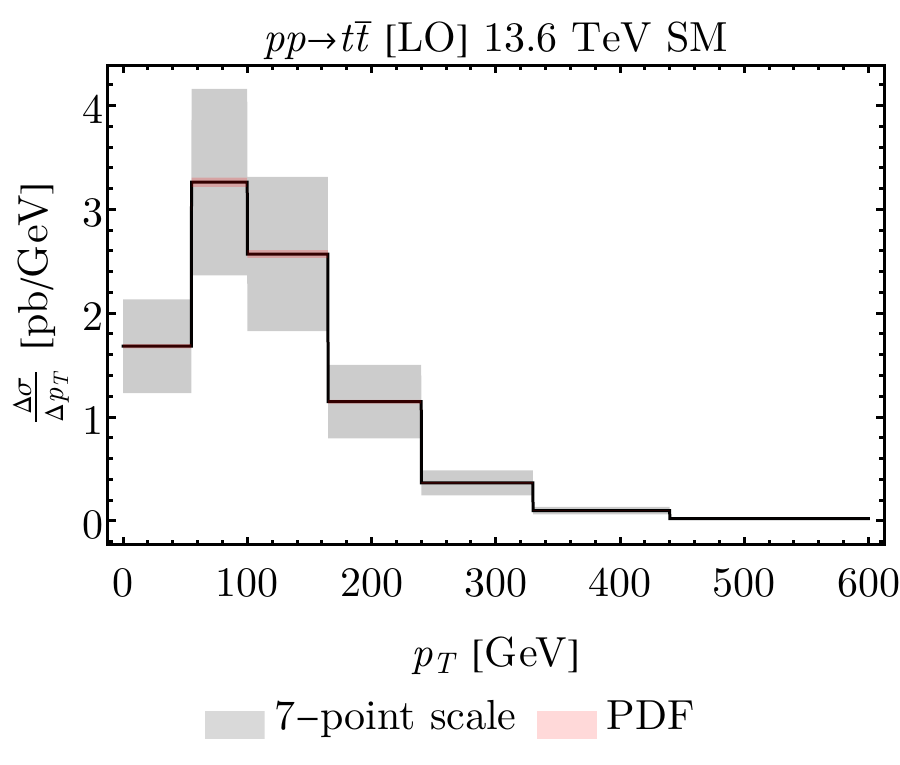}
    \includegraphics[width=0.32\linewidth]{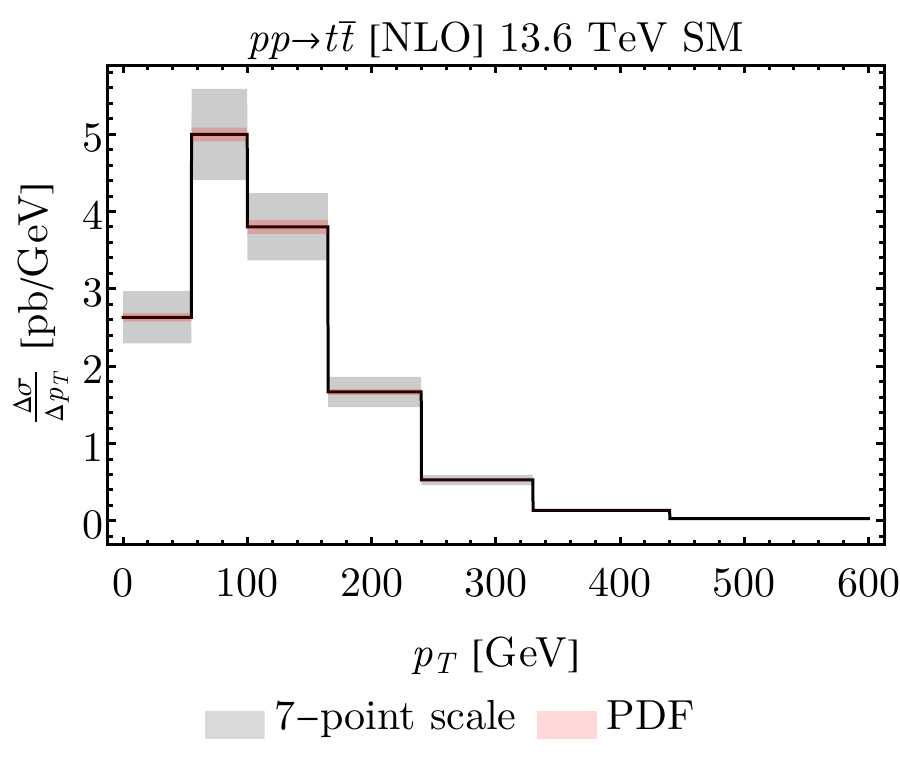}
    \includegraphics[width=0.32\linewidth]{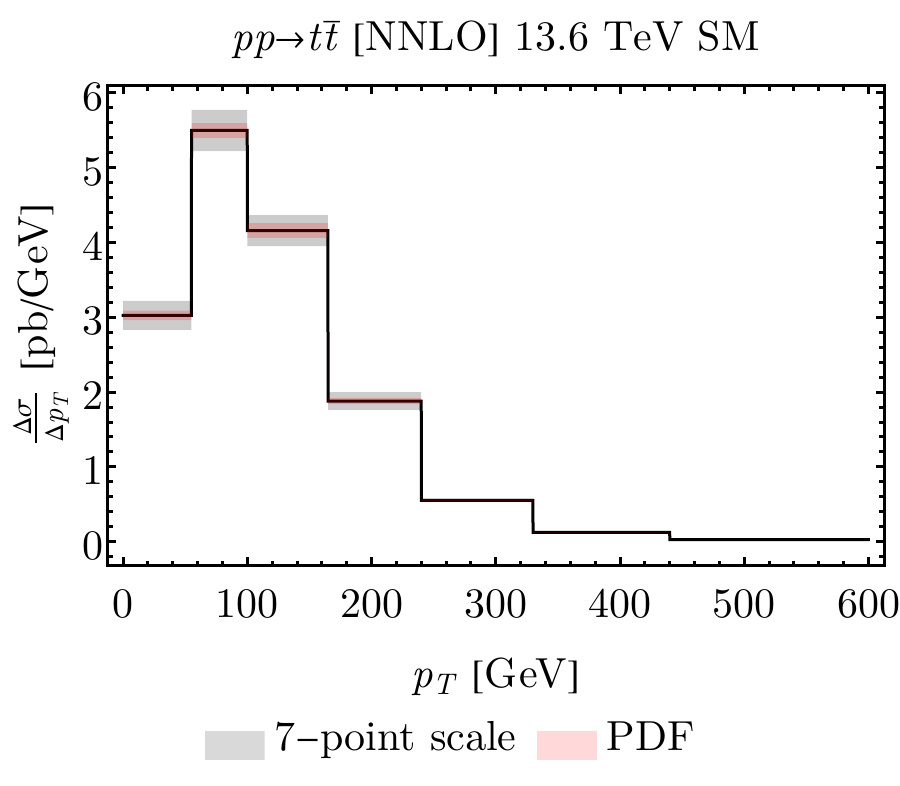}
    \caption{The top-quark single-differential $p_T$ distribution in $t\bar t$ production within the SM across the bins at LO (left), NLO (center), and NNLO (right) at 13 TeV (top) and 13.6 TeV (bottom).}
    \label{fig:pt_distr}
\end{figure}

\begin{figure}
    [H]
    \centering
    \includegraphics[width=0.32\linewidth]{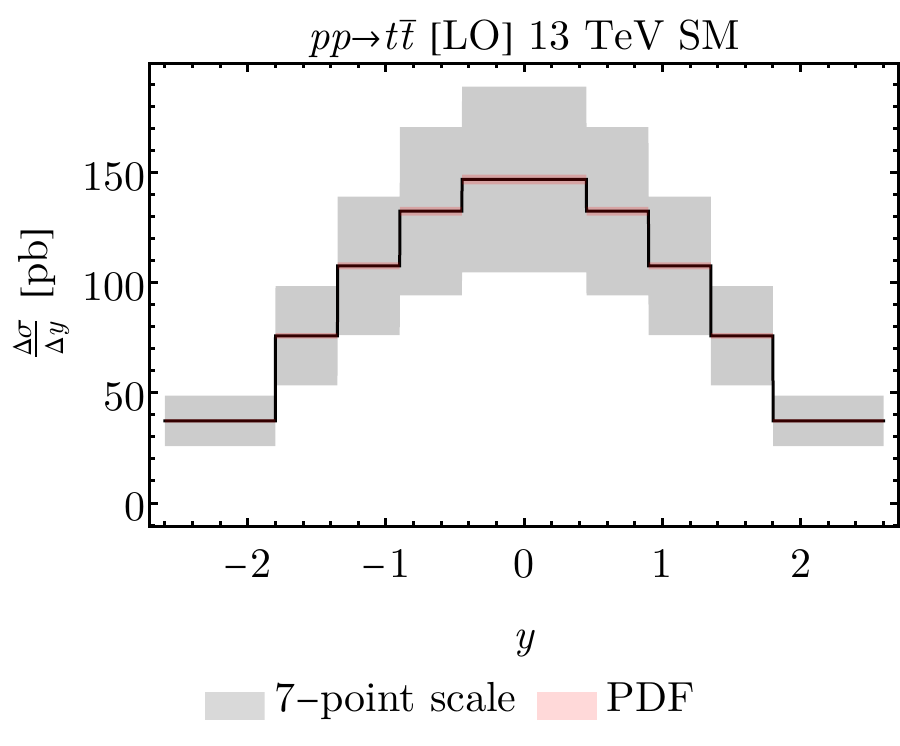}
    \includegraphics[width=0.32\linewidth]{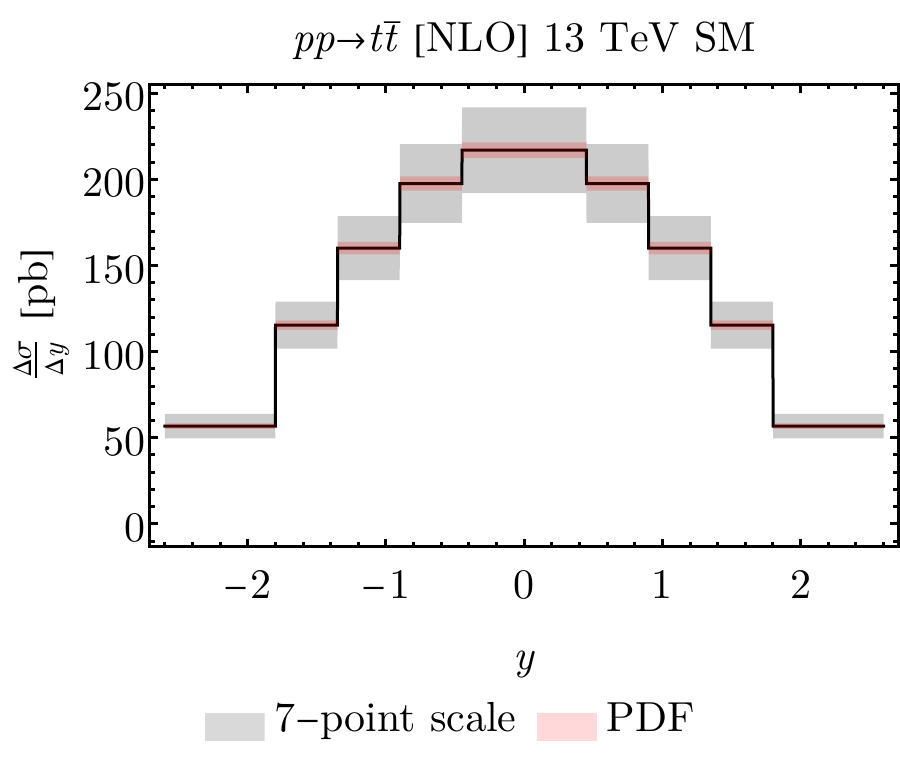}
    \includegraphics[width=0.32\linewidth]{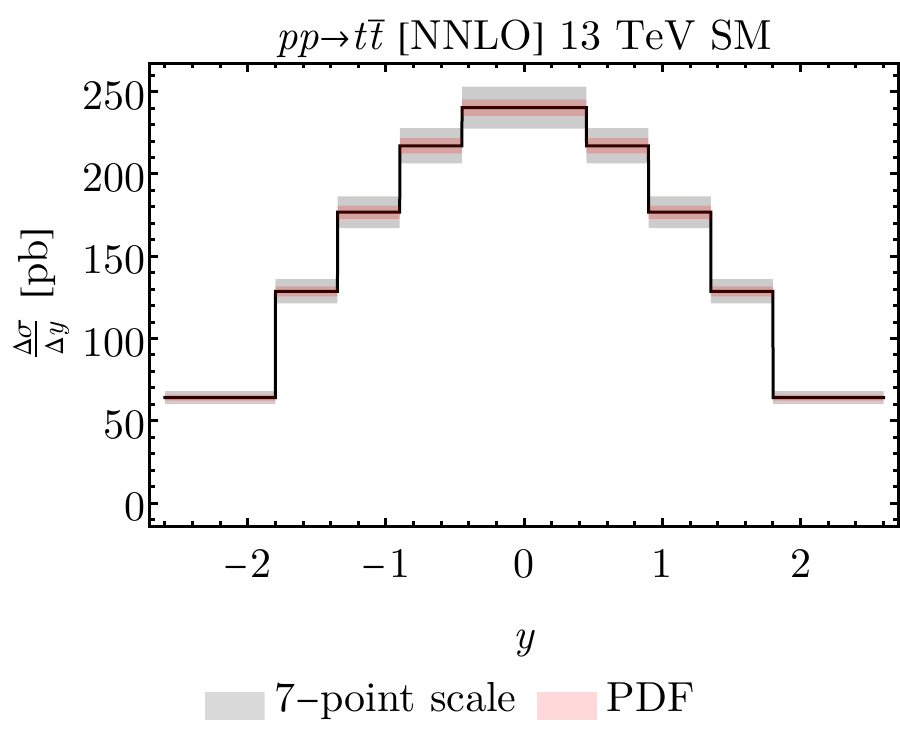}
    \includegraphics[width=0.32\linewidth]{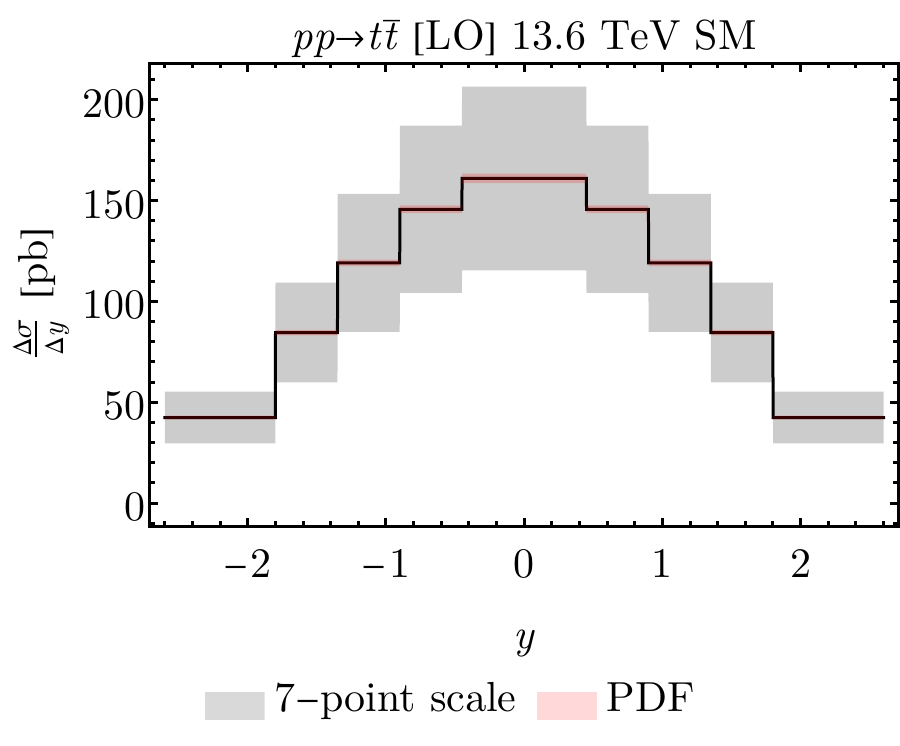}
    \includegraphics[width=0.32\linewidth]{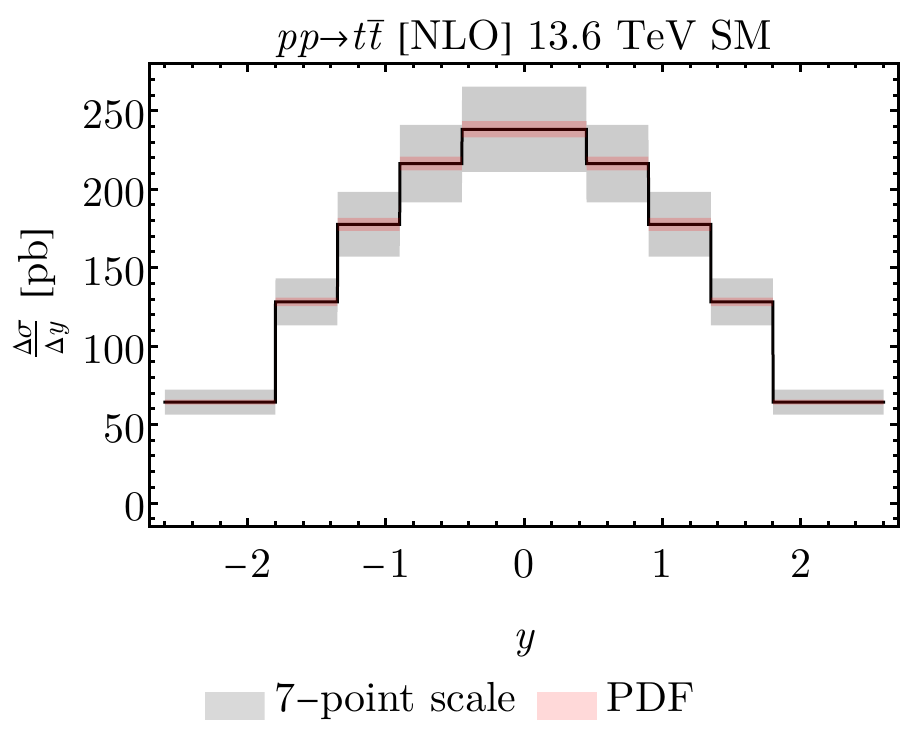}
    \includegraphics[width=0.32\linewidth]{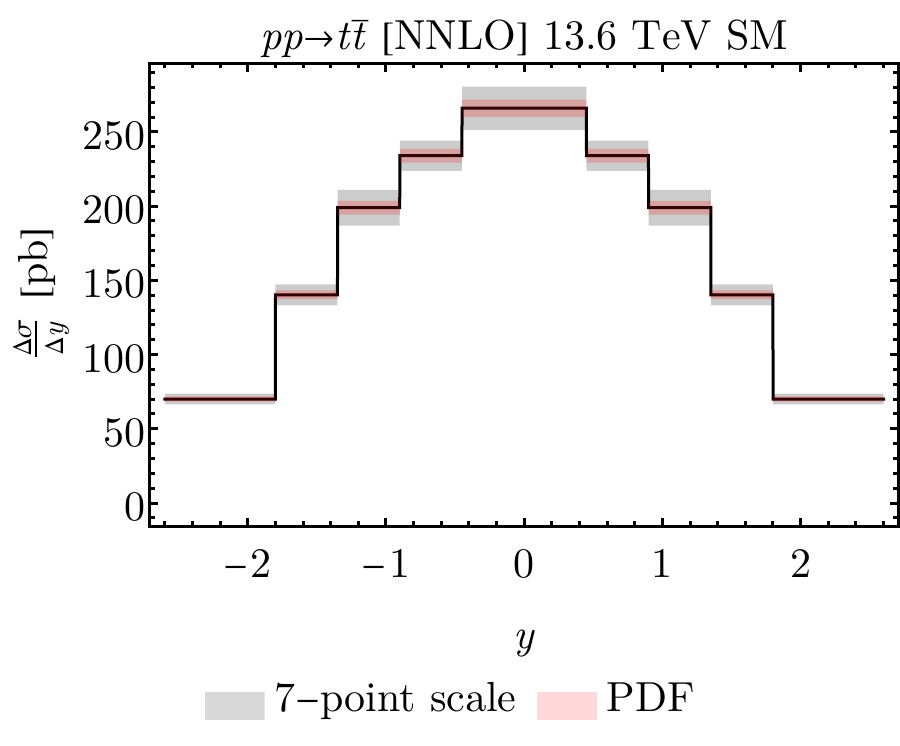}
    \caption{The same as Fig.~\ref{fig:pt_distr} but for the top-quark single-differential $y$ distribution.}
    \label{fig:y_distr}
\end{figure}

\begin{figure}
    [H]
    \centering
    \includegraphics[width=0.32\linewidth]{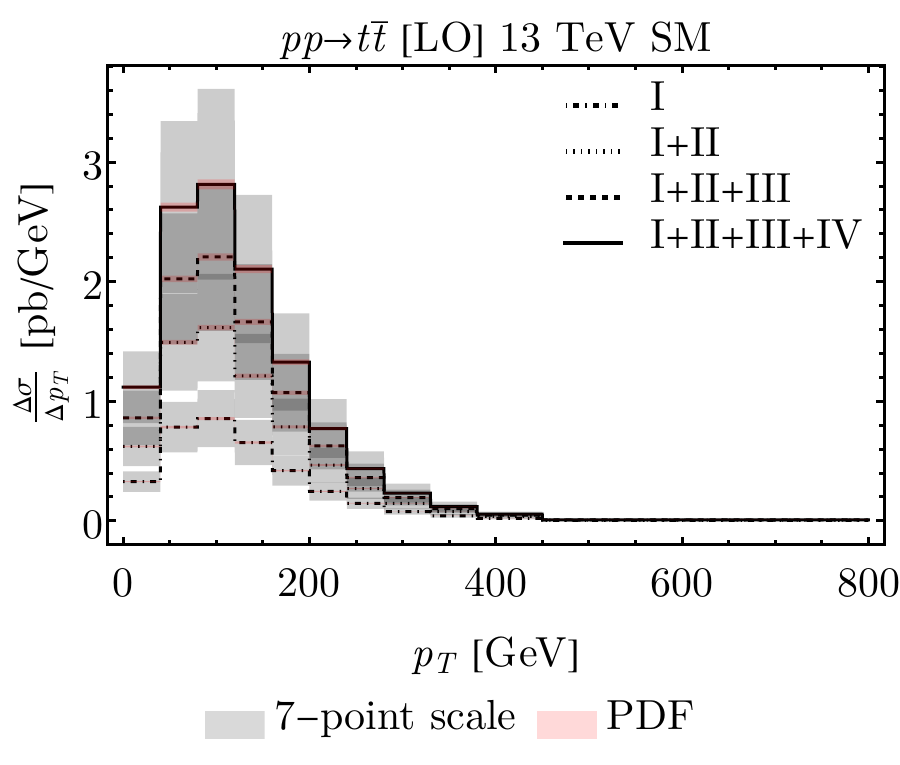}
    \includegraphics[width=0.32\linewidth]{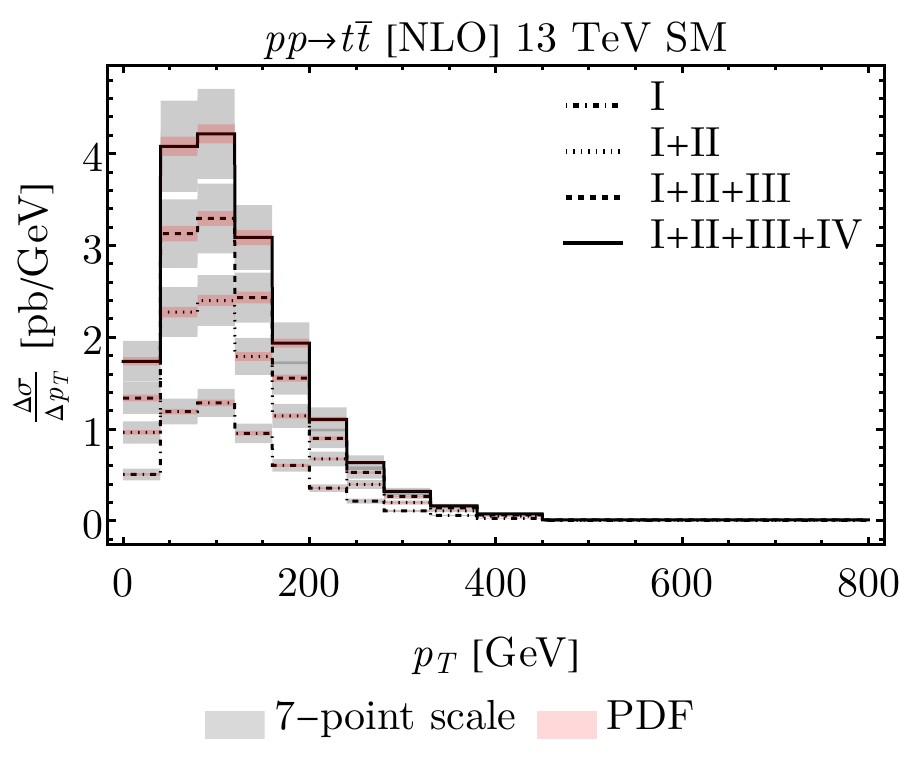}
    \includegraphics[width=0.32\linewidth]{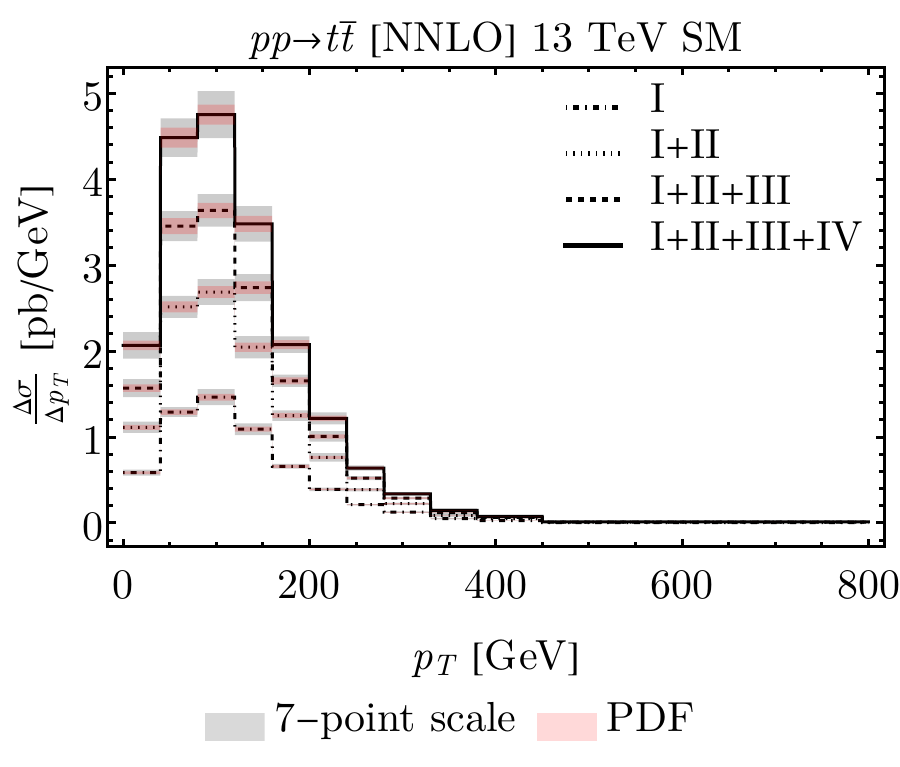}
    \includegraphics[width=0.32\linewidth]{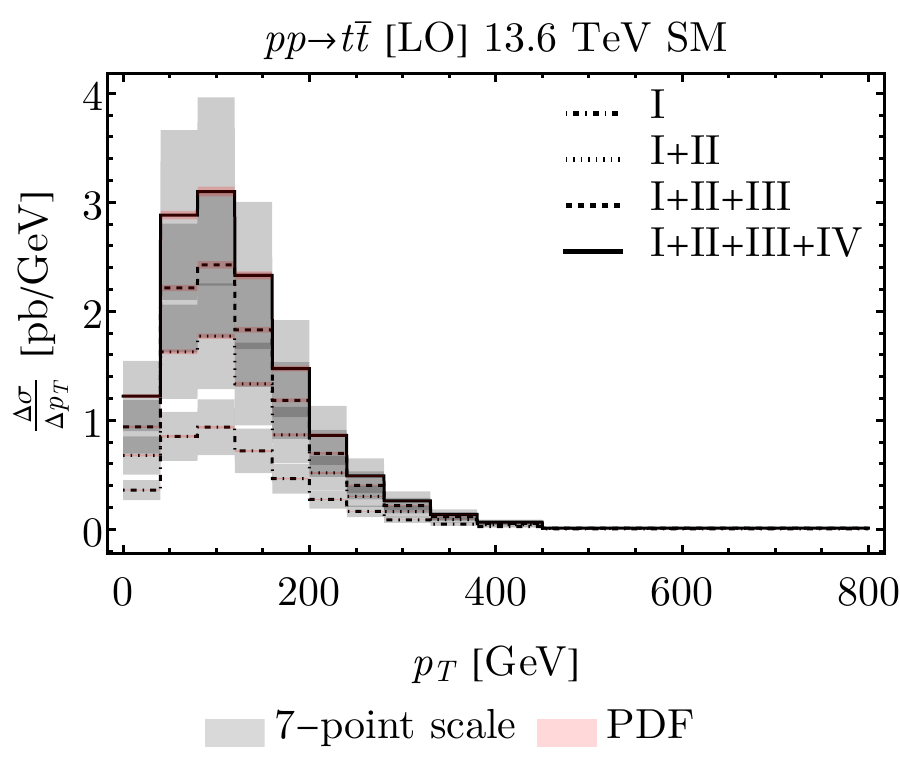}
    \includegraphics[width=0.32\linewidth]{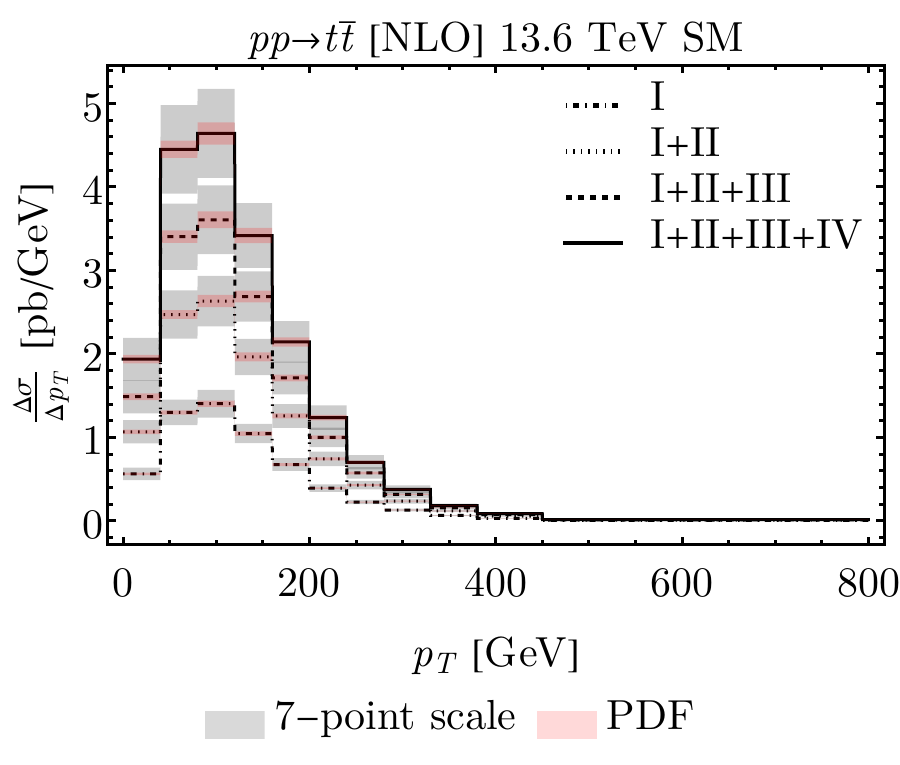}
    \includegraphics[width=0.32\linewidth]{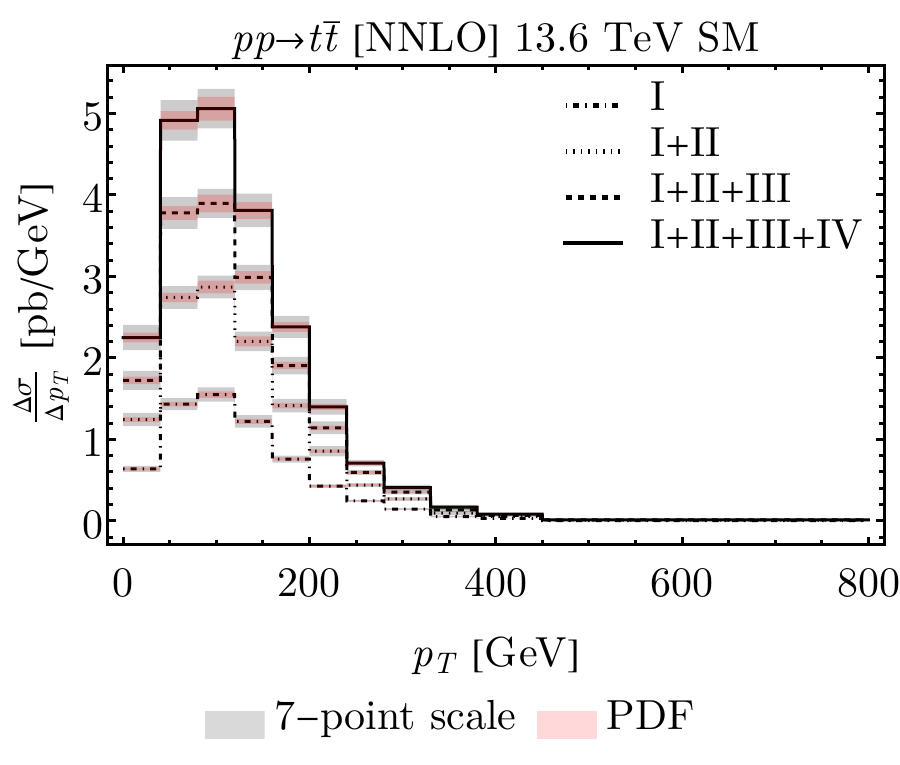}
    \caption{The same as Fig.~\ref{fig:pt_distr} but for the top-quark double-differential $p_T \times y$ distribution.}
    \label{fig:pty_distr}
\end{figure}

 
The top-quark single-differential $p_T$ and $y$ distributions and the double-differential $p_T \times y$ distribution across the bins in top-pair production in the SM together with 7-point scale variations, as well as PDF uncertainties, at LO, NLO, and NNLO at 13 TeV and 13.6 TeV are given in Figs.~\ref{fig:pt_distr}, \ref{fig:y_distr}, and \ref{fig:pty_distr}. In these figures, the black curves represent the central predictions, the gray bands show the 7-point scale uncertainty, and the red bands indicate the PDF uncertainty. For the NNLO distribution, the PDF uncertainty is obtained by carrying over the relative PDF variation evaluated at NLO. We have checked that this procedure reproduces the NNLO PDF sensitivity with excellent accuracy. In Fig.~\ref{fig:pty_distr}, the rapidity region I is denoted by dot-dashed lines, the combined I+II region by dotted lines, the combined I+II+III region by dashed lines, and the full rapidity spectrum I+II+III+IV by the solid lines. 
At 13 TeV, the scale uncertainty decreases from 27--38\% at LO to 12--13\% at NLO and 3.9--9.4\% at NNLO for the $p_T$ distribution. For the $y$ distribution, it decreases from 29--31\% to 11--12\% and then to 4.9--6.2\%. For the double-differential $p_T\times y$ distribution, the corresponding ranges are 26--43\%, 11--13\%, and 2.3--32\%. The PDF uncertainty is much smaller: at LO/NLO it is 1.3--2.3\% / 1.9--3.3\% for $p_T$, 1.5--2.2\% / 2.1--2.9\% for $y$, and 1.2--4.7\% / 2.0--6.5\% for $p_T\times y$. At 13.6 TeV, the pattern is very similar. For $p_T$, the scale uncertainty changes from 27--38\% at LO to 11--13\% at NLO and 4.4--11\% at NNLO. For $y$, it changes from 28--30\% to 11--12\% and then to 4.4--6.1\%. For $p_T\times y$, the corresponding ranges are 26--43\%, 11--14\%, and 3.7--28\%. The LO/NLO PDF uncertainty ranges are 1.2--2.3\% / 1.8--3.9\% for $p_T$, 1.4--2.2\% / 2.0--2.5\% for $y$, and 1.2--4.5\% / 1.9--5.8\% for $p_T\times y$.
\par
\begin{figure}
    [H]
    \centering
    \includegraphics[width=0.4875\linewidth]{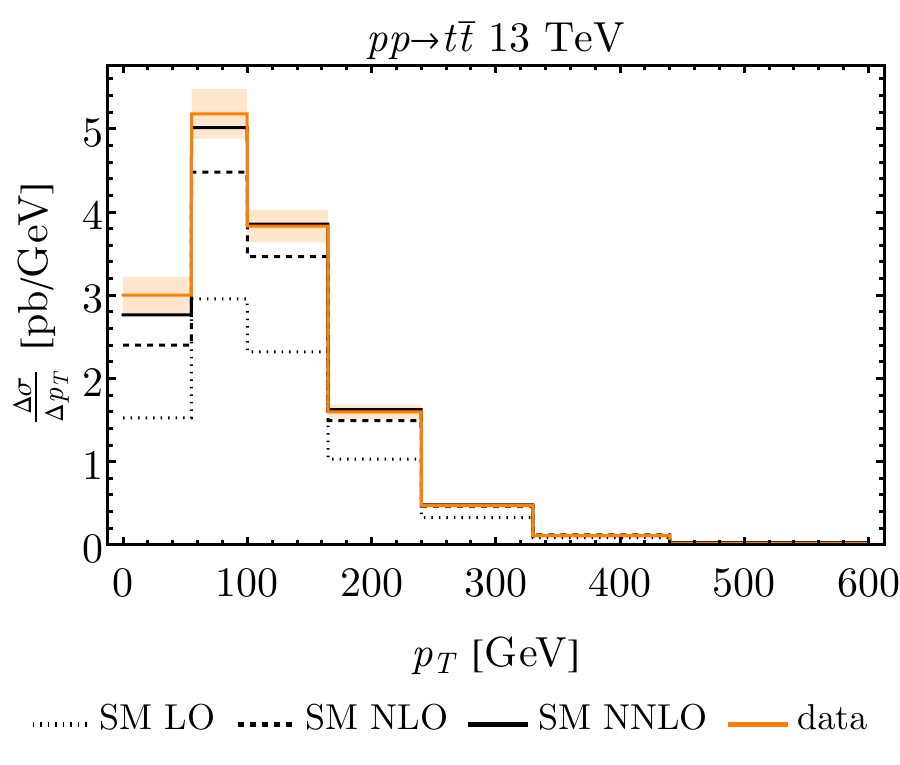}
    \includegraphics[width=0.4875\linewidth]{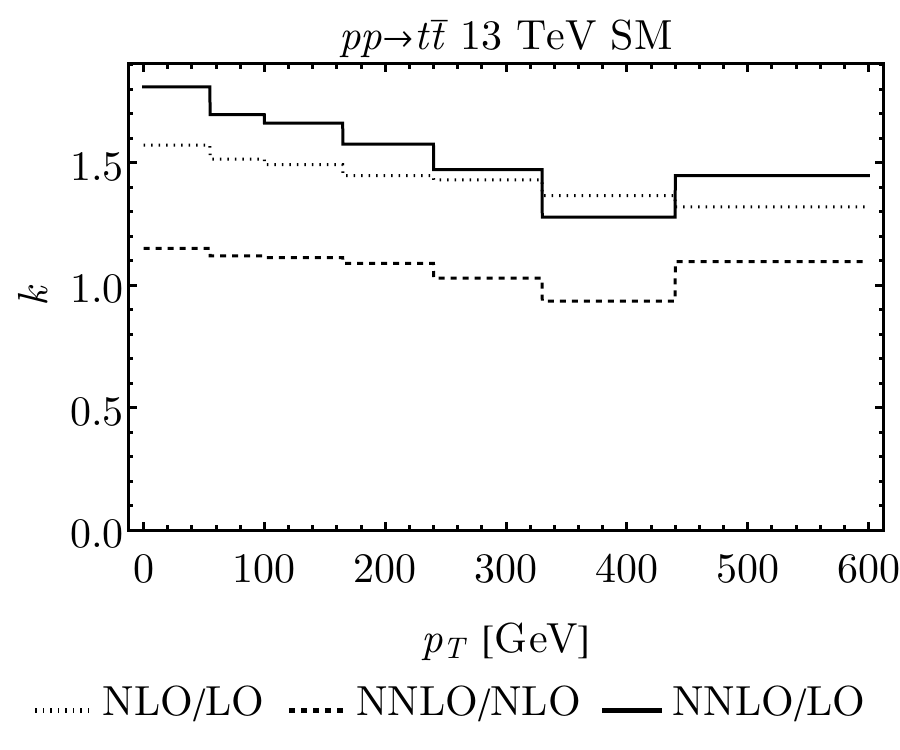}
    \includegraphics[width=0.4875\linewidth]{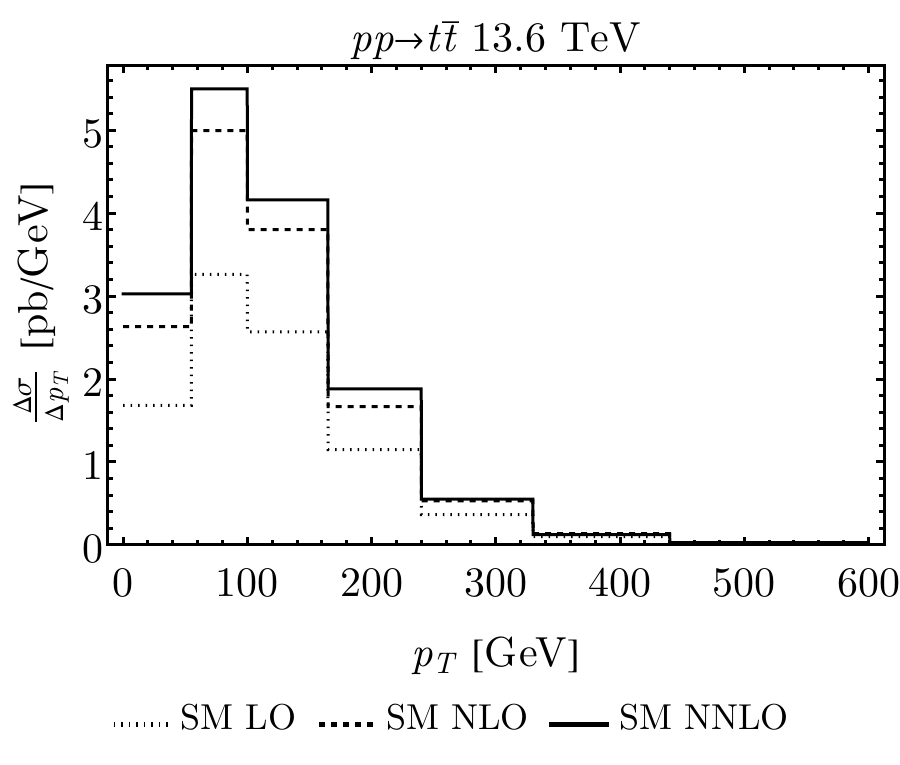}
    \includegraphics[width=0.4875\linewidth]{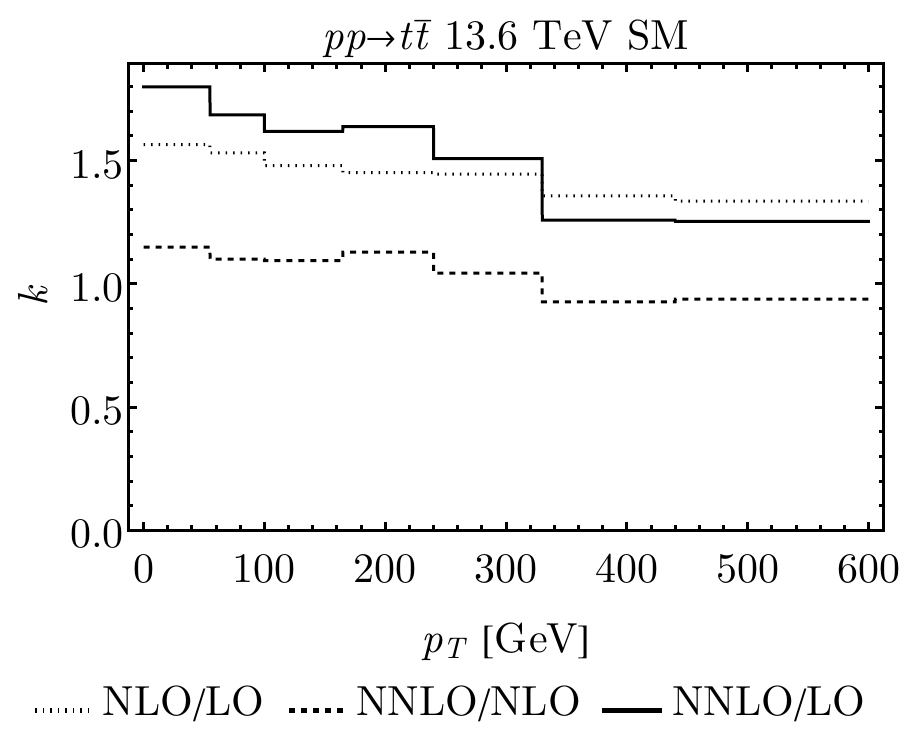}
    \caption{The top-quark single-differential $p_T$ distribution in $t\bar t$ production within the SM across the bins at LO, NLO, and NNLO (left) and the corresponding $k$-factors (right) at 13 TeV (top) and 13.6 TeV (bottom).}
    \label{fig:pt_distr_compare}
\end{figure}

\begin{figure}
    [H]
    \centering
    \includegraphics[width=0.4875\linewidth]{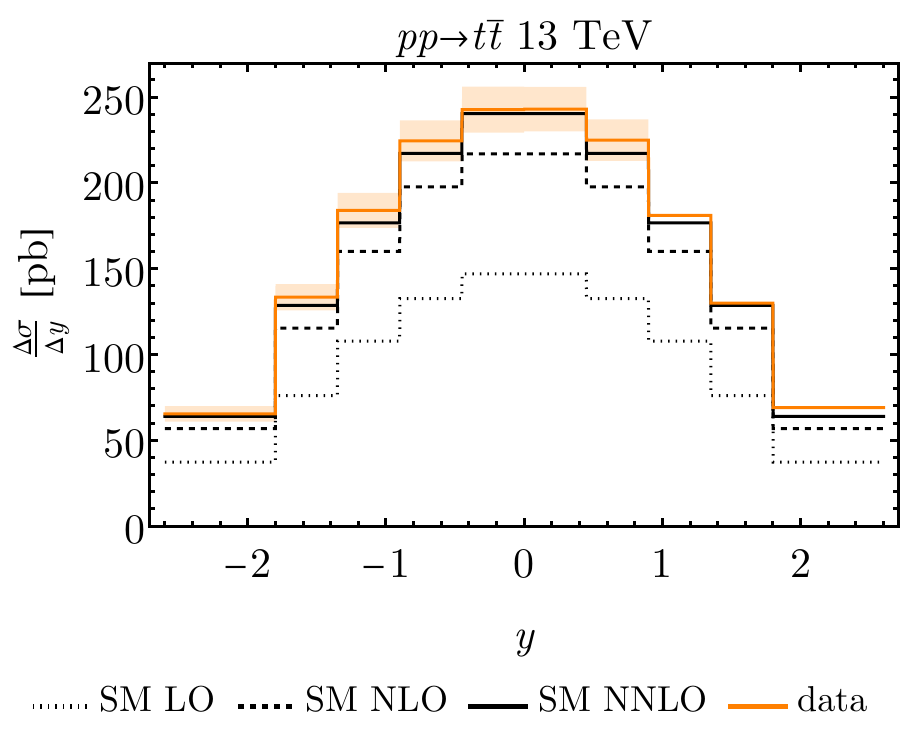}
    \includegraphics[width=0.4875\linewidth]{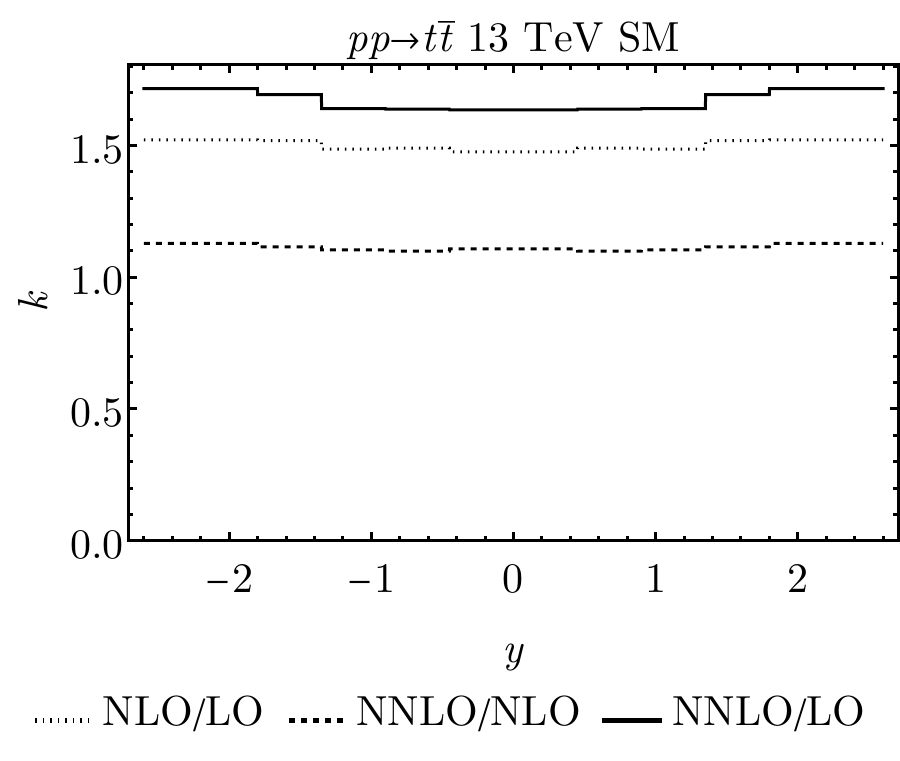}
    \includegraphics[width=0.4875\linewidth]{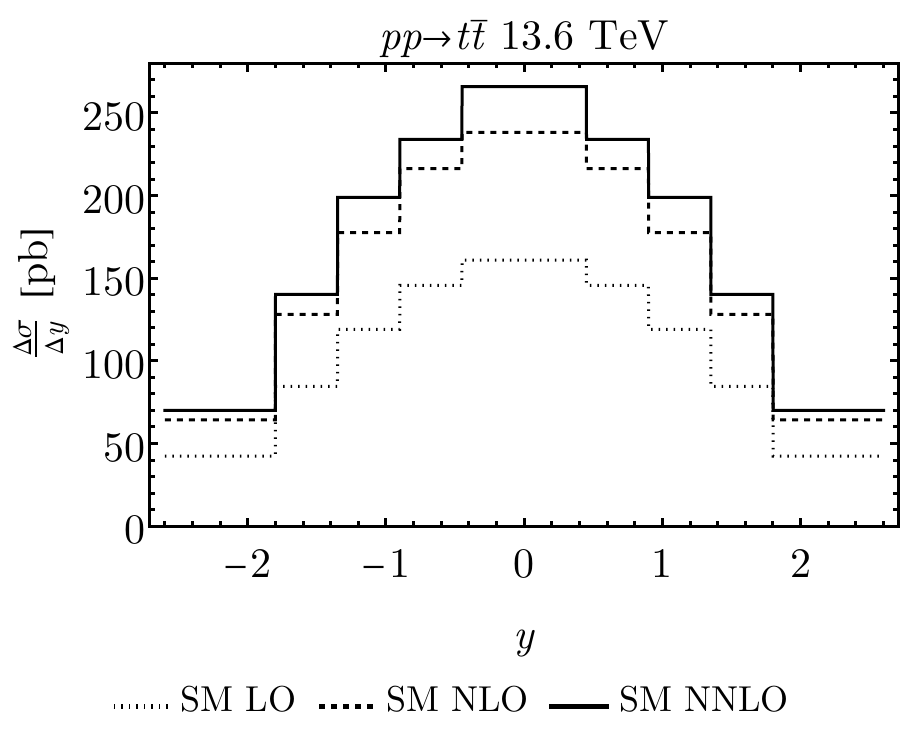}
    \includegraphics[width=0.4875\linewidth]{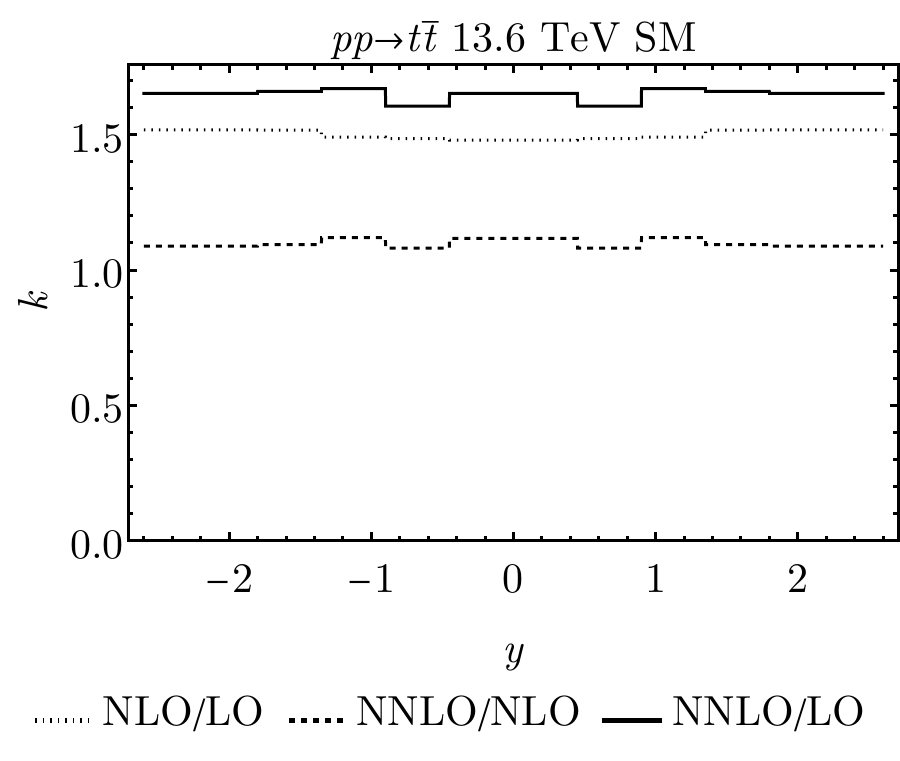}
    \caption{The same as Fig.~\ref{fig:pt_distr_compare} but for the top-quark single-differential $y$ distribution.}
    \label{fig:y_distr_compare}
\end{figure}

\begin{figure}
    [H]
    \centering
    \includegraphics[width=0.4875\linewidth]{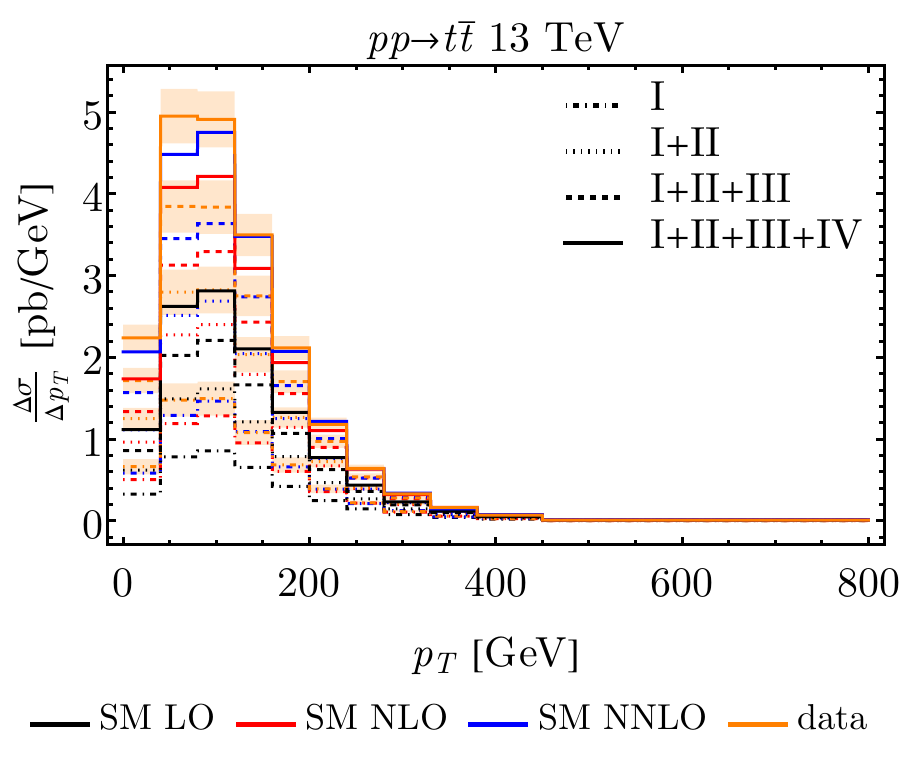}
    \includegraphics[width=0.4875\linewidth]{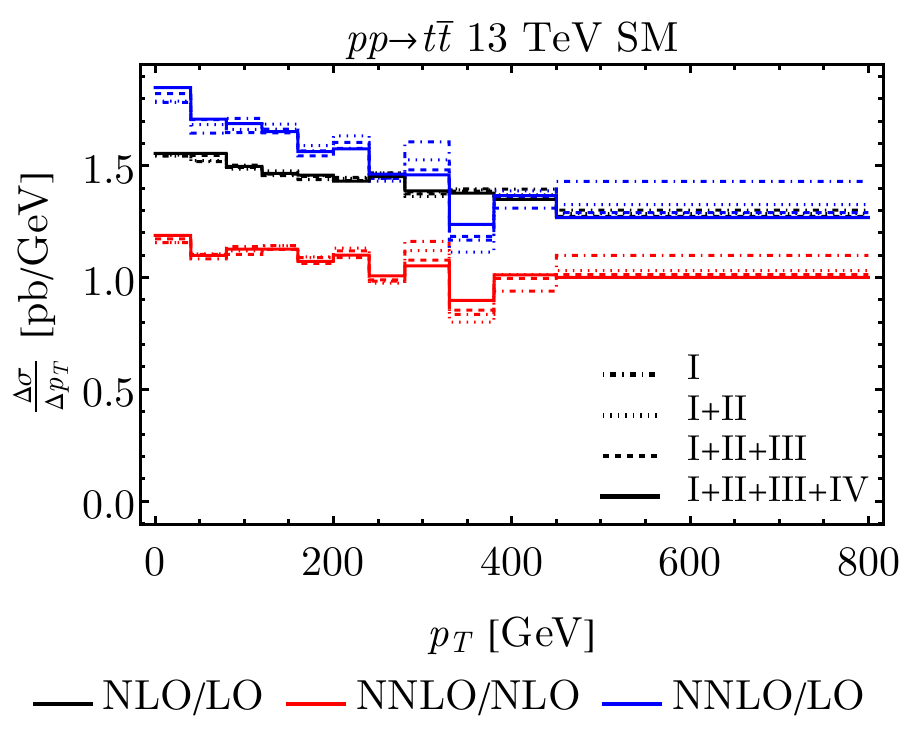}
    \includegraphics[width=0.4875\linewidth]{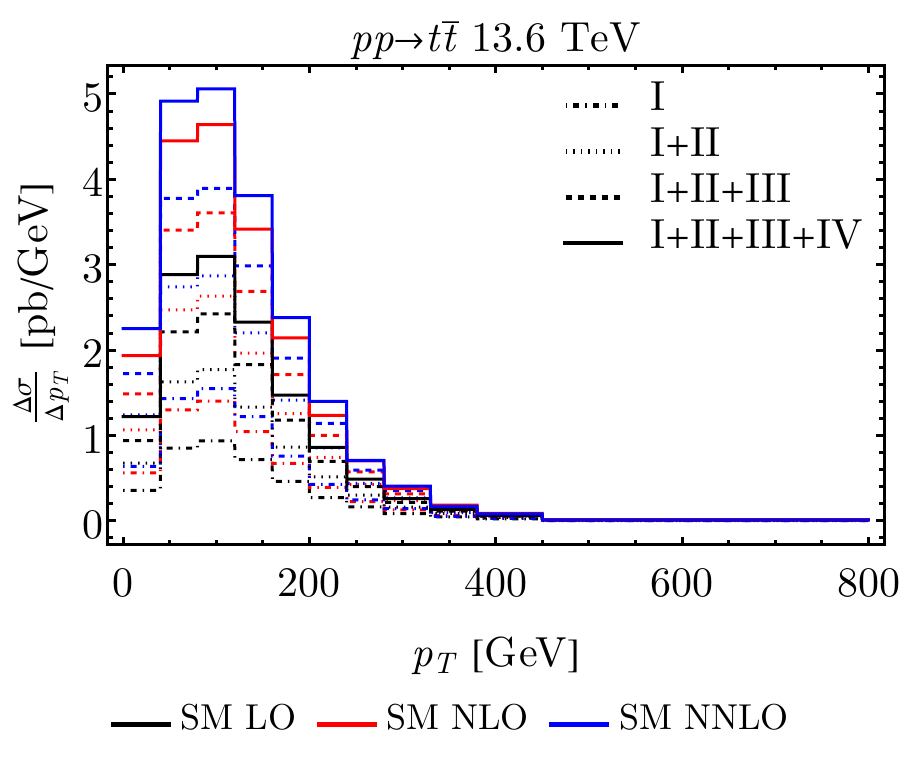}
    \includegraphics[width=0.4875\linewidth]{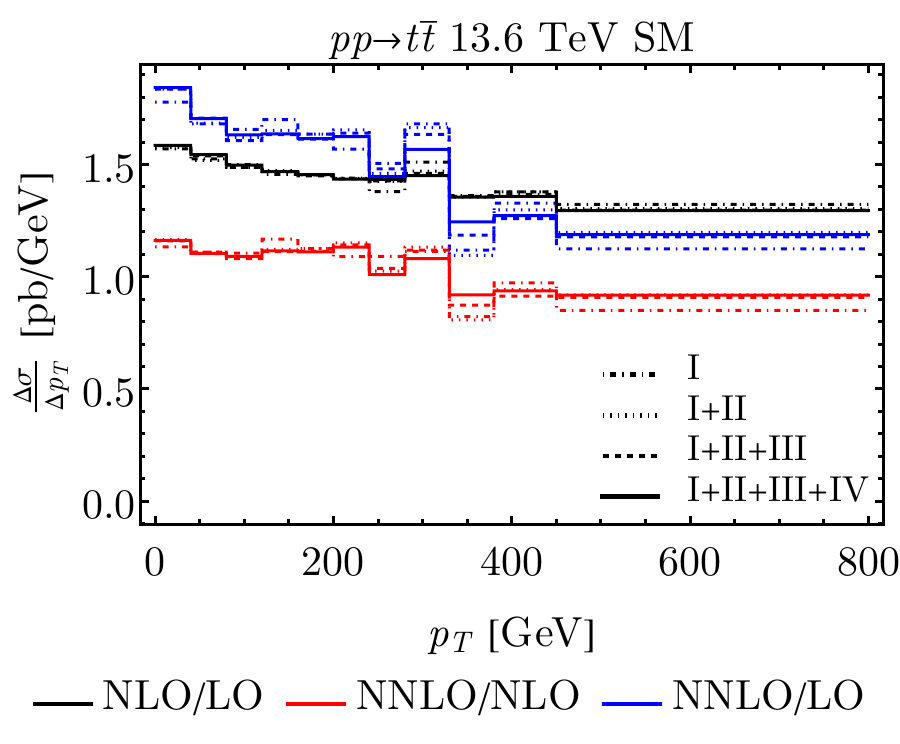}
    \caption{The same as Fig.~\ref{fig:pt_distr_compare} but for the top-quark double-differential $p_T \times y$ distribution.}
    \label{fig:pty_distr_compare}
\end{figure}

\begin{figure}
    [H]
    \centering
    \includegraphics[width=0.4875\linewidth]{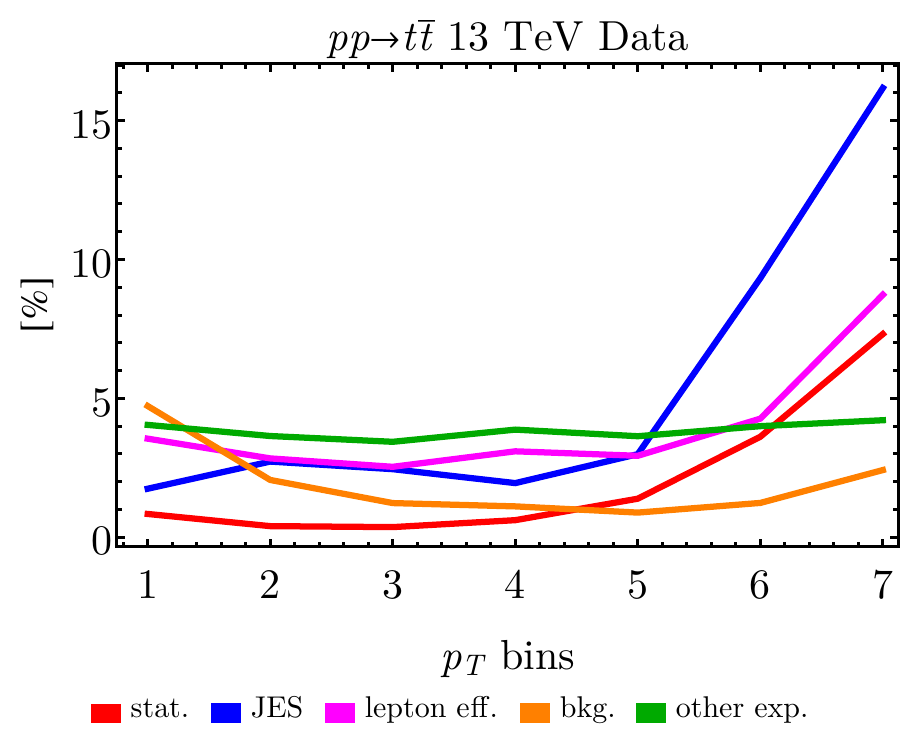}
    \includegraphics[width=0.4875\linewidth]{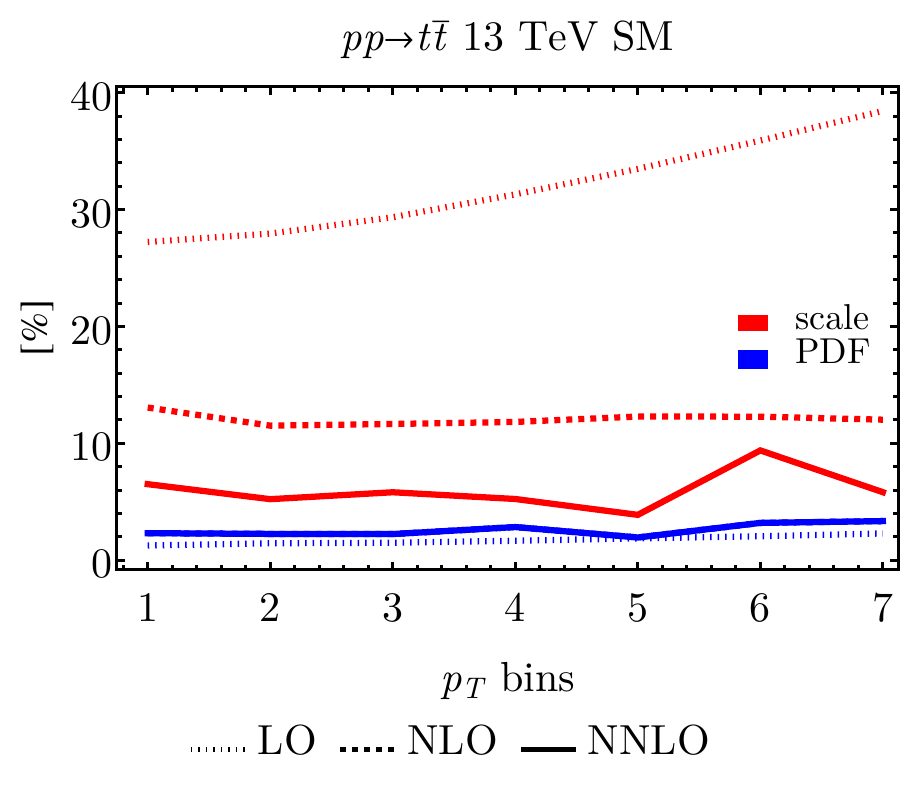}
    \includegraphics[width=0.4875\linewidth]{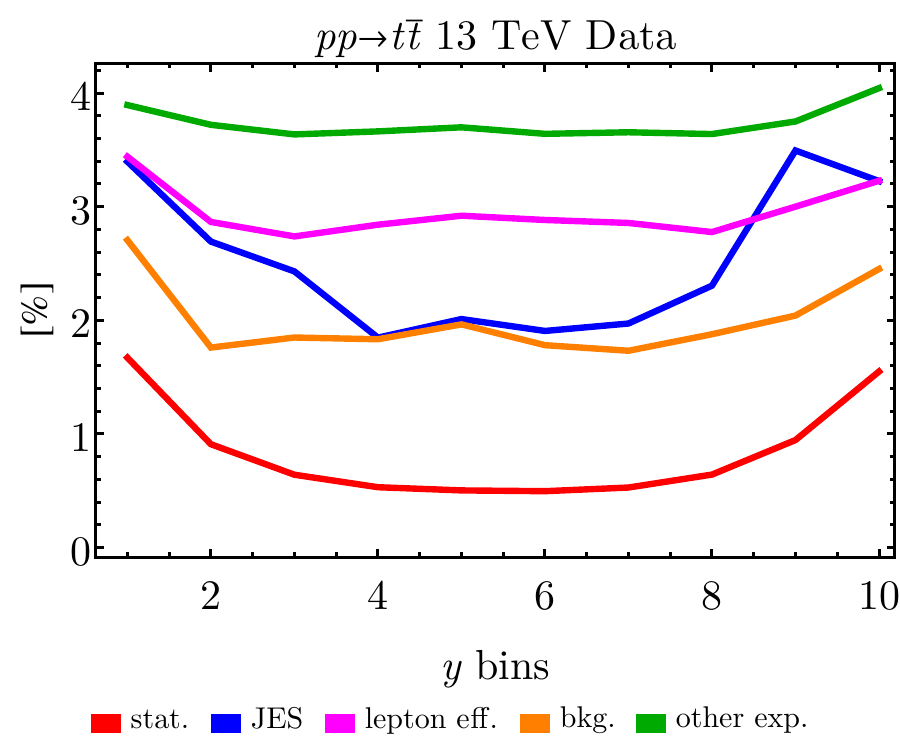}
    \includegraphics[width=0.4875\linewidth]{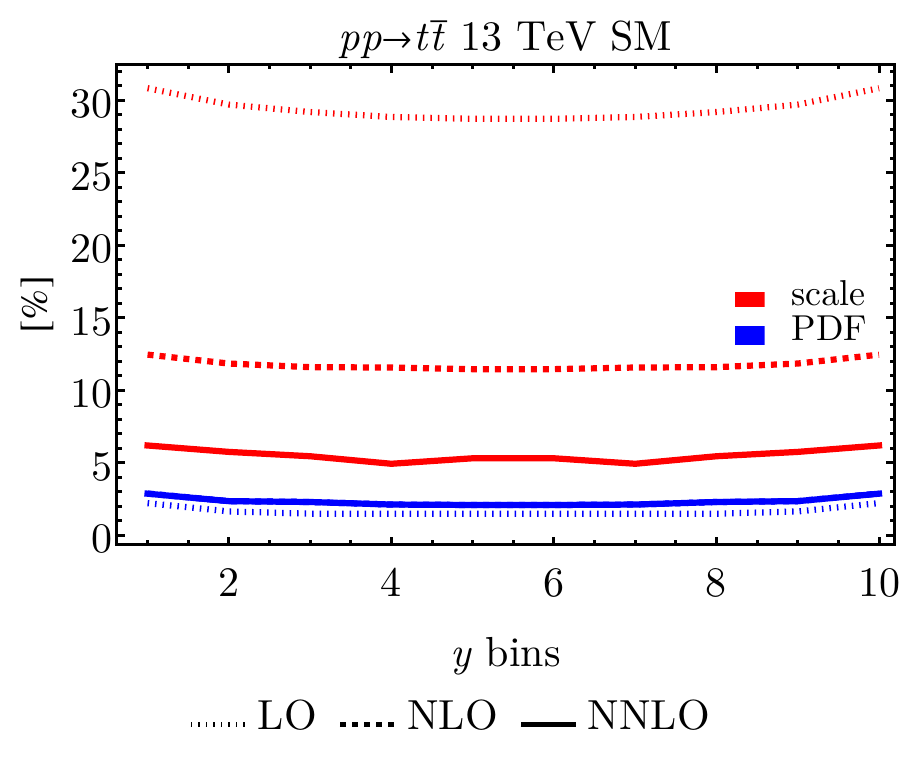}
    \includegraphics[width=0.4875\linewidth]{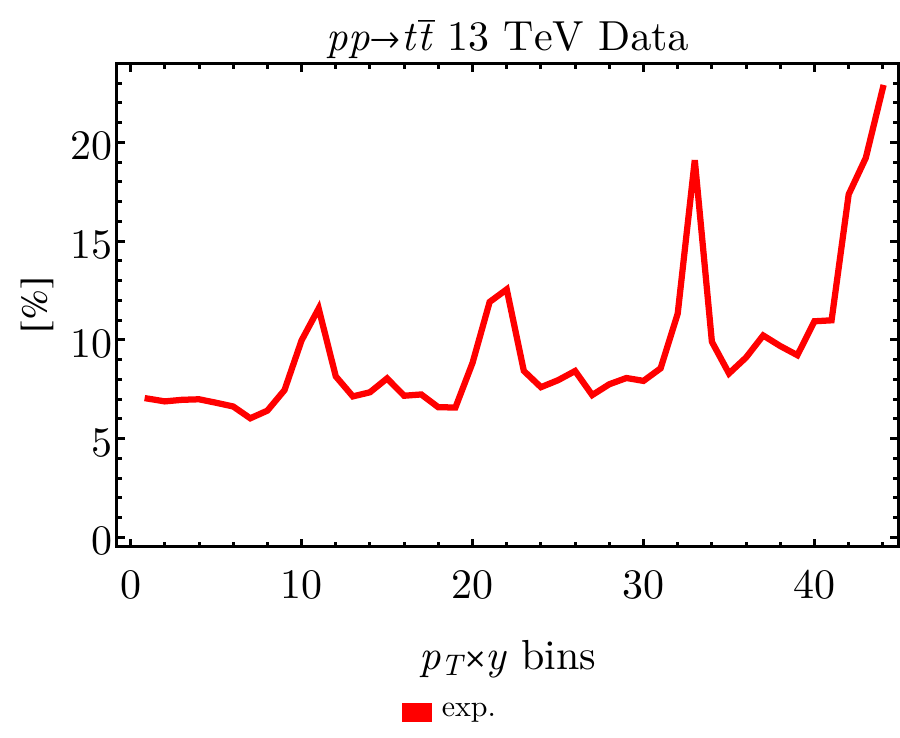}
    \includegraphics[width=0.4875\linewidth]{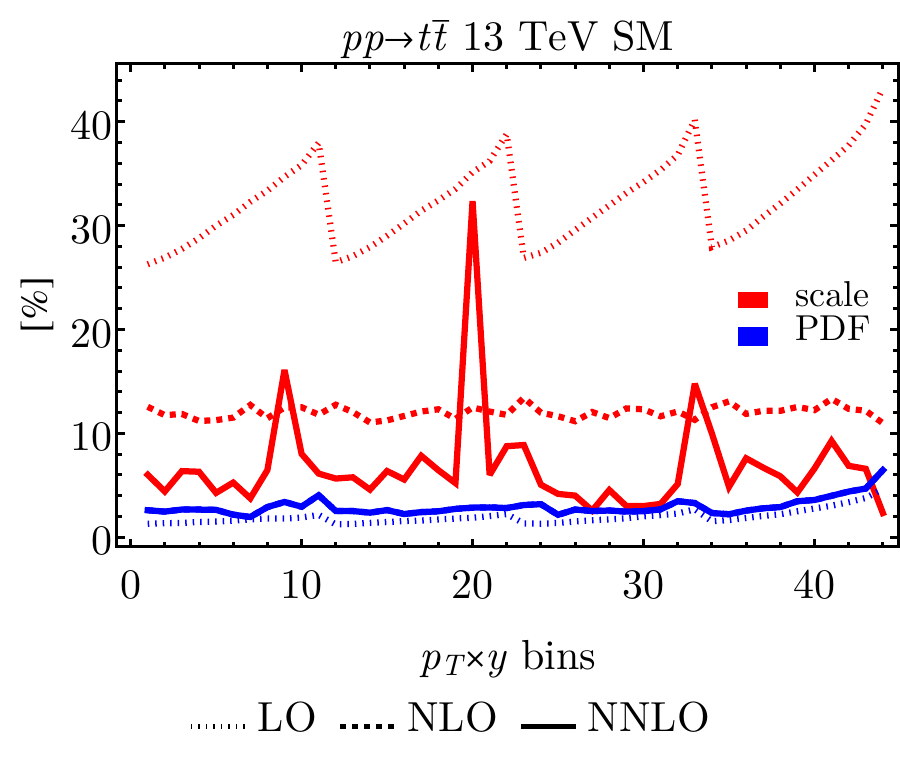}
    \caption{Relative uncertainty budget for the 13-TeV analysis: experimental uncertainties for the measured $p_T$ (top), $y$ (middle), and $p_T\times y$ (bottom) distributions, shown as percentages with respect to the data baseline (left), theoretical scale and PDF uncertainties, shown as percentages with respect to the LO, NLO, and NNLO SM predictions (right).}
    \label{fig:budget_13}
\end{figure}

 
In Figs.~\ref{fig:pt_distr_compare}, \ref{fig:y_distr_compare}, and \ref{fig:pty_distr_compare}, we compare the LO, NLO, and NNLO SM distributions and show the corresponding $k$-factors for the chosen bins at 13 TeV and 13.6 TeV. For the single-differential distributions, the dotted, dashed, and solid black lines represent the LO, NLO, and NNLO predictions, respectively. For the double-differential distributions, the dot-dashed, dotted, dashed, and solid lines correspond, for each $p_T$ bin, to the cumulative rapidity contributions labeled I, I+II, I+II+III, and I+II+III+IV. The perturbative orders LO, NLO, and NNLO are shown in black, red, and blue, respectively. Where available, the measured distribution is shown in orange, with the shaded band representing the total experimental uncertainty. For the single-differential distributions this is obtained by combining the statistical, background, jet energy scale, lepton efficiency, and other experimental uncertainties in quadrature, while for the double-differential distribution the total experimental uncertainty provided in the data file is used directly. The detailed breakdown of individual uncertainty contributions is shown in the error budget plots in Fig.~\ref{fig:budget_13}. In the $k$-factor panels for the single-differential distributions, the dotted, dashed, and solid lines denote the ratios NLO/LO, NNLO/NLO, and NNLO/LO, respectively. For the double-differential distributions, the same line styles are used for the cumulative rapidity contributions, while the colors black, red, and blue denote the ratios NLO/LO, NNLO/NLO, and NNLO/LO, respectively.
For the $p_T$ distribution, the NLO/LO $k$-factor is typically in the range 1.3--1.6 and decreases with increasing $p_T$, while the NNLO/NLO $k$-factor is nearly flat at about 1.1. Overall, the perturbative-order dependence is more pronounced in the $p_T$ distributions. By contrast, the $k$-factors for the rapidity distribution are essentially flat across the bins.
\par
\begin{figure}
    [H]
    \centering
    \includegraphics[width=0.4875\linewidth]{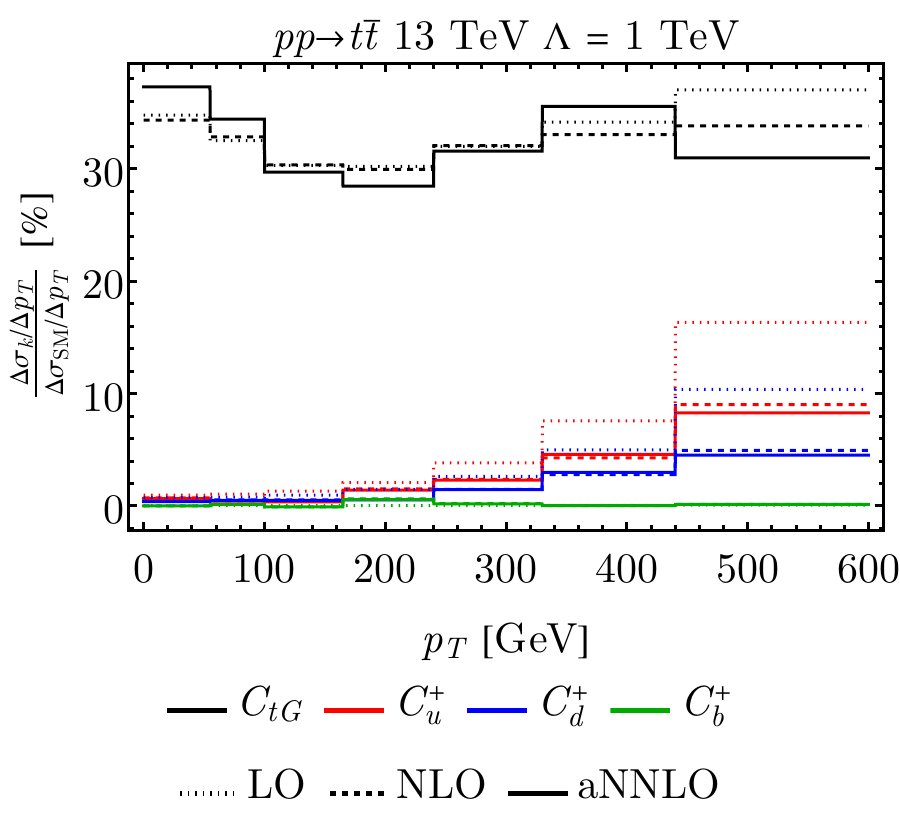}
    \includegraphics[width=0.4875\linewidth]{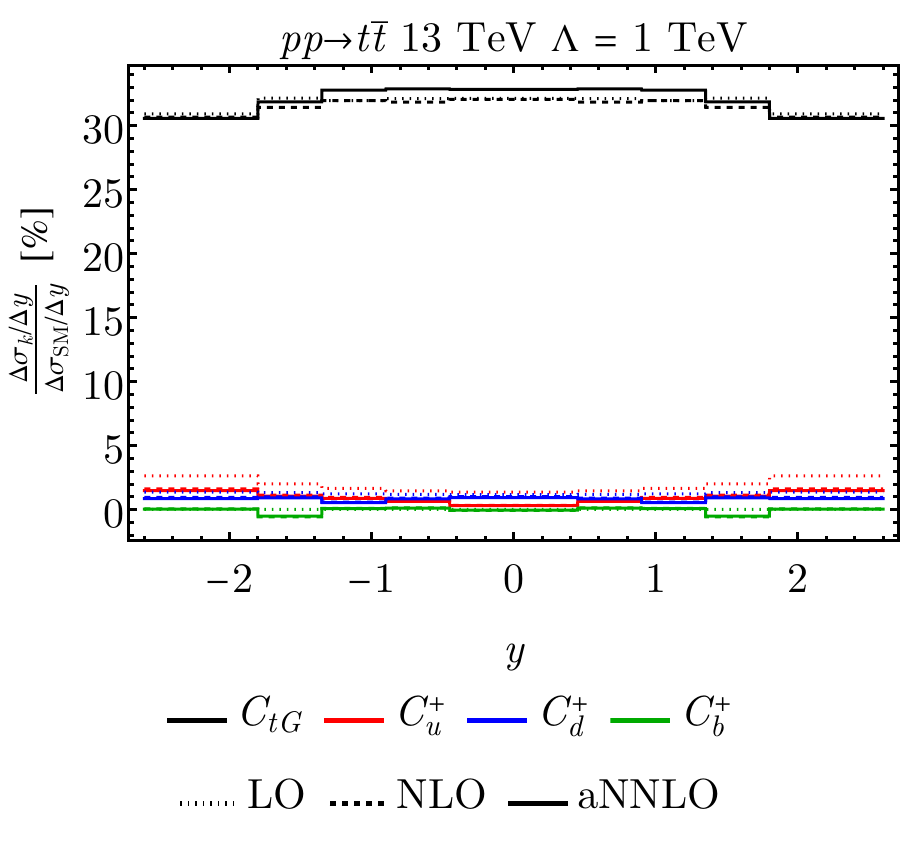}
    \includegraphics[width=0.4875\linewidth]{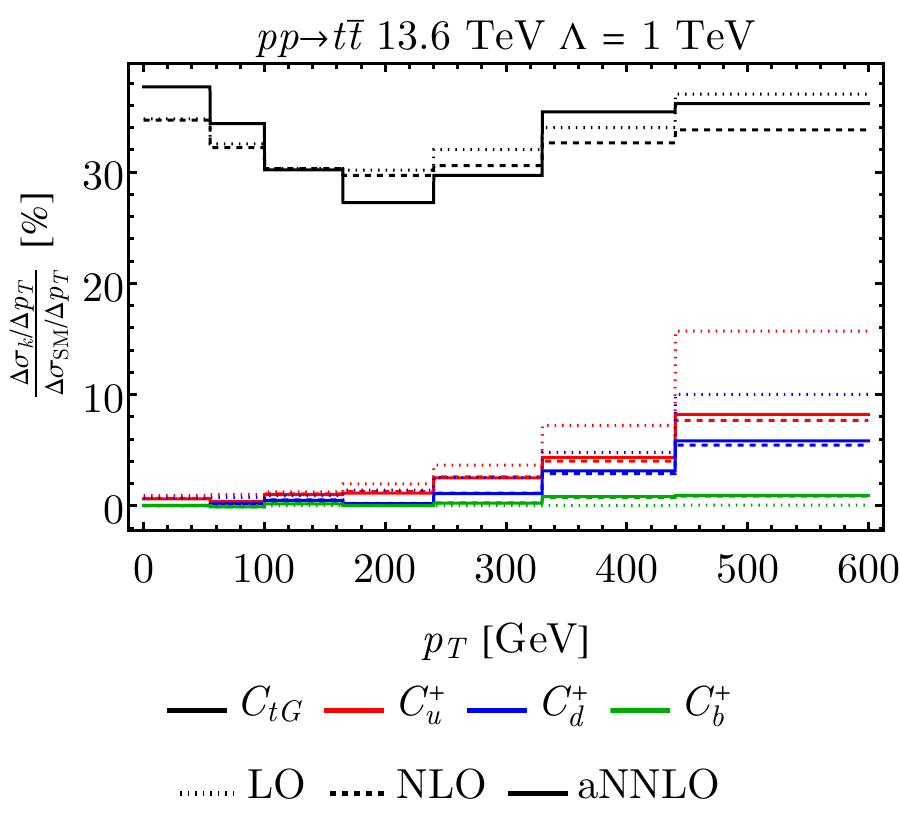}
    \includegraphics[width=0.4875\linewidth]{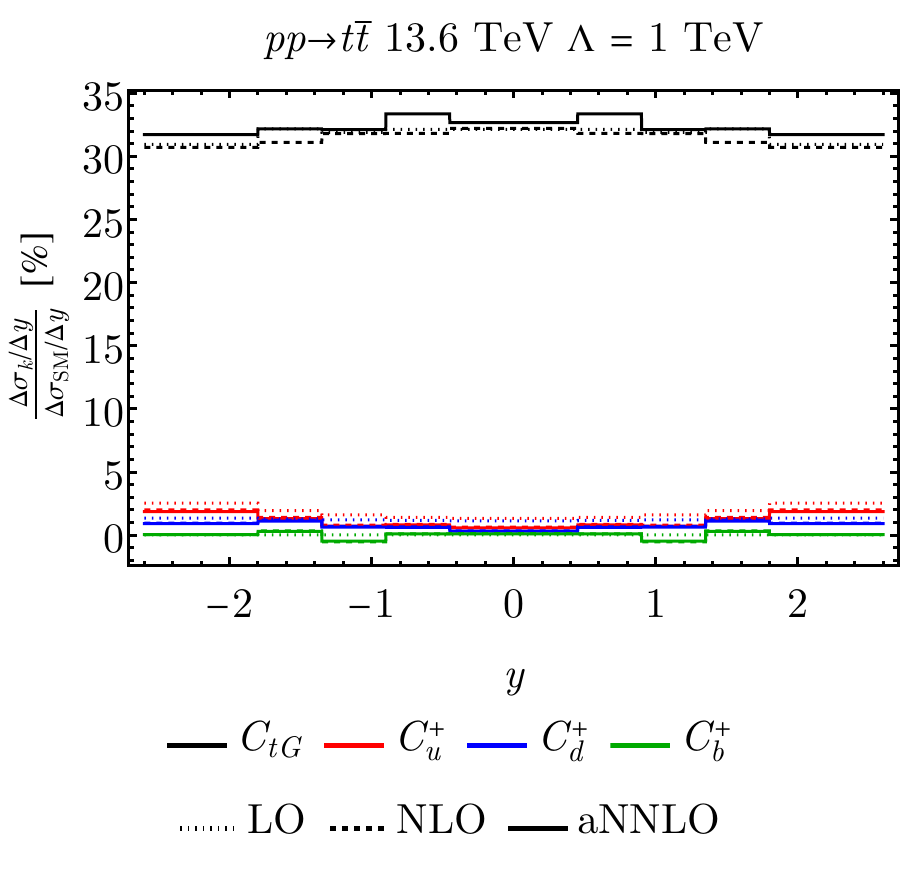}
    \caption{The SMEFT corrections to the top-quark single-differential $p_T$ (left) and $y$ distribution (right) characterized by $C_{tG}$, $C_u^+$, $C_d^+$, and $C_b^+$ relative to the SM in percent at LO, NLO, and aNNLO at 13 TeV (top) and 13.6 TeV (bottom).}
    \label{fig:pt_and_y_k_distr}
\end{figure}

\begin{figure}
    [H]
    \centering
    \includegraphics[width=0.32\linewidth]{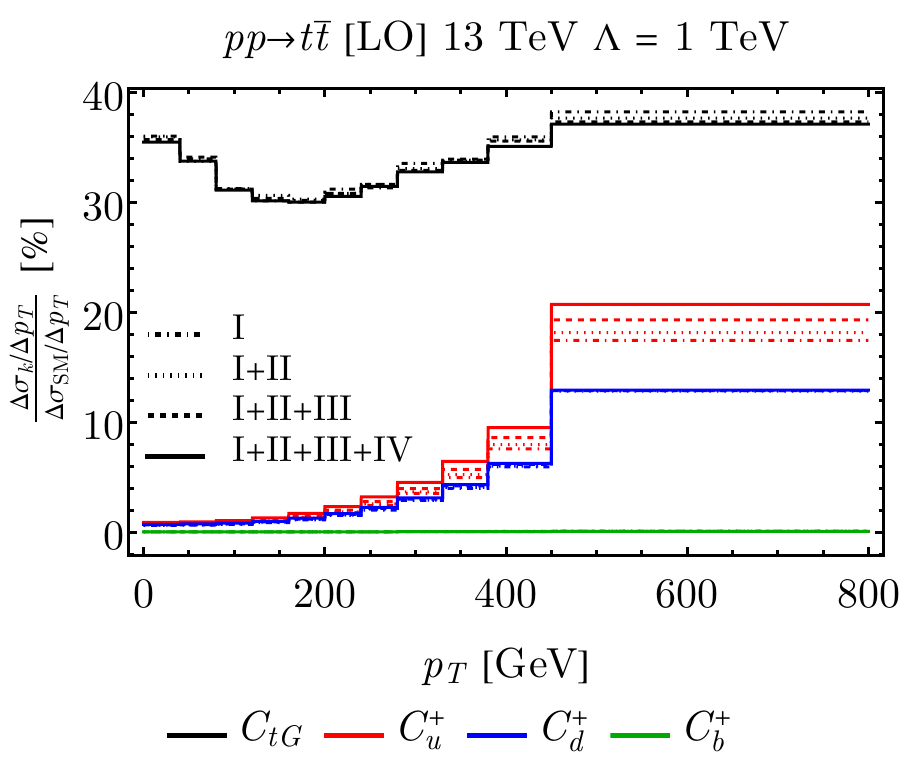}
    \includegraphics[width=0.32\linewidth]{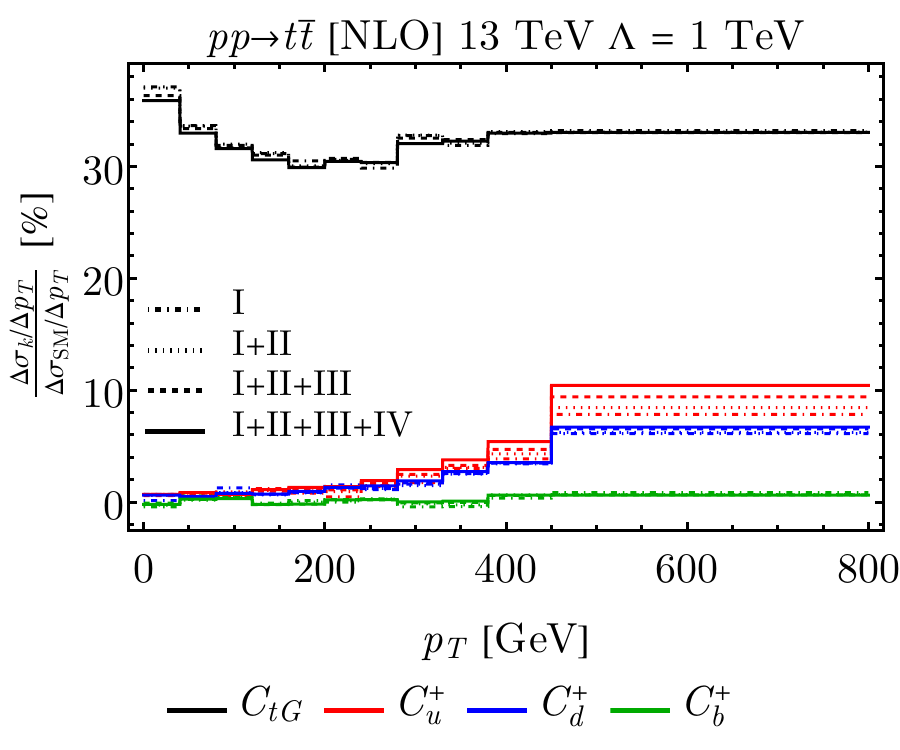}
    \includegraphics[width=0.32\linewidth]{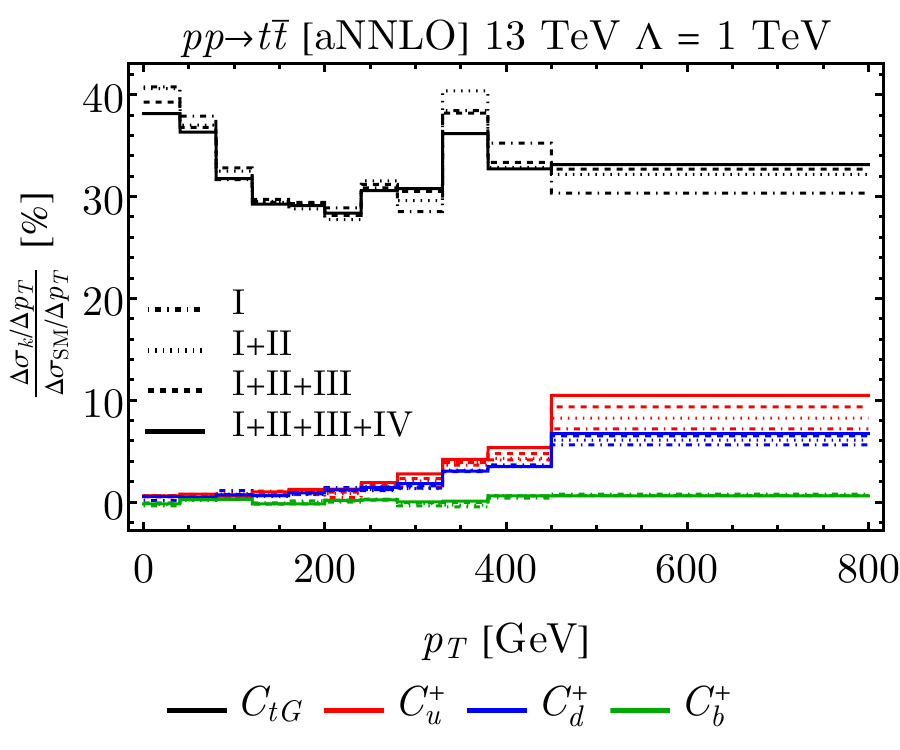}
    \includegraphics[width=0.32\linewidth]{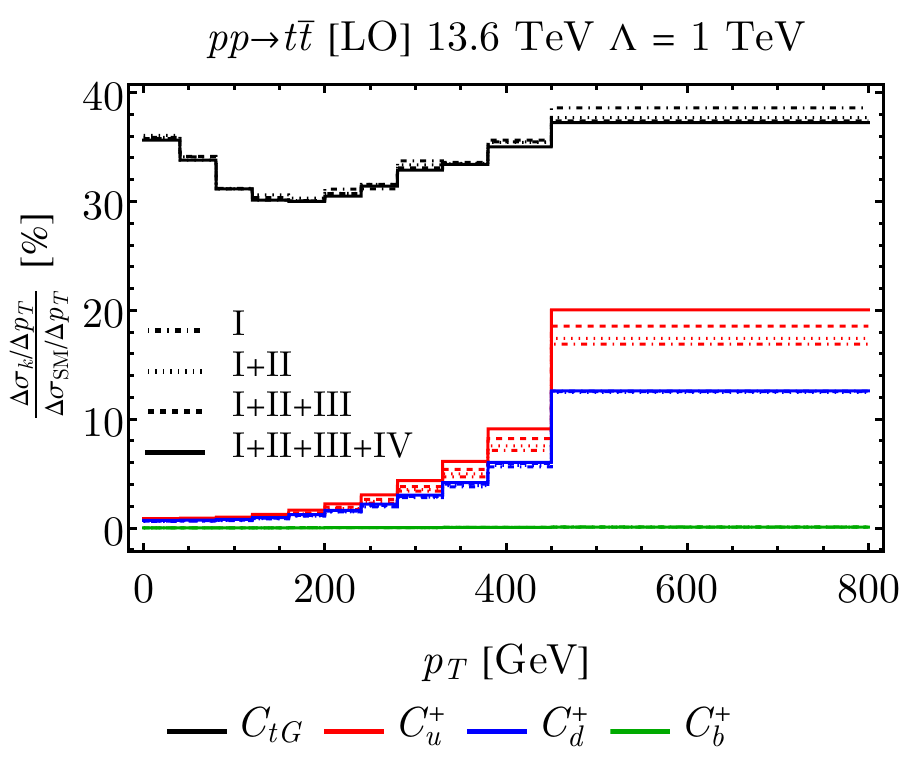}
    \includegraphics[width=0.32\linewidth]{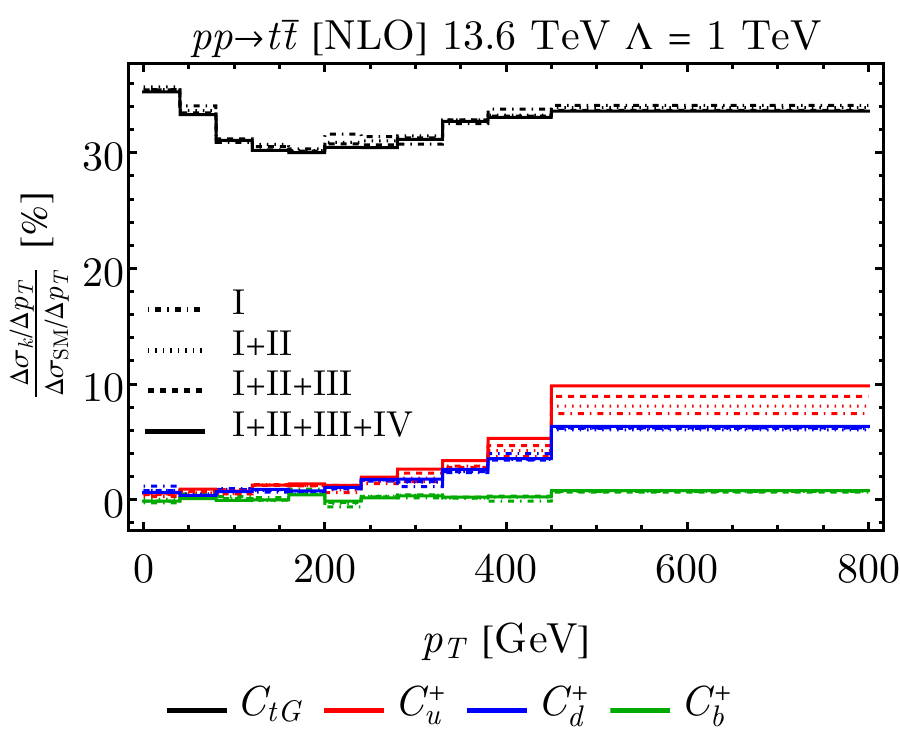}
    \includegraphics[width=0.32\linewidth]{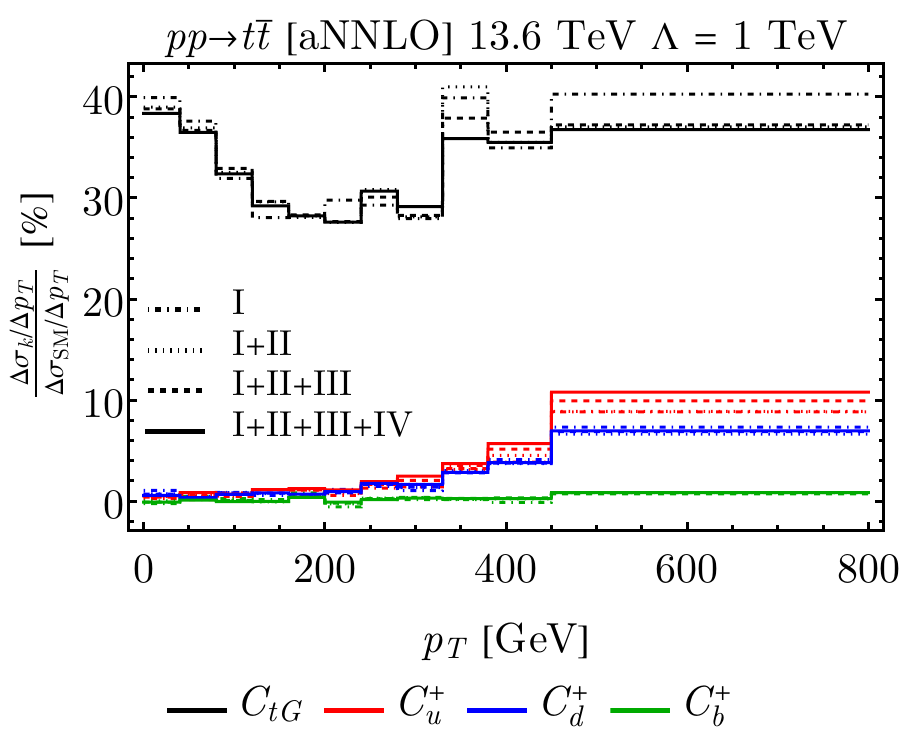}
    \caption{The SMEFT corrections to the top-quark double-differential $p_T \times y$ characterized by $C_{tG}$, $C_u^+$, $C_d^+$, and $C_b^+$ relative to the SM in percent at LO (left), NLO (middle), and aNNLO (right) at 13 TeV (top) and 13.6 TeV (bottom).}
    \label{fig:pty_k_distr}
\end{figure}

 
In Fig.~\ref{fig:pt_and_y_k_distr}, we present the linear SMEFT contributions to the single-differential top-quark $p_T$ and $y$ distributions as percent deviations from the SM at LO, NLO, and aNNLO for 13 TeV and 13.6 TeV. The corresponding linear SMEFT corrections to the double-differential $p_T \times y$ distribution are shown in the left, middle, and right panels of Fig.~\ref{fig:pty_k_distr} for LO, NLO, and aNNLO, respectively, for 13 TeV and 13.6 TeV. In these figures, black, red, blue, and green denote the contributions proportional to $C_{tG}$, $C_u^+$, $C_d^+$, and $C_b^+$, respectively. For the single-differential distributions, the dotted, dashed, and solid lines represent the LO, NLO, and aNNLO results, respectively. For the double-differential distributions, the same line styles as in the previous figures are used to indicate the cumulative rapidity contributions.
The SMEFT sensitivity plots show that the contributions from $C_{tG}$ are typically of order 30--40\% in both the $p_T$ and $y$ distributions, but they exhibit substantially stronger shape variation in $p_T$ than in $y$. The rapidity distributions are instead nearly flat, with only weak bin-to-bin dependence. The contributions from the four-quark operators grow with $p_T$, as expected, because these contact interactions are enhanced at larger partonic energies and thus become more important in the tails of the distributions. Their effects are ordered as $C_u^+ > C_d^+ > C_b^+$. We also observe that the relative size of each SMEFT contribution decreases at higher perturbative orders, indicating negative higher-order corrections to the normalized SMEFT shifts.
\par
Now, we describe our statistical treatment. We perform linear SMEFT fits at dimension six on the four Wilson coefficients of interest $C_{tG}$, $C_u^+$, $C_d^+$, and $C_b^+$. Our observables are the single-differential $p_T$ and $y$ distributions, as well as the double-differential $p_T\times y$ distribution. For our fitting, we perform standard $\chi^2$ test statistics. We define
\begin{gather}
    \chi^2 = (\mathcal O - \hat{\mathcal O})^\top \mathcal H (\mathcal O - \hat{\mathcal O}),
\end{gather}
where $\mathcal O$ is the data for the 13-TeV fits and the pseudodata for the 13.6-TeV projections, $\hat{\mathcal O}$ is the model function, namely the SMEFT observable linear in Wilson coefficients, and $\mathcal H = \mathcal E^{-1}$ is the inverse of the covariance matrix, $\mathcal E$. The covariance matrix has two parts, namely experimental and theoretical:
\begin{gather}
    \mathcal E = \mathcal E^{\rm exp.} + \mathcal E^{\rm theo.}.
\end{gather}
For the 13-TeV fits, the experimental part is directly given by the data:
\begin{gather}
    \mathcal E^{\rm exp.} = \mathcal E^{\rm stat.} + \mathcal E^{\rm JES} + \mathcal E^{\rm lepton\ eff.} + \mathcal E^{\rm bkg.} + \mathcal E^{\rm other exp.}
\end{gather}
for the single-differential distributions, where each component is built as a diagonal matrix in the absence of an official covariance matrix, and 
\begin{gather}
    \mathcal E^{\rm exp.} = \frac{1}{\mathcal B^2} \mathcal E^{\rm CMS}
\end{gather}
for the double-differential distribution, where the factor $\mathcal B$ has been explained above. For the 13.6-TeV projections, we build the experimental covariance matrix as
\begin{gather}
    \mathcal E^{\rm exp.}_{bb'} = I_{bb'} (\delta_b^{\rm stat.} \oplus \delta_b^{\rm uncorr.})^2 + \rho_{bb'} \delta_b^{\rm corr.} \delta_{b'}^{\rm corr.},
\end{gather}
where $b,b'=1,\ldots,N_{\rm bin}$ is the bin index, $N_{\rm bin} = 7, 10, 44$ is the number of bins for the $p_T$, $y$, and $p_T \times y$ distributions, respectively, $I$ is the identity matrix, $\delta_b^{\rm stat./uncorr./corr.}$ is the absolute statistical, uncorrelated, and correlated uncertainties at the $b$th bin, respectively, given by
\begin{gather}
    \delta_b^{\rm stat.} = {1 \over \sqrt{\sigma_{{\rm SM}, b} \mathcal L}} \times \sigma_{{\rm SM}, b}, \quad 
    \delta_b^{\rm uncorr.} = 5\% \times \sigma_{{\rm SM}, b}, \quad 
    \delta_b^{\rm corr.} = 1.7\% \times \sigma_{{\rm SM}, b},
\end{gather}
and $\oplus$ indicates sum in quadrature. We assume $\rho_{bb'} = 1$ for all bins. For the central SM value $\sigma_{{\rm SM}, b}$ for the projections, we use the NNLO predictions. 
\par 
In order to build the theoretical covariance matrix, we use the fully correlated PDF uncertainties and the bin-by-bin 7-point scale variations,
\begin{gather}
    \mathcal E^{\rm theo.} = \mathcal E^{\rm PDF} + \mathcal E^{\rm scale}.
\end{gather}
The PDF covariance matrix for Hessian sets is calculated as
\begin{gather}
    \mathcal E^{\rm PDF}_{bb'} = \frac14 \sum_{m=1}^{N_{\rm PDF}/2} (\mathcal O_{2m} - \mathcal O_{2m-1})_b (\mathcal O_{2m} - \mathcal O_{2m-1})_{b'},
\end{gather}
where $N_{\rm PDF}$ is the number of PDF members in a given set, and the scale variations are provided by \matrix~at each bin asymmetrically in the form $m_{-l}^{+u}$, which we symmetrize simply as $\delta^{\rm scale} = \sqrt{(u^2+l^2)/2}$, and using these values, 
\begin{gather}
    \mathcal E_{bb'}^{\rm scale} = I_{bb'} \pp{\delta_b^{\rm scale}}^2.
\end{gather}
We emphasize that the theoretical uncertainties are not built directly into the model function itself. Instead, the PDF and scale variations are evaluated at the SM reference point and propagated consistently throughout the analysis. Treating these uncertainties separately for the SMEFT contributions would effectively amount to double counting and would also induce an artificial dependence of the covariance matrix on the Wilson coefficient $C$. We checked the impact of this assumption by assigning to the bin-by-bin SMEFT corrections the same relative PDF and scale uncertainties as in the SM prediction, thereby introducing a controlled $C$-dependence in the theory covariance. The resulting Fisher matrices differ by less than 1\%, which translates into negligible changes in the extracted limits and correlation patterns. For this reason, we adopt the covariance matrix fixed at the SM point as our baseline choice.
\par 
For the data inputs, we use the measured central values reported in the official 13-TeV data tables. For the 13.6-TeV study, instead, we construct pseudodata by fluctuating the NNLO SM prediction according to the assumed experimental uncertainties:
\begin{gather}
    \mathcal O_b = \mathcal O_{{\rm SM}, b} + r_b (\delta_b^{\rm stat.} \oplus \delta_b^{\rm uncorr.}) + r' \delta_b^{\rm corr.},
\end{gather}
with $r_b, r' \in \mathcal N(0,1)$ denoting standard normal random variables. The correlated component is multiplied by the same random factor across all bins, so that it induces a coherent shift over the full distribution. Since a statistically meaningful projection requires an ensemble of pseudodata sets, we label each pseudoexperiment by an index $e$ and write the corresponding chi-squared function as $\chi_e^2$. The associated best-fit point, denoted by $\bar{\vec C}_e$, is obtained by requiring the gradient in parameter space to vanish,
\begin{gather}
    \vec \nabla \chi_e^2 (\bar{\vec C}_e) = \vec 0,
\end{gather}
while the Fisher matrix is given by the Hessian evaluated at that point,
\begin{gather}
    \mathcal F_e = \vec \nabla \vec \nabla \chi_e^2 (\bar{\vec C}_e).
\end{gather}
The SMEFT best-fit parameters averaged over the full ensemble of pseudoexperiments are then defined as
\begin{gather}
    \bar{\vec C} = \pp{\sum_{e=1}^{N_{\rm exp.}} \mathcal F_e}^{-1} \sum_{e=1}^{N_{\rm exp.}} \mathcal F_e \bar{\vec C}_e,
\end{gather}
and the corresponding averaged Fisher matrix is
\begin{gather}
    \mathcal F = {1 \over N_{\rm exp.}} \sum_{e=1}^{N_{\rm exp.}} \mathcal F_e,
\end{gather}
where $N_{\rm exp.}$ denotes the total number of pseudoexperiments. In the special case of a linear model function, which applies to observables linear in the Wilson coefficients, this procedure simplifies substantially. Because $\chi_e^2$ is then quadratic in the SMEFT parameters, the Fisher matrix is the same for every pseudoexperiment, which we denote simply by $\mathcal F$. In that case, the average best-fit point reduces to
\begin{gather}
    \bar{\vec C} = {1 \over N_{\rm exp.}} \sum_{e=1}^{N_{\rm exp.}} \bar{\vec C}_e,
\end{gather}
and, in the limit of infinitely many pseudoexperiments, the mean best-fit value approaches zero. This is exactly what one expects, since the pseudodata are generated by applying Gaussian fluctuations centered on the SM prediction, so the fitted SMEFT coefficients should average out to zero, while the main quantity of interest is their spread rather than any individual central value. As a result, for a given 13.6-TeV projection it is sufficient to perform only a single pseudoexperiment in order to extract the Fisher matrix. We emphasize that for the 13-TeV fit, we perform the $\chi^2$-minimization only once without running any pseudoexperiments. 
\par
For combined analyses, such as those including the $p_T$, $y$, and $p_T\times y$ distributions simultaneously, we treat the corresponding measurements as independent for simplicity and construct the total Fisher matrix by summing the individual ones.
\par
\begin{figure}
    [H]
    \centering
    \includegraphics[width=0.32\linewidth]{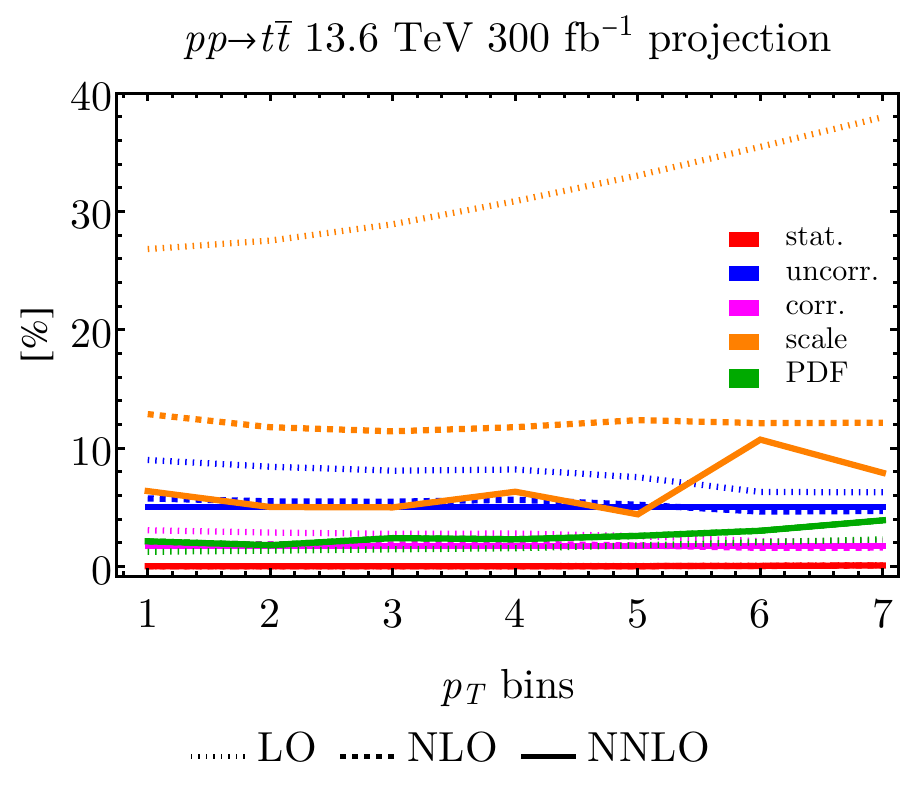}
    \includegraphics[width=0.32\linewidth]{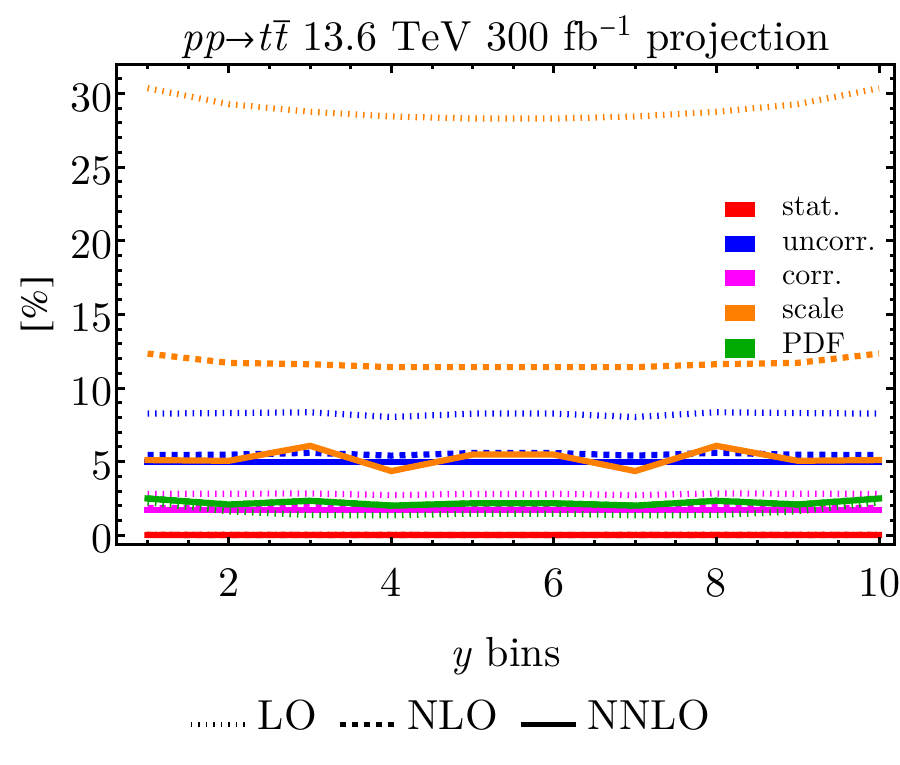}
    \includegraphics[width=0.32\linewidth]{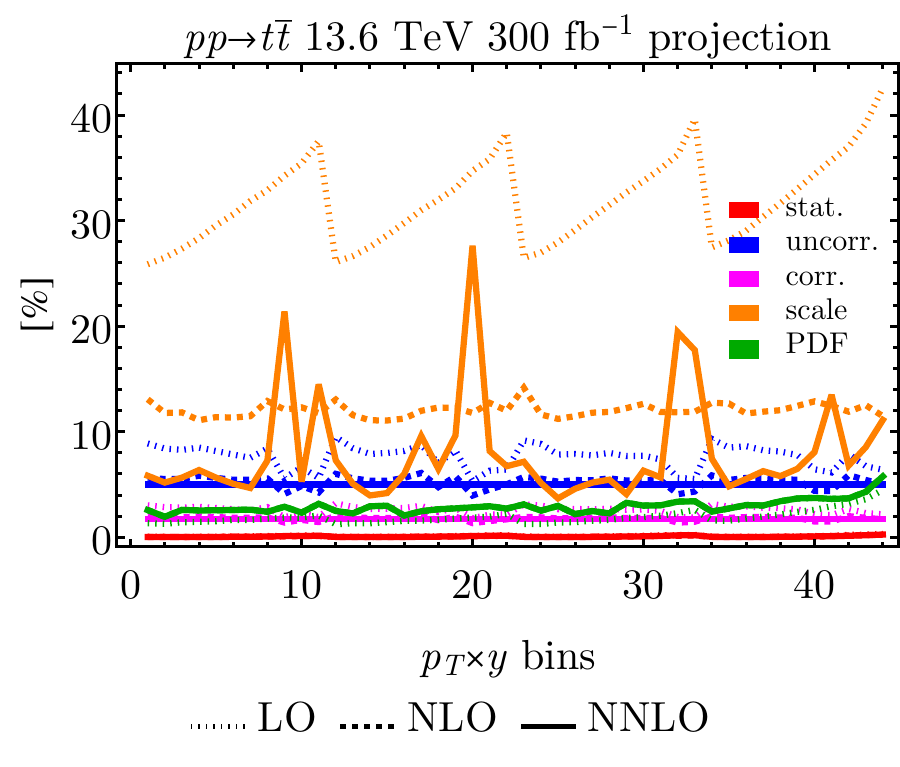}
    \caption{Relative uncertainty budget for the 13.6-TeV projections with $\mathcal L = 300~\fb^{-1}$: statistical, assumed uncorrelated and correlated experimental uncertainties, together with the scale and PDF uncertainties, shown as percentages with respect to the LO, NLO, and NNLO SM baselines for the $p_T$ (left), $y$ (middle), and $p_T\times y$ (right) distributions.}
    \label{fig:budget_13.6}
\end{figure}

 
Before presenting the fit results in the next section, we summarize the error budget in our fits and projections, namely the contributions of different kinds of uncertainties to the diagonal entries of the total covariance matrix. In Figs.~\ref{fig:budget_13} and \ref{fig:budget_13.6}, we show for each bin the percent uncertainty with respect to the assumed baseline at 13 TeV and 13.6 TeV, respectively. In Fig.~\ref{fig:budget_13}, we display the relative uncertainty budget bin by bin for the 13-TeV analysis, quoted as percentages with respect to the corresponding baseline. In the left panels, we show the experimental uncertainties for the measured $p_T$ and $y$ distributions relative to the data central values, while for the $p_T\times y$ distribution we show only the total experimental uncertainty per bin. In the right panels, we show the theoretical uncertainties relative to the LO, NLO, and NNLO SM predictions. For the experimental components in the $p_T$ and $y$ panels, the red curves denote statistical uncertainties, the blue curves JES effects, the magenta curves lepton-efficiency uncertainties, the orange curves background-related uncertainties, and the green curves the remaining experimental systematics. For the $p_T\times y$ case, only the total experimental uncertainty is shown, indicated by the red curve. On the theory side, red denotes scale uncertainty and blue denotes PDF uncertainty, while dotted, dashed, and solid lines correspond to LO, NLO, and NNLO, respectively. In the $p_T$ distribution, the statistical component is least important in the region where the signal is largest, and in that same region the other experimental systematics provide the dominant contribution; this pattern is even more visible in the $y$ distribution. For the theoretical uncertainties, the scale variation clearly dominates over the PDF contribution, and overall it provides the largest uncertainty in the fit. For the $p_T\times y$ distribution, the bins are ordered first in rapidity and then in transverse momentum, which gives rise to the oscillatory pattern seen in the plots. Whenever the largest $p_T$ bin is reached within a given rapidity interval, both experimental and theoretical uncertainties increase. Around the region of maximal sensitivity, the experimental uncertainties are typically of order $6$--$8\%$. The scale uncertainty changes substantially when going from LO to NLO and then to NNLO. At NNLO, the PDF uncertainty is typically around $2$--$4\%$, while the scale uncertainty is roughly $4$--$6\%$ near the peak region. In Fig.~\ref{fig:budget_13.6}, we show the corresponding uncertainty budget for the 13.6-TeV projections assuming an integrated luminosity of $300~\fb^{-1}$. Here the red curves represent statistical uncertainties, the blue curves the assumed $5\%$ fully uncorrelated experimental uncertainty, the magenta curves the assumed $1.7\%$ fully correlated luminosity uncertainty, the orange curves the scale variation, and the green curves the PDF uncertainty. As before, dotted, dashed, and solid lines correspond to LO, NLO, and NNLO predictions. We find that the statistical uncertainty is generally the smallest contribution, followed by the correlated luminosity term. The PDF uncertainty is also subleading, while the dominant contributions come from the scale variation and the assumed $5\%$ uncorrelated experimental component, which are comparable in size over much of the kinematic range.

\section{SMEFT fit results\label{sec:results}}
In this section, we present the SMEFT fit results. As a first step, we report the $\chi^2/{\rm dof}$ values obtained from the $p_T$, $y$, and $p_T\times y$ fits at 13 TeV for LO, NLO, and a/NNLO predictions. The number of degrees of freedom, ${\rm dof}$, is defined as the difference between the number of bins included in the fit and the number of fitted parameters. Here, a/NNLO denotes the setup in which the SM prediction is taken at NNLO, while the SM-SMEFT interference is evaluated at aNNLO.
\begin{align}
    \chi^2{}_{p_T}^{\rm LO}/{\rm dof} &= 2.17 C_{tG}^2+0.00435 C_b^+ C_{tG}+0.326 C_d^+ C_{tG}+0.486 C_u^+ C_{tG}-8.03 C_{tG}\nn+2.66\times 10^{-6} \left(C_b^+\right){}^2+0.0259 \left(C_d^+\right){}^2+0.0618 \left(C_u^+\right){}^2-0.00662 C_b^+\nn+0.000485 C_b^+ C_d^+-0.378 C_d^++0.000738 C_b^+ C_u^++0.0800 C_d^+ C_u^+\nn-0.544 C_u^++8.59, \\
    \chi^2{}_{p_T}^{\rm NLO}/{\rm dof} &= 9.91 C_{tG}^2+0.0983 C_b^+ C_{tG}+0.763 C_d^+ C_{tG}+1.22 C_u^+ C_{tG}-5.09 C_{tG}\nn+0.000842 \left(C_b^+\right){}^2+0.0299 \left(C_d^+\right){}^2+0.0876 \left(C_u^+\right){}^2-0.0180 C_b^+\nn+0.00326 C_b^+ C_d^++0.00956 C_d^++0.00669 C_b^+ C_u^++0.101 C_d^+ C_u^+\nn+0.0521 C_u^++1.62, \\
    \chi^2{}_{p_T}^{\rm a/NNLO}/{\rm dof} &= 23.3 C_{tG}^2+0.225 C_b^+ C_{tG}+1.42 C_d^+ C_{tG}+2.21 C_u^+ C_{tG}-0.999 C_{tG}\nn+0.00200 \left(C_b^+\right){}^2+0.0463 \left(C_d^+\right){}^2+0.133 \left(C_u^+\right){}^2+0.0125 C_b^+\nn+0.00660 C_b^+ C_d^++0.143 C_d^++0.0140 C_b^+ C_u^++0.154 C_d^+ C_u^++0.284 C_u^+\nn+0.605, \\
    \chi^2{}_{y}^{\rm LO}/{\rm dof} &= 1.72 C_{tG}^2+0.00259 C_b^+ C_{tG}+0.135 C_d^+ C_{tG}+0.193 C_u^+ C_{tG}-7.65 C_{tG}\nn+1.01\times 10^{-6} \left(C_b^+\right){}^2+0.00267 \left(C_d^+\right){}^2+0.00579 \left(C_u^+\right){}^2-0.00570 C_b^+\nn+0.000101 C_b^+ C_d^+-0.303 C_d^++0.000139 C_b^+ C_u^++0.00773 C_d^+ C_u^+\nn-0.439 C_u^++8.58, \\
    \chi^2{}_{y}^{\rm NLO}/{\rm dof} &= 7.94 C_{tG}^2-0.0302 C_b^+ C_{tG}+0.449 C_d^+ C_{tG}+0.453 C_u^+ C_{tG}-7.11 C_{tG}\nn+0.000623 \left(C_b^+\right){}^2+0.00659 \left(C_d^+\right){}^2+0.00814 \left(C_u^+\right){}^2+0.0127 C_b^+\nn-0.00119 C_b^+ C_d^+-0.202 C_d^+-0.00118 C_b^+ C_u^++0.0128 C_d^+ C_u^+\nn-0.219 C_u^++1.66, \\
    \chi^2{}_{y}^{\rm a/NNLO}/{\rm dof} &= 18.9 C_{tG}^2-0.0486 C_b^+ C_{tG}+0.957 C_d^+ C_{tG}+0.927 C_u^+ C_{tG}-3.27 C_{tG}\nn+0.00130 \left(C_b^+\right){}^2+0.0127 \left(C_d^+\right){}^2+0.0149 \left(C_u^+\right){}^2-0.000699 C_b^+\nn-0.00197 C_b^+ C_d^+-0.0807 C_d^+-0.00193 C_b^+ C_u^++0.0235 C_d^+ C_u^+\nn-0.0978 C_u^++0.215, \\
    \chi^2{}_{p_T\times y}^{\rm LO}/{\rm dof} &= 0.241 C_{tG}^2+0.000625 C_b^+ C_{tG}+0.0615 C_d^+ C_{tG}+0.0972 C_u^+ C_{tG}\nn-0.669 C_{tG}+5.86\times 10^{-7} \left(C_b^+\right){}^2+0.0104 \left(C_d^+\right){}^2+0.0328 \left(C_u^+\right){}^2\nn-0.000404 C_b^++0.000136 C_b^+ C_d^++0.00622 C_d^++0.000207 C_b^+ C_u^+\nn+0.0356 C_d^+ C_u^++0.0252 C_u^++0.927, \\
    \chi^2{}_{p_T\times y}^{\rm NLO}/{\rm dof} &= 0.580 C_{tG}^2+0.00893 C_b^+ C_{tG}+0.0957 C_d^+ C_{tG}+0.144 C_u^+ C_{tG}-0.0763 C_{tG}\nn+0.000878 \left(C_b^+\right){}^2+0.0145 \left(C_d^+\right){}^2+0.0473 \left(C_u^+\right){}^2+0.0105 C_b^+\nn+0.00301 C_b^+ C_d^++0.0847 C_d^++0.00468 C_b^+ C_u^++0.0474 C_d^+ C_u^++0.136 C_u^+\nn+0.388, \\
    \chi^2{}_{p_T\times y}^{\rm a/NNLO}/{\rm dof} &= 0.875 C_{tG}^2+0.0173 C_b^+ C_{tG}+0.116 C_d^+ C_{tG}+0.172 C_u^+ C_{tG}-0.169 C_{tG}\nn+0.00296 \left(C_b^+\right){}^2+0.0212 \left(C_d^+\right){}^2+0.0707 \left(C_u^+\right){}^2+0.0277 C_b^+\nn+0.00498 C_b^+ C_d^++0.0815 C_d^++0.00581 C_b^+ C_u^++0.0655 C_d^+ C_u^++0.108 C_u^+\nn+0.803.
\end{align}
We then present the Fisher matrices for the 13.6-TeV projections for the $p_T$, $y$, and $p_T\times y$ distributions at LO, NLO, and a/NNLO. In analogy with the 13-TeV case, the LO and NLO results correspond to predictions evaluated at the respective fixed orders, while at a/NNLO the SM contribution is taken at NNLO and the SM-SMEFT interference at aNNLO. Since the 13.6-TeV study is based on pseudodata rather than measured data, the Fisher matrices provide the most direct characterization of the projected sensitivity and parameter correlations:
\begin{align}
    \mathcal F{}_{p_T}^{\rm LO} &= \left(
\begin{array}{cccc}
 14.1 & 1.65 & 1.11 & 0.0149 \\
 1.65 & 0.444 & 0.288 & 0.00272 \\
 1.11 & 0.288 & 0.187 & 0.00179 \\
 0.0149 & 0.00272 & 0.00179 & 0.0000197 \\
\end{array}
\right), \\
    \mathcal F{}_{p_T}^{\rm NLO} &= \left(
\begin{array}{cccc}
 65.1 & 5.14 & 3.06 & 0.616 \\
 5.14 & 0.841 & 0.575 & 0.109 \\
 3.06 & 0.575 & 0.405 & 0.0761 \\
 0.616 & 0.109 & 0.0761 & 0.0155 \\
\end{array}
\right), \\
    \mathcal F{}_{p_T}^{\rm a/NNLO} &= \left(
\begin{array}{cccc}
 144. & 8.70 & 4.85 & 0.925 \\
 8.70 & 1.60 & 1.08 & 0.198 \\
 4.85 & 1.08 & 0.764 & 0.138 \\
 0.925 & 0.198 & 0.138 & 0.0272 \\
\end{array}
\right), \\
    \mathcal F{}_{y}^{\rm LO} &= \left(
\begin{array}{cccc}
 20.5 & 1.12 & 0.788 & 0.0158 \\
 1.12 & 0.0654 & 0.0438 & 0.000818 \\
 0.788 & 0.0438 & 0.0304 & 0.000597 \\
 0.0158 & 0.000818 & 0.000597 & 0.0000125 \\
\end{array}
\right), \\
    \mathcal F{}_{y}^{\rm NLO} &= \left(
\begin{array}{cccc}
 90.8 & 3.19 & 2.19 & 0.0471 \\
 3.19 & 0.142 & 0.0908 & 0.00719 \\
 2.19 & 0.0908 & 0.0633 & 0.00418 \\
 0.0471 & 0.00719 & 0.00418 & 0.0101 \\
\end{array}
\right), \\
    \mathcal F{}_{y}^{\rm a/NNLO} &= \left(
\begin{array}{cccc}
 216. & 7.04 & 4.78 & 0.290 \\
 7.04 & 0.313 & 0.193 & 0.0215 \\
 4.78 & 0.193 & 0.133 & 0.0143 \\
 0.290 & 0.0215 & 0.0143 & 0.0231 \\
\end{array}
\right), \\
    \mathcal F{}_{p_T\times y}^{\rm LO} &= \left(
\begin{array}{cccc}
 67.8 & 8.54 & 5.56 & 0.0718 \\
 8.54 & 3.50 & 1.95 & 0.0142 \\
 5.56 & 1.95 & 1.16 & 0.00931 \\
 0.0718 & 0.0142 & 0.00931 & 0.0000986 \\
\end{array}
\right), \\
    \mathcal F{}_{p_T\times y}^{\rm NLO} &= \left(
\begin{array}{cccc}
 214. & 16.8 & 11.7 & 0.955 \\
 16.8 & 6.56 & 3.46 & 0.483 \\
 11.7 & 3.46 & 2.11 & 0.238 \\
 0.955 & 0.483 & 0.238 & 0.101 \\
\end{array}
\right), \\
    \mathcal F{}_{p_T\times y}^{\rm a/NNLO} &= \left(
\begin{array}{cccc}
 351. & 16.5 & 12.6 & 0.641 \\
 16.5 & 7.88 & 4.32 & 0.593 \\
 12.6 & 4.32 & 2.85 & 0.293 \\
 0.641 & 0.593 & 0.293 & 0.171 \\
\end{array}
\right).
\end{align}
We summarize the main fit results at 13 TeV in Tables~\ref{tab:13_chi2_bestfit}--\ref{tab:13_correlations}. Table~\ref{tab:13_chi2_bestfit} reports the fit quality in terms of $\chi^2/{\rm dof}$ together with the corresponding best-fit values of the Wilson coefficients for each distribution and perturbative order. Table~\ref{tab:13_bounds} gives the resulting 95\% CL intervals in both the nonmarginalized and marginalized cases, while Table~\ref{tab:13_scales} presents the corresponding effective scales, expressed as $\Lambda/\sqrt{\Delta C_k}$ in TeV, where $\Delta C_k$ indicates the 95\% CL variation of Wilson coefficient $C_k$. Here, nonmarginalized bounds are obtained by varying one Wilson coefficient at a time with all others fixed to zero, whereas marginalized bounds are derived by allowing all fitted coefficients to vary simultaneously and integrating out the remaining parameter directions. Finally, Table~\ref{tab:13_correlations} collects the correlation coefficients among the fitted Wilson coefficients, which provide a compact measure of the parameter degeneracies induced by the different observables.
\begin{table}
    [H]
    \centering
    \caption{The $\chi^2$ per degree of freedom and the best-fit values of the Wilson coefficients $C_{tG}$, $C_u^+$, $C_d^+$, and $C_b^+$ at 13 TeV.}
    \label{tab:13_chi2_bestfit}
    \begin{tabular}{|c|c|c|c|c|c|c|c|}
        \hline
        Distribution & Order & $\chi^2_{\rm SM} / {\rm dof}$ & $\chi^2_{\rm min} / {\rm dof}$ & $C_{tG}$ & $C_u^+$ & $C_d^+$ & $C_b^+$ \\
        \hline 
        \multirow{3}{*}{$p_T$} & LO & 8.6 & 0.029 & 5.6 & $-2.1\times 10^2$ & $3.8\times 10^2$ & $-8.3\times 10^3$ \\ 
        \cline{2-8}
        & NLO & 1.6 & 0.11 & 0.76 & 13 & --30 & --26 \\
        \cline{2-8}
        & a/NNLO & 0.60 & 0.23 & 0.16 & --0.40 & --2.9 & --5.8 \\ 
        \hline 
        \multirow{3}{*}{$y$} & LO & 8.6 & 0.011 & --16 & 23 & $3.0\times 10^2$ & $7.6\times 10^3$ \\
        \cline{2-8}
        & NLO & 1.7 & 0.020 & 0.20 & 5.1 & 3.7 & 3.2 \\ 
        \cline{2-8}
        & a/NNLO & 0.22 & 0.043 & 0.022 & 2.7 & 0.020 & 2.7 \\ 
        \hline 
        \multirow{3}{*}{$p_T \times y$} & LO & 0.93 & 0.099 & 3.2 & --4.9 & 7.7 & $-1.4\times 10^3$ \\
        \cline{2-8} 
        & NLO & 0.39 & 0.19 & 0.43 & 0.45 & --5.0 & --0.82 \\ 
        \cline{2-8}
        & a/NNLO & 0.80 & 0.63 & 0.28 & 0.32 & --2.8 & --3.5 \\ 
        \hline 
    \end{tabular}
\end{table}
Table~\ref{tab:13_chi2_bestfit} makes clear that perturbative accuracy is crucial for the interpretation of the SMEFT fit. For the $p_T$ and $y$ distributions, the LO SM predictions are far from the data, with $\chi^2_{\rm SM}/{\rm dof}\simeq 8.6$, and even the NLO description is still substantially poorer than at a/NNLO. In such cases, the fit naturally drives the Wilson coefficients to large values in order to bridge the gap between the low-order SM prediction and the data. These large best-fit values should therefore not be interpreted as evidence for large SMEFT effects, but rather as the fit attempting to absorb missing higher-order QCD corrections. This is especially true for the four-quark directions, which in our analysis represent particular linear combinations of operators in the SMEFT@NLO top basis and are included mainly to monitor their interplay with the extraction of $C_{tG}$. At a/NNLO, the SM already provides a much better description of the data, and correspondingly the preferred Wilson coefficients become much smaller, with $C_{tG}$ close to zero for all three observables. The $p_T\times y$ distribution is less dramatic in this respect, since its SM $\chi^2/{\rm dof}$ is already acceptable at lower orders, so the fitted shifts are more stable across perturbative orders. The main lesson of this table is therefore that the physically meaningful information on $C_{tG}$ comes from the highest-order analysis, whereas the LO and NLO fits are included mainly for completeness and to illustrate the extent to which low-order fits can mimic SMEFT effects.
\begin{table}
    [H]
    \centering
    \caption{Nonmarginalized and marginalized 95\% CL bounds $\Delta C_k$ on Wilson coefficients $C_{tG}$, $C_u^+$, $C_d^+$, and $C_b^+$ at 13 TeV.}
    \label{tab:13_bounds}
    \begin{tabular}{|c|c|c|c|c|c|c|c|c|c|}
        \hline 
        \multirow{2}{*}{Distribution} & \multirow{2}{*}{Order} & \multicolumn{4}{c|}{Nonmarginalized} & \multicolumn{4}{c|}{Marginalized} \\
        \cline{3-10} 
        & & $\Delta C_{tG}$ & $\Delta C_u^+$ & $\Delta C_d^+$ & $\Delta C_b^+$ & $\Delta C_{tG}$ & $\Delta C_u^+$ & $\Delta C_d^+$ & $\Delta C_b^+$ \\
        \hline 
        \multirow{3}{*}{$p_T$} & LO & 0.54 & 3.2 & 5.0 & $4.9\times 10^2$ & 18 & $2.9\times 10^3$ & $5.0\times 10^3$ & $7.1\times 10^4$ \\ 
        & NLO & 0.25 & 2.7 & 4.6 & 28 & 0.61 & 31 & 56 & 53 \\ 
        & a/NNLO & 0.17 & 2.2 & 3.7 & 18 & 0.38 & 22 & 40 & 36 \\
        \hline 
        \multirow{3}{*}{$y$} & LO & 0.43 & 7.4 & 11 & $5.6\times 10^2$ & $1.2\times 10^3$ & $4.9\times 10^2$ & $2.0\times 10^4$ & $4.8\times 10^5$ \\ 
        & NLO & 0.20 & 6.3 & 7.0 & 23 & 1.3 & 14 & 41 & 26 \\
        & a/NNLO & 0.13 & 4.6 & 5.0 & 16 & 0.76 & 9.7 & 27 & 18 \\ 
        \hline 
        \multirow{3}{*}{$p_T \times y$} & LO & 0.45 & 1.2 & 2.1 & $2.9\times 10^2$ & 1.1 & 7.5 & 19 & $1.5\times 10^3$ \\ 
        & NLO & 0.29 & 1.0 & 1.8 & 7.4 & 0.34 & 2.4 & 4.7 & 8.2 \\ 
        & a/NNLO & 0.23 & 0.82 & 1.5 & 4.0 & 0.26 & 1.6 & 3.0 & 4.3 \\
        \hline  
        \multirow{3}{*}{Combined} & LO & 0.27 & 1.1 & 1.9 & $2.3\times 10^2$ & 0.84 & 7.2 & 18 & $1.4\times 10^3$ \\
        & NLO & 0.14 & 0.93 & 1.6 & 6.8 & 0.17 & 2.3 & 4.4 & 7.3 \\ 
        & a/NNLO & 0.094 & 0.76 & 1.3 & 3.8 & 0.11 & 1.5 & 2.9 & 4.0 \\ 
        \hline 
    \end{tabular}
\end{table}
Table~\ref{tab:13_bounds} shows that the strongest and most stable sensitivity is obtained for $C_{tG}$, whereas the bounds on the four-quark directions remain comparatively weak, especially after marginalization and particularly at lower perturbative orders. This pattern is expected in our setup, since the main purpose of including the four-quark combinations is to monitor their interplay with the extraction of $C_{tG}$ rather than to derive competitive standalone limits on them. For all observables, the bounds improve substantially when going from LO to NLO and then to a/NNLO, reflecting the increased constraining power of the higher-order predictions once the SM baseline is brought under better control. At LO, the difference between nonmarginalized and marginalized limits can be very large, especially for the $p_T$ and $y$ distributions, indicating strong parameter degeneracies and an unstable fit structure. This effect is progressively reduced at NLO and becomes much milder at a/NNLO, where the marginalized bounds remain reasonably close to the nonmarginalized ones, most notably for $C_{tG}$. The same trend is seen in the combined fit, which yields the strongest limits overall and shows that the higher-order analysis leads not only to tighter bounds but also to a more stable determination of the Wilson coefficients against correlations among the fitted directions.
\begin{table}
    [H]
    \centering
    \caption{Effective scales $\Lambda/\sqrt{\Delta C_k}$ in TeV corresponding to the bounds presented in Table~\ref{tab:13_bounds} at 13 TeV.}
    \label{tab:13_scales}
    \begin{tabular}{|c|c|c|c|c|c|c|c|c|c|}
        \hline 
        \multirow{2}{*}{Distribution} & \multirow{2}{*}{Order} & \multicolumn{4}{c|}{Nonmarginalized} & \multicolumn{4}{c|}{Marginalized} \\
        \cline{3-10} 
        & & ${\Lambda \over \sqrt{\Delta C_{tG}}}$ & ${\Lambda \over \sqrt{\Delta C_u^+}}$ & ${\Lambda \over \sqrt{\Delta C_d^+}}$ & ${\Lambda \over \sqrt{\Delta C_b^+}}$ & ${\Lambda \over \sqrt{\Delta C_{tG}}}$ & ${\Lambda \over \sqrt{\Delta C_u^+}}$ & ${\Lambda \over \sqrt{\Delta C_d^+}}$ & ${\Lambda \over \sqrt{\Delta C_b^+}}$ \\
        \hline 
        \multirow{3}{*}{$p_T$} & LO & 1.4 & 0.56 & 0.45 & 0.045 & 0.23 & 0.019 & 0.014 & 0.0038 \\
        & NLO & 2.0 & 0.61 & 0.46 & 0.19 & 1.3 & 0.18 & 0.13 & 0.14 \\
        & a/NNLO & 2.5 & 0.67 & 0.52 & 0.24 & 1.6 & 0.21 & 0.16 & 0.17 \\
        \hline 
        \multirow{3}{*}{$y$} & LO & 1.5 & 0.37 & 0.30 & 0.042 & 0.029 & 0.045 & 0.0071 & 0.0014 \\
        & NLO & 2.2 & 0.40 & 0.38 & 0.21 & 0.89 & 0.27 & 0.16 & 0.19 \\
        & a/NNLO & 2.8 & 0.46 & 0.45 & 0.25 & 1.1 & 0.32 & 0.19 & 0.23 \\
        \hline 
        \multirow{3}{*}{$p_T \times y$} & LO & 1.5 & 0.91 & 0.68 & 0.059 & 0.94 & 0.36 & 0.23 & 0.026 \\
        & NLO & 1.9 & 1.0 & 0.74 & 0.37 & 1.7 & 0.65 & 0.46 & 0.35 \\
        & a/NNLO & 2.1 & 1.1 & 0.82 & 0.50 & 2.0 & 0.80 & 0.57 & 0.48 \\
        \hline  
        \multirow{3}{*}{Combined} & LO & 1.9 & 0.95 & 0.72 & 0.066 & 1.1 & 0.37 & 0.24 & 0.027 \\
        & NLO & 2.7 & 1.0 & 0.78 & 0.38 & 2.5 & 0.66 & 0.48 & 0.37 \\
        & a/NNLO & 3.3 & 1.1 & 0.86 & 0.51 & 3.0 & 0.82 & 0.59 & 0.50 \\
        \hline 
    \end{tabular}
\end{table}
Table~\ref{tab:13_scales} recasts the bounds of Table~\ref{tab:13_bounds} in terms of the effective scale $\Lambda/\sqrt{\Delta C_k}$ and therefore makes the same pattern especially transparent. The strongest reach is again obtained for $C_{tG}$, whose effective scale is pushed into the multi-TeV regime and improves systematically with perturbative order, reaching about $3~\TeV$ in the combined marginalized a/NNLO fit. By contrast, the four-quark directions correspond to much lower effective scales, reflecting the comparatively weak sensitivity of the present observables to these operator combinations. The large drop from nonmarginalized to marginalized scales at LO, and to a lesser extent at NLO, is another manifestation of the strong parameter degeneracies induced by low-order fits. At a/NNLO this gap becomes noticeably smaller, showing that the higher-order analysis not only strengthens the limits but also stabilizes their interpretation in terms of effective scales.
\begin{table}
    [H]
    \centering
    \caption{Correlations $\rho(C_k, C_{k'})$ among Wilson coefficients $C_{tG}$, $C_u^+$, $C_d^+$, and $C_b^+$ at 13 TeV.}
    \label{tab:13_correlations}
    \begin{tabular}{|c|c|c|c|c|c|c|c|}
        \hline
        Distribution & Order & $\rho(C_{tG}, C_u^+)$ & $\rho(C_{tG}, C_d^+)$ & $\rho(C_{tG}, C_b^+)$ & $\rho(C_u^+, C_d^+)$ & $\rho(C_u^+, C_b^+)$ & $\rho(C_d^+, C_b^+)$ \\
        \hline 
        \multirow{3}{*}{$p_T$} & LO & --0.98 & 0.98 & --0.99 & --1.0 & 1.0 & --1.0 \\ 
        & NLO & 0.75 & --0.80 & --0.78 & --0.99 & --0.78 & 0.79 \\ 
        & a/NNLO & 0.75 & --0.80 & --0.77 & --0.99 & --0.81 & 0.80 \\
        \hline 
        \multirow{3}{*}{$y$} & LO & --0.89 & --1.0 & --1.0 & 0.88 & 0.90 & 1.0 \\ 
        & NLO & --0.40 & --0.94 & --0.44 & 0.092 & 0.18 & 0.45 \\
        & a/NNLO & --0.41 & --0.94 & --0.45 & 0.11 & 0.20 & 0.46 \\ 
        \hline 
        \multirow{3}{*}{$p_T \times y$} & LO & --0.64 & 0.70 & --0.86 & --0.97 & 0.81 & --0.90 \\
        & NLO & 0.11 & --0.34 & 0.035 & --0.87 & 0.052 & --0.23 \\ 
        & a/NNLO & 0.024 & --0.25 & --0.038 & --0.82 & 0.13 & --0.26 \\
        \hline 
        \multirow{3}{*}{Combined} & LO & --0.12 & 0.17 & 0.84 & 0.20 & --0.59 & 0.11 \\
        & NLO & 0.44 & 0.73 & 0.25 & 0.67 & 0.095 & --0.0056 \\
        & a/NNLO & 0.41 & 0.69 & 0.29 & 0.65 & 0.087 & --0.0059 \\
        \hline 
    \end{tabular}
\end{table}
Table~\ref{tab:13_correlations} confirms that the lower-order fits are strongly affected by parameter degeneracies, especially for the single-differential $p_T$ and $y$ observables, where several correlations are close to $\pm 1$ at LO. This is fully consistent with the large separation between nonmarginalized and marginalized bounds seen earlier. Once higher-order corrections are included, the correlation pattern becomes noticeably more stable and less extreme, although some sizable correlations remain, in particular among the four-quark directions and between $C_{tG}$ and the light-quark combinations. The $p_T\times y$ distribution is again better behaved, with a clear reduction in the magnitude of most correlations already at NLO and a similar pattern persisting at a/NNLO. In the combined fit, the correlation structure is substantially moderated relative to the individual distributions, showing that the different observables lift part of the degeneracy present in each channel separately. Overall, the table shows that higher perturbative accuracy does not merely tighten the bounds, but also leads to a more robust fit geometry, which is particularly important here since the four-quark combinations are included mainly to assess their interplay with the extraction of $C_{tG}$ rather than as primary targets of the analysis.
\par 
The corresponding projection results at 13.6 TeV are summarized in Tables~\ref{tab:13.6_bounds}--\ref{tab:13.6_correlations}. Since this analysis is based on pseudodata, we focus directly on the projected bounds, the associated effective scales, and the correlation structure. In analogy with the 13-TeV case, Table~\ref{tab:13.6_bounds} lists the nonmarginalized and marginalized 95\% CL intervals, Table~\ref{tab:13.6_scales} converts these bounds into effective scales, and Table~\ref{tab:13.6_correlations} gives the corresponding correlation coefficients among the SMEFT parameters.
\begin{table}
    [H]
    \centering
    \caption{Nonmarginalized and marginalized 95\% CL bounds $\Delta C_k$ on Wilson coefficients $C_{tG}$, $C_u^+$, $C_d^+$, and $C_b^+$ at 13.6 TeV.}
    \label{tab:13.6_bounds}
    \begin{tabular}{|c|c|c|c|c|c|c|c|c|c|}
        \hline 
        \multirow{2}{*}{Distribution} & \multirow{2}{*}{Order} & \multicolumn{4}{c|}{Nonmarginalized} & \multicolumn{4}{c|}{Marginalized} \\
        \cline{3-10} 
        & & $\Delta C_{tG}$ & $\Delta C_u^+$ & $\Delta C_d^+$ & $\Delta C_b^+$ & $\Delta C_{tG}$ & $\Delta C_u^+$ & $\Delta C_d^+$ & $\Delta C_b^+$ \\
        \hline 
        \multirow{3}{*}{$p_T$} & LO & 0.53 & 2.9 & 4.5 & $4.4\times 10^2$ & 17 & $2.8\times 10^3$ & $4.9\times 10^3$ & $6.6\times 10^4$ \\ 
        & NLO & 0.25 & 2.1 & 3.1 & 16 & 0.49 & 20 & 28 & 58 \\ 
        & a/NNLO & 0.17 & 1.6 & 2.3 & 12 & 0.26 & 12 & 16 & 42 \\
        \hline 
        \multirow{3}{*}{$y$} & LO & 0.44 & 7.7 & 11 & $5.6\times 10^2$ & $1.1\times 10^3$ & $4.0\times 10^2$ & $1.8\times 10^4$ & $4.3\times 10^5$ \\ 
        & NLO & 0.21 & 5.3 & 7.9 & 19 & 0.53 & 18 & 31 & 21 \\
        & a/NNLO & 0.14 & 3.6 & 5.5 & 13 & 0.31 & 11 & 19 & 14 \\ 
        \hline 
        \multirow{3}{*}{$p_T \times y$} & LO & 0.24 & 1.1 & 1.8 & $2.0\times 10^2$ & 0.85 & 7.5 & 19 & $1.4\times 10^3$ \\ 
        & NLO & 0.14 & 0.77 & 1.4 & 6.2 & 0.17 & 2.3 & 4.1 & 7.7 \\ 
        & a/NNLO & 0.11 & 0.70 & 1.2 & 4.7 & 0.12 & 1.8 & 3.0 & 5.6 \\
        \hline  
        \multirow{3}{*}{Combined} & LO & 0.20 & 0.98 & 1.7 & $1.7\times 10^2$ & 0.74 & 7.2 & 18 & $1.2\times 10^3$ \\
        & NLO & 0.10 & 0.72 & 1.2 & 5.5 & 0.13 & 2.2 & 3.7 & 7.0 \\ 
        & a/NNLO & 0.076 & 0.63 & 1.0 & 4.2 & 0.084 & 1.7 & 2.7 & 5.0 \\ 
        \hline 
    \end{tabular}
\end{table}
Table~\ref{tab:13.6_bounds} shows that the 13.6-TeV projections largely preserve the hierarchy already seen at 13 TeV; that is, the strongest sensitivity is obtained for $C_{tG}$, while the four-quark directions remain significantly less constrained and are more affected by marginalization. Relative to the 13-TeV results, the projected bounds improve in a moderate but fairly systematic way, with the clearest gains appearing in the a/NNLO analysis and in the combined fit. In particular, the marginalized a/NNLO bound on $C_{tG}$ improves from $0.11$ to $0.084$ in the combined fit, with similar improvements visible in the individual $p_T$, $y$, and $p_T\times y$ channels. By contrast, the projected changes in the four-quark bounds are less uniform, which is consistent with their weaker intrinsic sensitivity and stronger interplay with correlations. Overall, the main message is that the 13.6-TeV setup sharpens the reach on $C_{tG}$ without qualitatively changing the fit structure already established at 13 TeV.
\begin{table}
    [H]
    \centering
    \caption{Effective scales $\Lambda/\sqrt{\Delta C_k}$ in TeV corresponding to the bounds presented in Table~\ref{tab:13_bounds} at 13.6 TeV.}
    \label{tab:13.6_scales}
    \begin{tabular}{|c|c|c|c|c|c|c|c|c|c|}
        \hline 
        \multirow{2}{*}{Distribution} & \multirow{2}{*}{Order} & \multicolumn{4}{c|}{Nonmarginalized} & \multicolumn{4}{c|}{Marginalized} \\
        \cline{3-10} 
        & & ${\Lambda \over \sqrt{\Delta C_{tG}}}$ & ${\Lambda \over \sqrt{\Delta C_u^+}}$ & ${\Lambda \over \sqrt{\Delta C_d^+}}$ & ${\Lambda \over \sqrt{\Delta C_b^+}}$ & ${\Lambda \over \sqrt{\Delta C_{tG}}}$ & ${\Lambda \over \sqrt{\Delta C_u^+}}$ & ${\Lambda \over \sqrt{\Delta C_d^+}}$ & ${\Lambda \over \sqrt{\Delta C_b^+}}$ \\
        \hline 
        \multirow{3}{*}{$p_T$} & LO & 1.4 & 0.58 & 0.47 & 0.047 & 0.24 & 0.019 & 0.014 & 0.0039 \\
        & NLO & 2.0 & 0.68 & 0.57 & 0.25 & 1.4 & 0.22 & 0.19 & 0.13 \\
        & a/NNLO & 2.4 & 0.80 & 0.67 & 0.29 & 2.0 & 0.29 & 0.25 & 0.15 \\
        \hline 
        \multirow{3}{*}{$y$} & LO & 1.5 & 0.36 & 0.30 & 0.042 & 0.031 & 0.050 & 0.0074 & 0.0015 \\
        & NLO & 2.2 & 0.44 & 0.36 & 0.23 & 1.4 & 0.23 & 0.18 & 0.22 \\
        & a/NNLO & 2.7 & 0.53 & 0.43 & 0.28 & 1.8 & 0.30 & 0.23 & 0.27 \\
        \hline 
        \multirow{3}{*}{$p_T \times y$} & LO & 2.0 & 0.97 & 0.74 & 0.071 & 1.1 & 0.37 & 0.23 & 0.027 \\
        & NLO & 2.7 & 1.1 & 0.86 & 0.40 & 2.4 & 0.66 & 0.50 & 0.36 \\
        & a/NNLO & 3.0 & 1.2 & 0.93 & 0.46 & 2.9 & 0.74 & 0.58 & 0.42 \\
        \hline  
        \multirow{3}{*}{Combined} & LO & 2.3 & 1.0 & 0.77 & 0.076 & 1.2 & 0.37 & 0.23 & 0.028 \\
        & NLO & 3.1 & 1.2 & 0.90 & 0.43 & 2.8 & 0.68 & 0.52 & 0.38 \\
        & a/NNLO & 3.6 & 1.3 & 0.99 & 0.49 & 3.4 & 0.76 & 0.61 & 0.45 \\
        \hline 
    \end{tabular}
\end{table}
Table~\ref{tab:13.6_scales} shows the same qualitative picture in terms of effective reach. The projected 13.6-TeV analysis improves the sensitivity to $C_{tG}$ most clearly, pushing the marginalized effective scale in the combined a/NNLO fit from about $3.0~\TeV$ at 13 TeV to about $3.4~\TeV$. More moderate gains are also seen for the light-quark combinations, while the reach on $C_b^+$ remains comparatively limited. As at 13 TeV, the strongest scales are obtained in the combined fit and at the highest perturbative order, with the $p_T\times y$ distribution providing the most powerful individual channel. Overall, the 13.6-TeV projections do not change the hierarchy among operator directions, but they extend the effective reach in a systematic way, most notably for the chromomagnetic operator that is the primary target of this analysis.
\begin{table}
    [H]
    \centering
    \caption{Correlations $\rho(C_i, C_j)$ among Wilson coefficients $C_{tG}$, $C_u^+$, $C_d^+$, and $C_b^+$ at 13.6 TeV.}
    \label{tab:13.6_correlations}
    \begin{tabular}{|c|c|c|c|c|c|c|c|}
        \hline
        Distribution & Order & $\rho(C_{tG}, C_u^+)$ & $\rho(C_{tG}, C_d^+)$ & $\rho(C_{tG}, C_b^+)$ & $\rho(C_u^+, C_d^+)$ & $\rho(C_u^+, C_b^+)$ & $\rho(C_d^+, C_b^+)$ \\
        \hline 
        \multirow{3}{*}{$p_T$} & LO & --0.98 & 0.98 & --0.99 & --1.0 & 1.0 & --1.0 \\ 
        & NLO & --0.77 & 0.70 & --0.15 & --0.92 & 0.042 & --0.41 \\ 
        & a/NNLO & --0.70 & 0.59 & 0.064 & --0.87 & --0.18 & --0.30 \\
        \hline 
        \multirow{3}{*}{$y$} & LO & --0.80 & --1.0 & --1.0 & 0.78 & 0.81 & 1.0 \\
        & NLO & --0.14 & --0.48 & 0.27 & --0.76 & --0.14 & --0.082 \\
        & a/NNLO & --0.097 & --0.51 & 0.23 & --0.75 & --0.051 & --0.17 \\
        \hline 
        \multirow{3}{*}{$p_T \times y$} & LO & --0.69 & 0.77 & --0.92 & --0.97 & 0.81 & --0.91 \\
        & NLO & 0.18 & --0.39 & 0.036 & --0.89 & --0.35 & 0.10 \\ 
        & a/NNLO & 0.10 & --0.27 & 0.061 & --0.88 & --0.34 & 0.11 \\
        \hline 
        \multirow{3}{*}{Combined} & LO & --0.20 & 0.13 & 0.92 & 0.26 & --0.52 & 0.079 \\ 
        & NLO & 0.64 & 0.61 & 0.35 & 0.93 & 0.59 & 0.59 \\ 
        & a/NNLO & 0.55 & 0.52 & 0.23 & 0.91 & 0.56 & 0.55 \\
        \hline 
    \end{tabular}
\end{table}
Table~\ref{tab:13.6_correlations} shows that the projected 13.6-TeV fits follow the same broad pattern as at 13 TeV, but with some notable redistribution of the correlation structure across observables and perturbative orders. At LO, the single-differential fits again exhibit nearly maximal correlations, confirming that low-order projections remain strongly affected by parameter degeneracies. Once NLO and especially a/NNLO corrections are included, the individual $p_T$, $y$, and $p_T\times y$ channels become visibly better behaved, with several correlations involving $C_{tG}$ reduced in magnitude relative to the 13-TeV case. At the same time, sizable correlations persist among the four-quark directions, particularly between $C_u^+$ and $C_d^+$, which is consistent with their weaker individual sensitivity. In the combined fit, however, the projected correlations at NLO and a/NNLO remain substantial, indicating that the stronger overall reach at 13.6 TeV does not eliminate the residual interplay among operator directions. Overall, the table confirms that the projected fit remains stable at higher perturbative orders, while the dominant correlation effects continue to arise from the auxiliary four-quark combinations included to test the robustness of the extraction of $C_{tG}$.
\par 
Finally, we combine the information from the 13-TeV analysis and the 13.6-TeV projections and summarize the resulting sensitivity in Tables~\ref{tab:13_13.6_bounds_scales} and \ref{tab:13_13.6_correlations}. Table~\ref{tab:13_13.6_bounds_scales} reports the corresponding nonmarginalized and marginalized 95\% CL bounds together with the associated effective scales $\Lambda/\sqrt{\Delta C_k}$. Table~\ref{tab:13_13.6_correlations} collects the resulting correlation coefficients among the fitted Wilson coefficients, thereby quantifying the residual parameter degeneracies in the combined analysis.
\begin{table}
    [H]
    \centering
    \caption{Nonmarginalized and marginalized 95\% CL bounds $\Delta C_k$ on Wilson coefficients $C_{tG}$, $C_u^+$, $C_d^+$, and $C_b^+$, together with the corresponding effective scales $\Lambda/\sqrt{\Delta C_k}$ in TeV, in the combined projection of the 13-TeV and 13.6-TeV analyses.}
    \label{tab:13_13.6_bounds_scales}
    \begin{tabular}{|c|c|c|c|c|c|c|c|c|}
        \hline 
        \multirow{2}{*}{Order} & \multicolumn{4}{c|}{Nonmarginalized} & \multicolumn{4}{c|}{Marginalized} \\ 
        \cline{2-9}
        & $C_{tG}$ & $C_u^+$ & $C_d^+$ & $C_b^+$ & $C_{tG}$ & $C_u^+$ & $C_d^+$ & $C_b^+$ \\ 
        \hline 
        LO & 0.16 & 0.74 & 1.3 & $1.4\times 10^2$ & 0.55 & 5.0 & 13 & $9.1\times 10^2$ \\ 
        NLO & 0.083 & 0.57 & 0.98 & 4.3 & 0.10 & 1.6 & 2.8 & 5.0 \\ 
        a/NNLO & 0.059 & 0.48 & 0.81 & 2.8 & 0.066 & 1.1 & 1.9 & 3.1 \\ 
        \hline 
        & ${\Lambda \over \sqrt{\Delta C_{tG}}}$ & ${\Lambda \over \sqrt{\Delta C_u^+}}$ & ${\Lambda \over \sqrt{\Delta C_d^+}}$ & ${\Lambda \over \sqrt{\Delta C_b^+}}$ & ${\Lambda \over \sqrt{\Delta C_{tG}}}$ & ${\Lambda \over \sqrt{\Delta C_u^+}}$ & ${\Lambda \over \sqrt{\Delta C_d^+}}$ & ${\Lambda \over \sqrt{\Delta C_b^+}}$ \\
        \hline 
        LO & 2.5 & 1.2 & 0.89 & 0.085 & 1.3 & 0.45 & 0.28 & 0.033 \\ 
        NLO & 3.5 & 1.3 & 1.0 & 0.48 & 3.2 & 0.80 & 0.60 & 0.45 \\ 
        a/NNLO & 4.1 & 1.4 & 1.1 & 0.60 & 3.9 & 0.95 & 0.72 & 0.57 \\ 
        \hline 
    \end{tabular}
\end{table}
\begin{table}
    [H]
    \centering
    \caption{Correlations $\rho(C_k, C_{k'})$ among Wilson coefficients $C_{tG}$, $C_u^+$, $C_d^+$, and $C_b^+$ in the combined projection of the 13-TeV and 13.6-TeV analyses.}
    \label{tab:13_13.6_correlations}
    \begin{tabular}{|c|c|c|c|c|c|c|}
        \hline 
        Order & $\rho(C_{tG}, C_u^+)$ & $\rho(C_{tG}, C_d^+)$ & $\rho(C_{tG}, C_b^+)$ & $\rho(C_u^+, C_d^+)$ & $\rho(C_u^+, C_b^+)$ & $\rho(C_d^+, C_b^+)$ \\ 
        \hline 
        LO & --0.68 & 0.77 & --0.93 & --0.97 & 0.79 & --0.90 \\ 
        NLO & 0.14 & --0.36 & 0.067 & --0.88 & --0.17 & --0.63 \\ 
        a/NNLO & 0.070 & --0.28 & 0.053 & --0.86 & --0.079 & --0.12 \\
        \hline 
    \end{tabular}
\end{table}
Taken together, Tables~\ref{tab:13_13.6_bounds_scales} and \ref{tab:13_13.6_correlations} show that combining the 13-TeV results with the 13.6-TeV projections leads to the strongest overall sensitivity, with the most significant improvement again occurring for $C_{tG}$. In the highest-order analysis, the combined marginalized bound reaches $\Delta C_{tG}=0.066$, corresponding to an effective scale of about $3.9~\TeV$, while the four-quark directions remain less constrained, although they also improve relative to the separate analyses. The gap between nonmarginalized and marginalized limits is now relatively modest for $C_{tG}$, indicating that its extraction is fairly stable in the combined fit. This is consistent with the correlation matrix, where most correlations involving $C_{tG}$ are small at NLO and a/NNLO, while the strongest residual degeneracy remains between $C_u^+$ and $C_d^+$. Overall, the combined analysis strengthens the reach without changing the qualitative picture established earlier.
\par 
We do not display two-parameter confidence ellipses, since in the present setup only $C_{tG}$ is constrained with meaningful precision, while the four-quark directions are included primarily to monitor correlation effects and remain comparatively weakly bounded. As a result, such contours would be dominated by broad, weakly constrained directions and would add little beyond the information already contained in the marginalized bounds and correlation matrices.
\par 
To assess the regime of validity of the EFT expansion, we assign to each fitted $p_T$ bin the hardness proxy
\begin{gather}
    Q(p_{T,\max}) = 2\sqrt{m_t{}^2+p_{T,\max}{}^2},
\end{gather}
where $p_{T,\max}$ is the upper edge of the bin. This choice is motivated by the fact that, for a given top transverse momentum, the minimal hard scale of the underlying $t\bar t$ system is set by twice the top transverse mass, while the rapidity is not itself a hardness variable. We then define
\begin{gather}
    \epsilon = \frac{Q}{\Lambda_{\rm eff}} = \frac{Q\sqrt{\Delta C_k}}{\Lambda},
\end{gather}
with $\Lambda=1~\TeV$ and $\Lambda_{\rm eff}=\Lambda/\sqrt{\Delta C_k}$ as above. Using the combined marginalized effective scales, we find $\Lambda_{\rm eff}\simeq 3.9~\TeV$ for the best-constrained direction, namely $C_{tG}$, and $\Lambda_{\rm eff}\simeq 0.57~\TeV$ for the weakest one, namely $C_b^+$. This gives $\epsilon \simeq 0.091$--$0.42$ for $C_{tG}$ across the fitted $p_T$ range, indicating that the EFT expansion is parametrically under good control for the operator that is the main target of this analysis. By contrast, for $C_b^+$ we obtain $\epsilon \simeq 0.62$--$2.9$, showing that the weakest four-quark direction cannot be interpreted as a robust stand-alone EFT reach up to the highest bins. We emphasize that this is not a tension with the main goal of the fit on the grounds that the combinations $C_u^+$, $C_d^+$, and $C_b^+$ are included primarily to quantify correlation effects in the extraction of $C_{tG}$, rather than because the present dataset is expected to determine each of them sharply on its own. Since our analysis is performed at linear order in the SMEFT expansion, this test should be viewed as an \textit{a posteriori} consistency check, and it supports the use of the extracted bound on $C_{tG}$ as the physically meaningful result of the fit.
\par 
Before closing this section, we place our result in the context of several representative determinations in the literature, collected in Table~\ref{tab:lit_comparison}. These results are not directly identical in setup or scope and should therefore be compared with appropriate care. Our analysis is based on differential $t\bar t$ information and a four-parameter SMEFT fit, with the additional four-quark directions included mainly to assess their impact on the extraction of $C_{tG}$. By contrast, the HEPfit~\cite{Cornet-Gomez:2025jot} and SMEFiT~\cite{Celada:2024mcf} results come from broad global analyses with substantially larger operator spaces and many complementary datasets, so part of their constraining power comes from the ability to lift degeneracies across different sectors. The SFitter result~\cite{Brivio:2019ius} is a dedicated global top-sector analysis at NLO in QCD including quadratic SMEFT terms across 22 parameters. The $tW$ result~\cite{Kidonakis:2026fle} corresponds to projections using the top-quark $p_T\times y$ double-differential distribution, while the $t\bar t$ cross section~\cite{Kidonakis:2023htm} represents a dedicated one-parameter extraction from an inclusive top-pair cross-section measurement~\cite{ATLAS:2023gsl, CMS:2021vhb}. The comparison is therefore meant to provide orientation on the relative size of current constraints on $C_{tG}$, rather than a strict like-for-like ranking of analyses.
\begin{table}
    [H]
    \centering
    \caption{Comparison of representative 95\% CL constraints on the top chromomagnetic dipole operator $C_{tG}$ from this work and selected recent analyses in the literature, together with the corresponding effective scales in the \smeftatnlo~convention.}
    \label{tab:lit_comparison}
    \begin{tabular}{|l|c|c|c|c|c|}
        \hline
         & \begin{tabular}{@{}c@{}}SMEFT\\order\end{tabular} & Accuracy & \begin{tabular}{@{}c@{}}Number of\\parameters\end{tabular} & $\Delta C_{tG}$ & $\Lambda/\sqrt{\Delta C_{tG}}~[\TeV]$ \\
        \hline
        This work & Linear & a/NNLO & 4 & 0.066 & 3.9 \\
        \hline
        Global fit with HEPfit~\cite{Cornet-Gomez:2025jot} & Linear & NLO & 22 & 0.32 & 1.8 \\
        \hline
        Global fit with SMEFiT~\cite{Celada:2024mcf} & Quadratic & NLO & 50 & 0.081 & 3.5 \\
        \hline
        Global fit with SFitter~\cite{Brivio:2019ius} & Quadratic & NLO & 22 & 0.34 & 1.7 \\
        \hline
        $tW$ $p_T\times y$ distribution~\cite{Kidonakis:2026fle} & Quadratic & aNNLO & 3 & 0.60 & 1.3 \\
        \hline
        $t\bar t$ cross section~\cite{Kidonakis:2023htm} & Quadratic & aNNLO & 1 & 0.25 & 2.0 \\
        \hline
    \end{tabular}
\end{table}
At the level of a direct comparison, Table~\ref{tab:lit_comparison} shows that our combined a/NNLO result yields the strongest quoted constraint on $C_{tG}$ among the entries listed, with a marginalized bound of $0.066$ corresponding to an effective scale of about $3.9~\TeV$. This is substantially stronger than the representative HEPfit global result, the dedicated $tW$ differential fit, and the inclusive $t\bar t$ cross-section extraction, and is also somewhat stronger than the SMEFiT bound. The SFitter bound is weaker, which is consistent with the fact that their analysis relies on a broader operator space and does not include the higher-order QCD corrections that stabilize the extraction of $C_{tG}$ in our fit. As a simple estimate of complementarity, one may combine independent single-parameter sensitivities through the harmonic average of the squared $1\sigma$ marginalized bounds. Under this approximation, our result would improve the HEPfit bound by about $60\%$ and the SMEFiT bound by about $24\%$. Although this estimate is only illustrative and does not replace a genuine global combination, it nonetheless indicates that the differential $t\bar t$ information considered here provides nontrivial additional sensitivity to $C_{tG}$.

\section{Conclusion\label{sec:conclusion}}
In this work, we studied the hadronic process $pp \to t\bar t$ within the dimension-six SMEFT, focusing on the top chromomagnetic operator together with the three four-quark directions $C_u^+$, $C_d^+$, and $C_b^+$ that arise as the relevant linear combinations in the \smeftatnlo\ top-basis description of unpolarized $t\bar t$ production. Our analysis was based on the single-differential top-quark $p_T$ and $y$ distributions, as well as the double-differential $p_T\times y$ distribution, at 13~TeV, together with a projection to 13.6~TeV using the same binning. For the perturbative predictions, we combined NNLO SM results with aNNLO SM-SMEFT interference corrections, and we also presented the corresponding LO and NLO results in order to assess the impact of perturbative accuracy on the extracted SMEFT sensitivity.
\par 
On the experimental side, we used the published 13-TeV information as directly as possible. For the single-differential distributions, where no unfolded covariance matrix is currently available, we constructed the experimental covariance from the reported uncertainty components under a bin-by-bin uncorrelated approximation. For the double-differential measurement, we used the published covariance matrix and rescaled it consistently to an approximate inclusive normalization within the narrow-width approximation. For the 13.6-TeV study, we adopted the same observables and bin definitions and completed the projection by constructing pseudodata from the NNLO SM baseline together with a simple assumed experimental covariance model. In all cases, PDF and scale uncertainties were incorporated through a separate theory covariance matrix, with the PDF contribution treated as fully correlated and the scale contribution estimated from 7-point variations.
\par 
Our results show clearly that perturbative accuracy is essential for a meaningful SMEFT interpretation of $t\bar t$ differential data. At low perturbative order, especially for the single-differential $p_T$ and $y$ observables, the fit tends to absorb missing higher-order QCD effects into shifts of the Wilson coefficients, leading to artificially large best-fit values and strong parameter degeneracies. Once the SM baseline is upgraded to NNLO and the linear SMEFT contribution to aNNLO, the fit becomes substantially more stable. The preferred values of the Wilson coefficients move close to zero, the separation between nonmarginalized and marginalized limits is reduced, and the correlation pattern becomes significantly better behaved. In this sense, the highest-order analysis is the only one that should be regarded as quantitatively robust, while the LO and NLO fits are mainly useful for illustrating how strongly low-order predictions can mimic SMEFT effects.
\par 
The strongest and most stable sensitivity is obtained for the chromomagnetic coefficient $C_{tG}$, which is the primary target of the analysis. In the combined 13-TeV and 13.6-TeV study, our highest-order fit yields a marginalized 95\% CL bound $\Delta C_{tG} = 0.066$ which corresponds to an effective scale of 3.9~TeV. This is the main phenomenological result of the paper. By contrast, the three four-quark directions remain much more weakly constrained and should not be viewed as competitive standalone determinations. Their main role here is to quantify the extent to which such additional SMEFT directions can affect the extraction of $C_{tG}$ through correlations, and to test the robustness of the chromomagnetic bound in a multiparameter setting.
\par 
We also assessed the EFT-expansion regime \textit{a posteriori} through a simple hardness estimate based on the fitted $p_T$ range. This check indicates that the extracted sensitivity to $C_{tG}$ lies in a parametrically controlled regime over the bins relevant to the fit, whereas the weakest four-quark direction extends into a much less reliable EFT domain in the highest-energy bins. This again supports the interpretation that the physically meaningful outcome of the present analysis is the bound on $C_{tG}$, while the auxiliary four-quark combinations are primarily included as correlation directions rather than precision targets in their own right.
\par 
Finally, when compared with representative recent determinations in the literature, our result is numerically very competitive. In particular, our combined highest-order bound on $C_{tG}$ is substantially stronger than the representative HEPfit result and also somewhat stronger than the quoted SMEFiT sensitivity after translating to a common convention. The sensitivity obtained here would naively improve the representative HEPfit constraint by about $60\%$ and the representative SMEFiT constraint by about $24\%$. While this estimate is only indicative and does not replace a genuine global combination, it nonetheless shows that differential $t\bar t$ information at high perturbative accuracy provides nontrivial additional leverage on the top chromomagnetic operator.
\par 
Overall, our study shows that differential $t\bar t$ production, treated at the level of NNLO SM predictions and aNNLO SM-SMEFT interference, provides a powerful and theoretically controlled probe of the top chromomagnetic interaction. The main lesson is that once perturbative QCD effects are brought under sufficient control, the available and projected LHC differential data can constrain $C_{tG}$ very strongly, with sensitivity reaching the multi-TeV regime. This makes high-precision $t\bar t$ phenomenology an important ingredient in the broader SMEFT program and a useful source of complementary information for future global analyses.

\textbf{Acknowledgments:} This material is based upon work supported by the National Science Foundation under Grant No. PHY 2412071. K\c{S} is supported by the Kennesaw State University Office of Research Postdoctoral Fellowship Program. This work was supported in part by research computing resources and technical expertise via a partnership between Kennesaw State University's Office of the Vice President for Research and the Office of the CIO and Vice President for Information Technology.  

\textbf{Data availability:} The data associated with this article are available
at Ref.~\cite{Simsek2026-os}.

\bibliography{refs}

\end{document}